\documentstyle[aps,epsf,graphicx]{revtex}
\newcommand{\kcrit}{\mbox{$\kappa_{\rm crit}$}}
\newcommand{\ksea}{\mbox{$\kappa^{\rm sea}$}}
\newcommand{\kval}{\mbox{$\kappa^{\rm val}$}}

\newcommand{\kvala}{\mbox{$\kappa^{\rm val}_a$}}
\newcommand{\kvalb}{\mbox{$\kappa^{\rm val}_b$}}
\newcommand{\ltap}{\;\raisebox{-.5ex}{\rlap{$\sim$}} \raisebox{.5ex}{$<$}\;}
\newcommand{\gtap}{\;\raisebox{-.5ex}{\rlap{$\sim$}} \raisebox{.5ex}{$>$}\;}
\newcommand{\e}{{\rm e}}

\def\fm{{\rm fm}}
 
\def\psibar{\overline{\psi}}
 
\def\csw{c_{\rm sw}}

\newcommand{\be}{\begin{equation}}
\newcommand{\ee}{\end{equation}}
\newcommand{\bea}{\begin{eqnarray}}
\newcommand{\eea}{\end{eqnarray}}

\newcommand{\plus}{\makebox[15pt][c]{$+$}}
\newcommand{\minus}{\makebox[15pt][c]{$-$}}
\newcommand{\figurebox}[2]{\fbox{\vbox to#2in{\hbox to #1in{\hfil}\vfil}}}
\newcommand{\err}[2]{\raisebox{0.08em}{\scriptsize
                          {$\;\begin{array}{@{}l@{}}
                          \plus\makebox[0.55em][r]{#1} \\[-0.12em]
                          \minus\makebox[0.55em][r]{#2}
                        \end{array}$}}}
\newcommand{\texp}{\tau^{\rm exp}}
\newcommand{\tint}{\tau^{\rm int}}

\newcommand{\bm}[1]{\mbox{\boldmath ${#1}$}}

\newcommand{\rvec}{\bm{r}}

\newcommand{\kvec}{\bm{k}}
\newcommand{\gtaeq}{\raisebox{-.6ex}{$\stackrel{\textstyle{>}}{\sim}$}}
\newcommand{\Tmin}{t_{\rm min}}
\newcommand{\Tmax}{t_{\rm max}}
\newcommand{\mps}{M_{\rm PS}}
\newcommand{\mv}{M_{\rm V}}

\newcommand{\half}{\frac{1}{2}}
\newcommand{\Tr}{\mathop{\rm Tr}}

\newcommand{\chinfty}{\chi^\infty}
\newcommand{\chiqu}{\chi^{\rm qu}}

\def\lsi{\raise0.3ex\hbox{$<$\kern-0.75em\raise-1.1ex\hbox{$\sim$}}}
\def\gsi{\raise0.3ex\hbox{$>$\kern-0.75em\raise-1.1ex\hbox{$\sim$}}}

\newcommand{\MeV}{\mathop{\rm MeV}}
\newcommand{\eff}{f}
\begin{document}

\begin{titlepage}
 
\begin{flushright}
Cambridge Preprint DAMTP--2001--15\\
Edinburgh Preprint 2001/09\\
Liverpool Preprint LTH509\\
Oxford Preprint OUTP--01--37P\\
Swansea Preprint SWAT/307\\
%
%DRAFT $Date: 2001/11/28 09:44:34 $
%\\

\end{flushright}
 
\vspace*{3mm}
 
\begin{center}
{\Huge 
Effects of non-perturbatively improved dynamical fermions
in QCD at fixed lattice spacing
}\\[12mm]
{\large\it UKQCD Collaboration}\\[3mm]

{\bf C.R.~Allton}$^1$, 
{\bf S.P.~Booth}$^2$, 
{\bf K.C.~Bowler}$^3$,
{\bf J.~Garden}$^3$, 
{\bf A.~Hart}$^{3,4}$, 
{\bf D.~Hepburn}$^3$, 
{\bf A.C.~Irving}$^5$,
{\bf B.~Jo\'o}$^3$, 
{\bf R.D.~Kenway}$^3$, 
{\bf C.M.~Maynard}$^3$,
{\bf C.~McNeile}$^5$,
{\bf C.~Michael}$^5$,
{\bf S.M.~Pickles}$^3$, 
{\bf J.C.~Sexton}$^6$,
{\bf K.J.~Sharkey}$^5$,
{\bf Z.~Sroczynski}$^3$, 
{\bf M.~Talevi}$^3$,
{\bf M.~Teper}$^7$,
{\bf H.~Wittig}$^{7,5}$ \\

\vspace{8mm}

$^1$\,Department of Physics, University of Wales Swansea, Swansea
SA2~8PP, Wales\\

$^2$ Edinburgh Parallel Computing Centre, University of
Edinburgh, Edinburgh EH9~3JZ, Scotland\\

$^3$\,Department of Physics \& Astronomy, University of Edinburgh,
Edinburgh EH9~3JZ, Scotland\\

$^4$\,Department of Applied Mathematics and Theoretical Physics, 
University of Cambridge,
Cambridge CB3~0WA, England\\

$^5$\,Division of Theoretical Physics, Department of Mathematical
Sciences, University of Liverpool, Liverpool L69~3BX, England\\

$^6$\,School of Mathematics, Trinity College, Dublin~2,
and
Hitachi Dublin Laboratory, Dublin~2, Ireland\\

$^7$\,Theoretical Physics, University of Oxford, Oxford OX1~3NP, England\\
\end{center}

\vspace{3mm}
\begin{abstract}
  We present results for the static inter-quark potential, lightest
  glueballs, light hadron spectrum and topological susceptibility
  using a non-perturbatively improved action on a $16^3\times 32$
  lattice at a set of values of the bare gauge coupling and bare
  dynamical quark mass chosen to keep the lattice size fixed in
  physical units ($\sim 1.7$ fm).  By comparing these measurements
  with a matched quenched ensemble, we study the effects due to two
  degenerate flavours of dynamical quarks. With the greater control
  over residual lattice spacing effects which these methods afford, we
  find some evidence of charge screening and some minor effects on the
  light hadron spectrum over the range of quark masses studied
  ($M_{PS}/M_{V}\ge0.58$).  More substantial differences between
  quenched and unquenched simulations are observed in measurements of
  topological quantities.

\end{abstract}

\end{titlepage}

\section{Introduction}
\label{sec:intro}
Over recent years, considerable effort has gone into probing QCD
beyond the quenched approximation. For recent reviews see
\cite{Guesken_lat97,Burkhalter:1998wu,Mawhinney:2000fw,Aoki:2000kp} and for
results using a different improvement scheme see\cite{AliKhan:2001tx}.
Because of the impressive agreement of the quenched approximation (see
e.g.~\cite{Kenway:1998ew}) with experiment for the spectrum and other
easily accessible quantities, the effects of dynamical quarks in these
are expected to be quite small. It is difficult to isolate physical
effects which are unambiguously due to their inclusion, in part
because of the need for high statistics. On currently available
machines this requires coarse lattices. The use of $O(a)$
non-perturbatively improved fermions has been suggested as a means of
controlling and reducing discretisation errors~\cite{Jansen:1998mx}.
In an earlier paper~\cite{Allton:1998gi}, first results of the {\it
  UKQCD Collaboration} using a preliminary value of the improvement
coefficient $\csw$ were presented.  It was found that the effective
lattice spacing, as measured by Sommer's intermediate scale parameter
$r_0$~\cite{Sommer:1994ce}, depended quite strongly on the bare quark
mass at fixed gauge coupling.  However, the effect of dynamical quarks
on easily accessible physical observables was very weak and difficult
to disentangle from those induced by other changes in the simulation
parameters.  Eventually, one might hope to perform detailed studies
over the full space of parameters including bare gauge coupling, quark
mass(es) and lattice volume. In the meantime, less ambitious studies
may still serve as a guide to those regions of parameter space where
physical effects may be found.

In this paper, we present results of further simulations over a range
of sea quark masses.  For these simulations, we have used the final
published values of the $O(a)$ improvement coefficient
$\csw$~\cite{Jansen:1998mx} and have attempted to reduce variations
due to residual discretisation errors and finite volume effects by
working at fixed lattice spacing.  In order to achieve the latter, we
have used matching techniques described in an earlier
work~\cite{Irving:1998yu} to help obtain ensembles of configurations
whose lattice spacings, as defined by the scale $r_0$, are as closely
matched as practicable.  We present results for the spectrum and
potential on, or close to, a single fixed $r_0$ trajectory in the
$(\beta,\kappa)$ plane which extends from quenched configurations
($\kappa=0$) to the lightest accessible sea quark mass.  We choose
$r_0$ to set the scale since it has no valence quark complications and
is determined by intermediate scale properties of the static
potential.  These properties are expected to be less sensitive to
charge screening (short range) and string-breaking (long range)
effects arising from dynamical light quarks.

We interpret our results in the spirit of partial quenching.  That is,
we study chiral extrapolation in the valence quark masses of light
hadron masses using both quenched and partially unquenched
configurations.  We find that, with the available statistics, the
quality of these valence extrapolations is uniformly good. By studying
the spectra so obtained we search for evidence of the influence of
light dynamical quarks. We also study the behaviour of the topological
susceptibility in the presence of dynamical quarks.  Our data sample
includes measurements made with equal valence and sea quark masses.

The plan of the rest of the paper is as follows.
Section~\ref{sec:simul} contains brief details of the simulation
methods and parameters.  In section~\ref{sec:match}, we review the
matching techniques used to set up simulations at similar lattice
spacings.  We present results in section~\ref{sec:pot} for the static
potential in QCD and use it to define a lattice scale.  In
section~\ref{sec:spectrum}, we present results for the light hadron
spectrum including some measurements of the lightest glueball masses.
Section~\ref{sec:topol} contains results from topological charge and
susceptibility measurements.  
Finally, our conclusions are summarised in section~\ref{sec:conc}.

Some preliminary results from these analyses have 
been presented 
elsewhere
\cite{Garden:1999hs,Hart:1999hy,Hart:2000wr,Irving:2000hs,hart_wien,Allton:2000cy}.
%
% Simulations
%
\section{Simulations with improved Wilson fermions}
\label{sec:simul}
Details of our implementation of the Hybrid Monte Carlo simulation 
algorithm~\cite{Duane:1987de}
and its performance can be found in our earlier paper~\cite{Allton:1998gi}. 
Here, we summarise for convenience 
some key features. For the lattice action we used  a standard Wilson
action for the gauge fields 
together with 
the Sheikholeslami-Wohlert
$O(a)$-improved Wilson gauge-fermion action~\cite{Sheikholeslami:1985ij}:
\be
   S[U,\psibar,\psi] = S_G[U]+S_F[U,\psibar,\psi],
\ee
where
\be
   S_G[U] = \beta W_{\Box}
        =\beta\sum_P\left(1-\frac{1}{3}{\rm Re\,Tr\,}U_P\right)
\ee
and
\be
   S_F[U,\psibar,\psi] = S_F^W[U,\psibar,\psi] +
   \csw\frac{i\kappa}{2} \sum_{x,\mu,\nu}
   \psibar(x)\sigma_{\mu\nu}F_{\mu\nu}(x)\psi(x).
\ee
Here, $U_P$ is the usual directed product of gauge link variables and
$S_F^W$ is the standard Wilson fermion action,
\be
S_F^W = \sum_x \psibar(x)\psi(x) -\kappa \sum_{x,\mu} \left(
\psibar(x) (r-\gamma_\mu) U_\mu(x) \psi(x+\hat{\mu}) + 
\psibar(x+\hat{\mu}) (r+\gamma_\mu) U_\mu^\dagger(x) \psi(x)
\right)
\ee
with the Wilson parameter chosen as $r=1$. The spin matrix is
$\sigma_{\mu \nu} = \frac{i}{2} \left[ \gamma_\mu,\gamma_\nu \right]$,
and $F_{\mu\nu}(x)$ is the field strength tensor
\be 
F_{\mu\nu}(x) = \frac{1}{8} \left(f_{\mu\nu}(x) -
  f^\dagger_{\mu\nu}(x) \right)
\ee
where $f_{\mu\nu}(x) = U_{\mu \nu}(x) + U_{\nu,-\mu}(x) +
    U_{-\nu,-\mu}(x) + U_{-\nu,\mu}(x)$ is the sum of four
        similarly oriented (open) plaquettes around a site, 
	$x$~\cite{Sheikholeslami:1985ij}.
        
Beyond tree level, the improvement coefficient $\csw$ is a
function of the gauge coupling $\beta$($\equiv 6/g^2$). In the
studies reported here, we have used those values determined
non-perturbatively by the Alpha Collaboration and summarised
by an interpolation formula~\cite{Jansen:1998mx}. For example,
at $\beta=5.20$ we have used $\csw=2.0171$~%
\footnote{Although the effect of $O(a)$ improvement 
  is not expected to be as sensitive as the quoted
  number of significant figures suggests, the action and lattice
  observables do depend quite strongly on this parameter.  For reasons
  of reproducibility we have therefore used a 4 decimal place
  representation of the $\csw$ formula in generating configurations.}.

We have used two degenerate flavours of dynamical quarks in these
simulations. The bare quark mass is controlled by the hopping
parameter $\kappa$. Restoration of (spontaneously broken) chiral
symmetry, requires extrapolation in $\kappa$ to the critical value
$\kcrit$ at which the pion is effectively massless. As discussed
above, we will often discuss the situation encountered in the quenched
approximation, where the dynamical (sea) quark mass parameter
($\ksea$) is fixed (at $0$ in the quenched case) while the chiral
extrapolation is performed in the valence mass parameter ($\kval$)
only.  This is often referred to as a partially quenched
approximation.  It is particularly relevant where the dynamical quark
mass is still quite heavy and where there is no realistic prospect of
approaching the (degenerate) light quark chiral limit in both
parameters.

\subsection{Simulation parameters}
Since these simulations were the first to be done on a reasonably
large lattice ($16^3\times32$) using the fully improved value of
$\csw$, there was little guidance available on the choice of
simulation parameters. We chose $\beta=5.20$ as the lowest value at
which a reliable value of $\csw$ was available~\cite{Jansen:1998mx}.
The aim was to obtain as large a physical volume as practicable with
the available computing resource. The use of an improved action was
expected to offset (at least partially) the relatively coarse lattice
spacing which this implied. Equilibration was carried out through a
sequence of dynamical quark masses: $\ksea=0.13000$, $0.13350$,
$0.13400$, $0.13450$ to $0.13500$. The first production run was then
carried out at $\ksea=0.13500$ starting at trajectory number 10010,
where trajectories were of unit length.  Configurations were stored
after every 10 trajectories although a larger separation was used for
most operator measurements (see below).
 
Further simulations at higher quark masses ($\ksea =0.13450$, and
$0.13400$) and slightly shifted $\beta$ were then performed. The
shifts in $\beta$ were estimated using the methods described in
section~\ref{sec:match} and were designed to maintain a constant
lattice spacing as defined by $r_0$.

To complete the comparison of unquenching effects, we performed pure
gauge simulations using a standard update algorithm, heat-bath with
over-relaxation.  Again, the $\beta$ value was chosen to keep $r_0$ at
the value measured on the ensemble obtained at $(\beta,\ksea)
=(5.20,0.13500)$. The only exceptional configuration found was within
the quenched configurations, and this was only apparent for one of the
$\kval$ studied This configuration was excluded from further analyses.

An additional substantial, but unmatched, simulation was then
performed at $(\beta,\ksea) =(5.20,0.13550)$. This ensemble of
configurations was analysed along with the matched ensembles providing
further information on behaviour at light quark mass. A simulation at
even lighter quark mass ($\ksea=0.13565$) was begun.  Where relevant,
some preliminary results are presented here.
Table~\ref{tb:config_sum} contains a summary of the run parameters
for each ensemble.

The bulk of the simulations were carried out in double precision. This
followed initial concerns over the effect of rounding errors on
reversibility.  Detailed analysis of these and related effects have
been carried out and have been reported elsewhere~\cite{Joo:2000dh}.
This work shows that, at least for present volumes and step lengths,
the algorithm is reversible and stable for all practical purposes,
even when implemented in single precision.

\subsection{Autocorrelations}
We made autocorrelation measurements from the average plaquette value
measurements on every trajectory.  The methods used were those
described in detail in our earlier paper~\cite{Allton:1998gi}.  As
shown in Table~\ref{tb:autocor}, the observable autocorrelation (from
the plaquette) is of order 20 and so we have adopted a separation of
40 trajectories as standard in the analysis which follows.  
Nevertheless we keep in mind that subtle longer-term autocorrelations,
not directly measurable, may still be present and so we have done
additional checks on our statistical error estimates by re-binning the
measurements.  In the present data sample, we have not found any
evidence of such correlations.

Further measurements of the integrated autocorrelation time have been
attempted for the potential (section~\ref{sec:pot}) and the scalar
glueball (section~\ref{sec:gball}). At the lightest quark mass
($\ksea=0.13565$) autocorrelations were estimated from effective mass
(potential energy) measurements made every 20 trajectories at various
lattice distances ($r/a=1-5$) and Euclidean times ($t/a=3-5$). The
measured integrated autocorrelation times varied from 10 to 20
trajectories with large errors (typically $\pm 8$).  For the scalar
glueball, the integrated autocorrelation time for effective masses was
in the range $25$ to $30$ at $\ksea=0.13500$ and $0.13550$.

It is noteworthy that the autocorrelation is significantly {\em less}
in the current simulations than in our previous
runs~\cite{Allton:1998gi} at comparable quark masses but different
$\csw$.  The current simulations use the fully non-perturbatively
improved value of $\csw$. It is further noted that $\tint$ appears to
{\em decrease}, if anything, with decreasing quark mass. This is
contrary to the simple expectation that, as the lattice correlation
length (typically given by the inverse pion mass) increases, then so
should the correlation in computer time. A similar effect is evident
in the decorrelation properties of the topological charge (see
section~\ref{sec:topol}). It is possible to reproduce such behaviour
in simple models. The integrated autocorrelation time, which
determines the size of the errors, can decrease even in the presence
of increasingly long correlation modes simply due to increased noise
induced by dominant short correlation modes. 

To illustrate this point, consider first the
following simple model 
consisting of a single 
Markov chain $x(t),t=0,1,2,\dots$
\be
x(t) = ax(t-1) + z(t)\, ,\qquad x(0)=0\, , 
\label{eq:model1}
\ee
where the $z(t)$ are uncorrelated Gaussian noise of unit variance
and $0<a<1$.
It is simple to show that, for sufficiently long chains,
\bea
\rho_x(t)& =& a^t\equiv e^{-t/\texp(x)}\,\quad \hbox{and so}\\
\quad\texp(x)& =& -\ln a\, , \quad
\tint(x)=\half {{1+a}\over{1-a}}\, .
\eea
Here $\rho_x(t)$ is the normalised autocorrelation function for
the observable $x$.
The corresponding results for finite length chains are also 
calculable so one can study the effects
of using limited statistics to estimate autocorrelation
times. Here we stick to the infinite chain approximation.
For $a\lesssim 1$ (i.e. for large $\texp(x)$), 
\be \tint(x)=\texp(x)+1/(12\texp(x))
+{\cal O}((\texp)^{-3})\, .
\ee
Real HMC data for $\rho(t)$ do not, of course, 
show a simple exponential behaviour
and so it is useful to consider the next simplest model which contains
two independent correlation modes with relative coupling strength $r$:
\bea
\label{eq:model2}
 X(t) & =& x_1(t)+rx_2(t)\, ,\\
 x_i(t) & =& a_ix_i(t-1) + z_i(t)\qquad (i=1,2)\, .\nonumber 
\eea
The integrated autocorrelation time for $X(t)$ is given
by
\bea
 \tint(X)& =& \eta_1\tint(x_1)+\eta_2\tint(x_2)\, ,
\quad\hbox{where}\\
 \eta_1+\eta_2& =& 1\, ,\quad 
\eta_2={{r^2(1-a_1^2)}\over{1-a^2_2+r^2(1-a_1^2)}}\, .
\eea
Thus the relation between the integrated autocorrelation time
and the actual correlations present in $X_t$ is no longer 
straightforward. There may be quite long correlations present 
($\texp(x_2)\gg\texp(x_1)$) but, depending on the
relative strength of the modes (given by $r$), 
the \lq weighted average\rq{}
represented by the above formula can give a result bearing
no relation to either $\texp(x_1)$ or $\texp(x_2)$.

The possibility of such behaviour makes it essential
to check decorrelation for individual observables explicitly using binning
techniques. 

\subsection{Finite-size effects}
In retrospect, the value of $\ksea =0.13500$ turned out to be somewhat
conservative, in that the corresponding ratio of $\mps/\mv$ is quite
large ($0.70$, see section~\ref{sec:spectrum}).  The choice was based
on preliminary estimates of the limiting algorithm performance and on
measurements of the effective lattice spacing as described in
section~\ref{sec:pot}. It was felt that decreasing the quark mass
further would decrease the effective lattice size to a point where
finite-size effects would become a problem. In our earlier analysis of
finite-size effects (at least as far as they affect the potential and
light hadron spectrum for $\mps/\mv\gtaeq0.7$) we found that such
effects were negligible provided
\be
  L/r_0 ~\gtaeq~ 3.2\, .
\label{EQ_bound}
\ee
This corresponds to a spatial extent of around $1.6$ fm and is
satisfied by all but our lightest quark mass data set, as shown in
Table~\ref{tab_fin_vol}. Further investigations may be called for,
given the concerns over the baryon mass spectrum noted
in~\cite{AliKhan:2001tx} (see sec.VC).

%
% matching
%

\section{Matching simulation parameters}
\label{sec:match}
In a previous paper~\cite{Irving:1998yu} we have described techniques
which allow one to use unbiased stochastic estimates of the
logarithm of the fermion determinant to determine, approximately, curves
of constant observable in the space of simulation parameters.

\subsection{Determination of fixed observable curves}
The approximate character of the formalism arises from two sources.
First, the log of the fermion determinant is only 
determined stochastically on
each configuration and the corresponding fluctuations are
proportional to the
lattice volume.  Second, a linear approximation is used when dealing
with small changes so that these curves may only be determined
locally.  In the present application, the parameter space of interest
is the $(\beta,\kappa)$ plane and the observable of interest is the
QCD static potential scale parameter $r_0$ (see
section~\ref{sec:pot}).

To first order in small parameter changes $(\delta\beta,\delta\kappa)$, 
the shift in the  lattice operator $F$ is
given by~\cite{Irving:1998yu}

\be
\langle\delta F\rangle =
[\langle\tilde{F}\tilde{W}_{\Box}\rangle  +
\langle\tilde{F} {{\partial\tilde{T}}\over{\partial c_{\rm sw}}}\rangle \dot{c}_{\rm sw}]
\delta\beta
+ 
\langle\tilde{F}{{\partial\tilde{T}}\over{\partial\kappa}}\rangle 
\delta\kappa
.
\label{eq:dF}
\ee
The quantity
\[
\dot{c}_{\rm sw}={{dc_{\rm sw}}\over{d\beta}}
\]
is well determined~\cite{Jansen:1998mx}
and so the
identification of constant $F$ curves 
\be
\langle\delta\tilde{F}\rangle =0
\label{eq:constF}
\ee
reduces to measuring
correlations of the form
\be
\langle\tilde{F}\tilde{W}_{\Box}\rangle \quad{\rm and}\quad
\langle\tilde{F}\delta\tilde{T}\rangle \, .
\label{eq:dfcor} 
\ee
Here, $\langle\tilde{A}\rangle $ denotes the connected part 
$\langle A-\langle A\rangle \rangle $  of the operator $A$.
We refer readers to~\cite{Irving:1998yu} for a detailed
discussion of the stochastic evaluation of 
$T\equiv {\mathrm TrLn} M^\dagger M$.
Here, $M$ is the fermion matrix including the $O(a)$ improvement term.
The methods are based on a L\'anczos implementation of Gaussian 
quadrature~\cite{BFG95}. Recent progress in understanding the nature
of roundoff errors in the finite arithmetic L\'anczos process assures us
that this application of the L\'anczos process, unlike the basic algorithm 
itself, is highly stable with respect to roundoff~\cite{Cahill:1999yx}.

\subsection{Matching $r_0$}
Detailed tests of the matching procedures have been carried out using
the average plaquette, which is very accurately measured, and a
variety of Wilson loops~\cite{Irving:1998yu}. Some tests using $r_0$
were also carried out successfully on modest-sized lattices.  The
present work represents the first application, in earnest, to
production-size lattices. Since the fluctuations in $T$ are extensive
quantities, we expect there will be a limit on the size of lattices
where usefully accurate matching estimates may be made with a given
amount of work. The work required has been analysed in some detail
in~\cite{Irving:1998yu}.

The correlations (\ref{eq:dfcor}) require measurements of $F$ on each
configuration. These are available for operators such as Wilson loops
but not for physical quantities such as hadron masses and $r_0$.
Rather than determine the fixed $r_0$ curve directly using
(\ref{eq:constF}), we use (\ref{eq:dF}) to estimate the required gauge
correlators at nearby points in parameter space.  We then extract the
potential, and hence $r_0/a$, at the nearby parameter values from
these `shifted correlators'.  This allows estimates of the partial
derivatives with respect to $\beta$ and $\kappa$ and hence the shift
$\delta\beta_{\kappa}$ required to compensate a particular change in
$\ksea$.  
\be \delta\beta_{\kappa}=- {{\partial F}\over{\partial
    \ksea}}/ {{\partial F}\over{\partial \beta}} \delta\ksea
\label{eq:dbeta}
\ee
where, in the present application, 
$F=\hat{r}_0\equiv r_0/a$.\footnote{Here, and in what follows,
we use the notation $\hat{A}$ to denote a physical quantity $A$
expressed in lattice units.} 

Using an ensemble of 100 configurations at 
$(\beta,\kappa)=(5.2,0.1350)$ for all correlator measurements, we 
estimated that a shift of 
$$\delta\beta= 0.057\pm 0.033$$
would be required so as to match the value of $\hat{r}_0$ at
$(5.2+\delta\beta,0.1345)$
with that at
$(\beta,\kappa)=(5.2,0.1350)$.
A simulation run at
$(\beta,\kappa)=(5.26,.1345)$ confirms that $r_0$, and hence the
effective lattice spacing, is indeed well matched (see
Table~\ref{tb:r0}). 

However, it is clear from the size of the statistical errors that
estimates of $\delta\beta$ obtained on these configurations 
cannot be relied upon, in general, to predict matched parameters 
with great accuracy without further checks. The level of success in
achieving $\hat{r}_0$ matching can be gauged from Table~\ref{tb:r0}.

The above methods for matching parameters are only applicable for
small shifts.  To obtain the shift for matching quenched simulations
($\delta\kappa=-\kappa$) we have used tabulated
values~\cite{Edwards:1997xf} of $\hat{r}_0$ (lattice spacing $a$) to
provide an initial estimate.  Since there are systematic differences
arising from slightly differing methods for extracting the potentials
and $\hat{r}_0$ (see section~\ref{sec:pot}), we used this only as an
initial guide.  Following direct measurement of $\hat{r}_0$ with our
own techniques, we then made a further small shift in $\beta$.  The
results are shown in the next section.

%
% static potential
%

\section{The QCD static potential}
\label{sec:pot}
We have determined the static inter-quark potential $V(\rvec)$ using
standard methods and used it to search for signs of charge screening
and string breaking, as well as to determine the physical scale.

\subsection{Extraction of the potential} 
The methods follow those originally proposed by Michael and
collaborators~\cite{Michael:1985ne,Perantonis:1989uz}.  A variational
basis of generalised Wilson loops is constructed from gauge links
which are `fuzzed' in the spatial directions~\cite{Albanese:1987ds}.
The spatial paths between the static sources include a limited number
of off-axis directions as well as those along the lattice axes (see
the lower half of Table~\ref{tb:Vr}).  A transfer matrix formalism is
then used to extract the Euclidean time energy eigenstates which are
related to solutions of the generalised eigenvalue equation
\be
  W_{ij}(\rvec,t)\,\phi(\rvec)_j^{(k)} = \lambda^{(k)}(\rvec;t,t_0)\,
  W_{ij}(\rvec,t_0)\,\phi(\rvec)_j^{(k)},\quad i,j,k=0,1.
\ee
Here, we have used two levels of `fuzzing' so giving a $2\times 2$
eigenvalue equation. We used level 0 (unfuzzed) and level 16 which
means sixteen transformations of the spatial links. The link/staple
weighting used was 2.5. This choice of fuzzing parameters was made so
as to give a satisfactory variational basis with a modest amount of
computational effort.  Initial tuning experiments were made using 20
configurations at $(\beta,\ksea)=(5.2,0.13500)$ and repeated on a
corresponding matched ensemble of quenched configurations. Expanding
the basis to three levels of fuzzing did not significantly improve the
resulting effective mass plateaux extracted as described below.

In principle, one could use the largest eigenvalue
$\lambda^{(0)}(\rvec;t,t_0)$ for large $t,t_0$ to estimate the
potential. In practice, however, the eigenvalue system becomes
unstable at large $t$, particularly when modest numbers of
configurations are used, as is often the case in dynamical fermion
studies.  Instead, we used the leading eigenvector
$\phi(\rvec)^{(0)}$, corresponding to $\lambda^{(0)}(\rvec;t,t_0)$ at
$t=1,\,t_0=0$, to project onto the approximate ground
state\cite{Sommer:1994ce,Wittig:1994id}.  The resulting correlator
$\tilde{W}_0(t)$was then used to form effective mass estimates for the
approximate ground state
\be
  \tilde{E}_0(\rvec,t)={\rm ln}\Bigl [{{\tilde{W}_0(\rvec,t)}
        \over{\tilde{W}_0(\rvec,t+1)}}
        \Bigr]\, .
\ee
The ratio of the first two transfer matrix eigenvalues 
\be
R_{1,0}=\lambda^{(1)}(\rvec;1,0)/\lambda^{(0)}(\rvec;1,0)
\ee
was used to help obtain improved estimates of the ground state energy
with reduced contamination from the first excited state. To do this,
the correlator $\tilde{W}_0(t)$ was modelled as a sum of of two
exponential terms
\be
  \tilde{W}_0(\rvec,t)\propto 
        [\lambda^{(0)}(\rvec;1,0)]^t+ \epsilon_{1,0}
        [\lambda^{(1)}(\rvec;1,0)]^t\, .
\ee
One can easily show that, provided the contamination from the first
excited state ($\epsilon_{1,0}$) is small, the true ground state
energy in such a model is given by
\be
E_0(\rvec)= -{\rm ln} \lambda^{(0)}(R_{1,0})
\approx
{{\tilde{E}_0(\rvec,t)-R_{1,0}\tilde{E}_0(\rvec,t-1)}
\over{1-R_{1,0}}}\, .
\label{eq:Vextrap}
\ee
Rather than search for plateaux in this quantity, we used a weighted
mean of values from $t_{\rm min}$ to $t_{\rm max}$ where the weighting
was inversely proportional to the statistical error (estimated via
simple jackknife).  To obtain the final quoted values we used $(t_{\rm
  min},t_{\rm max})=(4,5)$. In all cases, the difference due to
increasing or decreasing the cut-offs by one was less than the
statistical errors quoted.  Overall statistical errors were estimated
by bootstrap sampling.

We also studied double exponentials fits to the effective mass
$\tilde{E}_0(\rvec,t)$ using time-slices up to $t=8$ and exponential
fits to the full $2\times2$ matrix correlator. The fits, where stable,
yielded results compatible with those obtained by the above methods.

In Figure~\ref{fg:Veff}, we show examples of the effective mass
and corresponding extrapolated energy~(\ref{eq:Vextrap}) used to determine
$V(\rvec)$. The lattice potential values are collected in
Table~\ref{tb:Vr}.

\subsection{Determination of $r_0/a$}
\label{sec:r0}

The potential $V(\rvec)$ can be used to determine the force between a
static quark-antiquark pair separated by a distance $r=|\rvec|$ and
hence to extract the Sommer scale parameter $r_0$. This is a
characteristic scale at which one may match the inter-quark force with
phenomenological potential models describing
quarkonia~\cite{Sommer:1994ce}. Specifically, it is defined by the
solution of the relation
\be
r_0^2{{dV}\over{dr}}\big|_{r_0}=1.65\, .
\label{eq:r0def}
\ee
Physically, $r_0\simeq0.49\,\fm$. and we adopt this latter value when
physical units are required.  This definition of the physical scale
has the advantage that one needs to know the potential only at
intermediate distances. An extrapolation of the potential to large
separation, which is conventionally performed to extract the string
tension, is thus avoided. Hence, the procedure is well-suited to the
case of full QCD for which the definition of a string tension, as the
limiting value of the force, is not applicable.  The string is, of
course, expected to break at some characteristic distance~$r_b$.

Our determination of $r_0/a$ follows the procedures originally
described in~\cite{Michael:1992nj} and recently adapted to provide a
comprehensive study of the scale parameter in quenched
QCD~\cite{Edwards:1997xf}.  That is, we perform fits to the
parametrisation
\bea
V(\rvec)=V_0+\sigma r 
        - e \bigl[{1\over \rvec}\bigr]  
        + f \bigl(\bigl[{1\over \rvec}\bigr]-{1\over r}\bigr)
\label{eq:Vr}
\eea
where $[\frac{1}{\rvec}]$ is the tree-level
lattice Coulomb term
\be
 \bigl[{1\over \rvec}\bigr]   = 
  4\pi\int_{-\pi}^{\pi}\frac{d^3k}{(2\pi)^3}\,
 \frac{\cos(\kvec\cdot\rvec)}{4\sum_{j=1}^3\sin^2(k_j/2)}\, .
\label{eq:Vcoul}
\ee
The parameter $f$ is introduced so as to model further lattice
corrections beyond tree level. We find that, for the coarse lattice
spacings considered in this work, a tree-level parametrisation gives a
poor description of the data for $r\ll r_0$.

Following~\cite{Edwards:1997xf}, we use fits of the form(\ref{eq:Vr})
to provide a good description of the intermediate range potential. We
then identify the fitted parameters as reliable estimates (up to
$O(a^2)$) of the corresponding continuum version which, from the
definition of $r_0$, satisfies
\be
\sigma r_0^2+e=c\equiv 1.65
\ee
and hence we extract our estimate of $r_0$ as
\be
r_0=\sqrt{{c-e}\over \sigma}\, .
\ee
Both on and off-axis measurements of the potential were used (see
Table~\ref{tb:Vr}).  We confirm the observation~\cite{Edwards:1997xf}
that the value of $r_0$ extracted in this way is remarkably
insensitive to changes in the fit range used.  The individual
parameters such as $e$ and $f$ are, however, quite sensitive. The
point at $\rvec=(a,0,0)$ was omitted from all fits since its inclusion
was found to give an unacceptably high contribution to $\chi^2$.  The
inclusion of data at the largest $|\rvec|$ ($>8a$) played little role
in the determination of $r_0$.  Since a limited range of $\rvec$ is
used to determine the parameters of (\ref{eq:Vr}), one should treat
the value of $\sigma$ with some caution.  It does not represent a
careful determination of the string tension which of couse is a large
distance property and, strictly speaking, only meaningful in the heavy
sea-quark limit.

We present a summary of the results for $r_0$ in Table~\ref{tb:r0}.
The systematic error estimates (shown as $(+x-y)$) were determined by
variations in the fitting range used for $\rvec$ and in the number of
parameters used in the fit. The central values quoted were obtained
using all potential data satisfying $\sqrt{2)}\leq \rvec\leq 8$.  As
described in the next subsection, a term proportional to $1/r^2$ was
tried.  The systematic error estimates also include the effects of
varying $t_{\rm min}$ by one unit in the evaluation of the potential
(see above).  It is seen that, for the ensembles at $(5.20,0.1350)$,
$(5.26,0.1345)$ and $(5.93,0)$, the matching in $\hat{r}_0$ (and hence
in effective lattice spacing) is very good (well within statistical
errors) while that at $(5.29,0.1340)$ is only slightly off (just over
one standard deviation). The unmatched simulation at the lightest
quark mass has a significantly smaller lattice spacing (seven standard
deviations).

It is worth noting (Table~\ref{tb:Vr}) that the absolute values of the
potential are not matched even when $r_0$ is. The same is true for the
average plaquette and the generalised Wilson loops themselves which go
into the potential determination. All of these loop operators have
large ultraviolet-sensitive contributions.  In
section~\ref{sec:spectrum}, we will comment further on the extent to
which matching is observed in other physical quantities.

The value $\hat{r}_0=4.714(13)$ for the quenched measurements at
$\beta=5.93$ may be compared to previous high statistics measurements
in quenched simulations. The interpolating parametrisations
of~\cite{Edwards:1997xf} and~\cite{Guagnelli:1998ud} respectively
suggest $4.757$ and $4.741(18)$ in fair agreement with, 
but slightly larger than, 
our determination of this quantity at this particular value of $\beta$. 
The slight discrepancy of our result with that of ref. [27] amounts to
about one standard deviation.'

The JLQCD collaboration have presented preliminary results
from an $N_f=2$ simulation using the same action as the present work
at $\beta=5.2$, $\csw=2.02$ and $\kappa=0.1350,0.1355$~\cite{Aoki:2000yi}.
The values of $\hat{r}_0$ in this case are slightly smaller than
those presented in Table~\ref{tb:r0}. Note that the value of $\csw$ used
by JLQCD is very slightly different from ours. The methods
used to extract the potential and $\hat{r}_0$ apparently have much in 
common with those described above but we have not been able
to check all the details. In particular, 
the errors so far presented by JLQCD are statistical only.

As mentioned above we have used both on-axis and off-axis Wilson loops
in our determination of $\hat{r_0}$. However, different spatial
orientations of Wilson loops differ by lattice artefacts of
order~$a^2$. Thus, if on-axis loops are used exclusively to extract
$\hat{r_0}$, then the result may not be consistent with a
determination using other orientations, provided that the statistical
accuracy is large enough to expose these discrepancies. For our
$N_f=2$ simulations the level of precision is about 1\%, so that any
significant discrepancy in $\hat{r_0}$ due to different orientations
will be hard to detect. In future high-statistics simulations
with dynamical quarks, a cleaner procedure might be to define
$\hat{r_0}$ consistently for one particular orientation and to extract
$\hat{r_0}$ from local interpolations of the force between static
quarks. This is the approach used in refs. [9,25,27,8]. It has also been
used for some of the ensembles presented here and for $r_0$ itself
makes little difference (within the statistical errors).

\subsection{Charge screening}
\label{sec:chscreen}
In Figure~\ref{fg:r0V} we plot the static potential in 
units of $r_0$. 
The zero of the potential has been set at $r=r_0$.  Overall, the
presence of dynamical fermions makes little difference when plotted in
physical units.  The data are apparently well described by the
universal bosonic string model potential~\cite{Luscher:1981ac} which
predicts
\be
  \big[V(r)-V(r_0)\big]r_0 = (1.65-e)\left(\frac{r}{r_0}-1\right)
  -e\left(\frac{r_0}{r}-1\right)\, .
\label{EQ_potstring}
\ee
Of course, the fact that the scaled potential measurements all have the same
value and slope at $r=r_0$ simply reflects the definition of $r_0$.
In Figure~\ref{fg:PotDiff} we show the deviations from this
model potential. Here $e=\pi/12$~\cite{Luscher:1981ac}. 
We note the following points:
\begin{itemize} 
  
\item At the shortest distances (see the points where
  $|\rvec|<0.5r_0$) there are indeed deviations from the string model.
  
\item The large fluctuations as a function of $\rvec/r_0$ indicate
  strong violations of rotational symmetry (see Table~\ref{tb:Vr} for
  a list of separations used).
  
\item There is some slight evidence that the deviations depend
  systematically on the quark mass -- compare the quenched points with
  those for the lightest values of $\ksea$.
  
\item The matching of the data ensembles allows a clean comparison of
  the data at different quark masses.
  
\item There are no indications of string breaking, but we note that
  the distance probed at light quarks masses is not large.  ($\rvec <
  1.3$ fm at the lightest quark masses used.)

\end{itemize}

As discussed above, the parametrisation (\ref{eq:Vr}) is not
particularly efficient at describing the short-range interactions on
the lattice. This is the case even though it allows, in a
model-dependent way ($f\neq 0$), for lattice artifacts beyond those
expected at tree level. The fits for the effective charge $e$ and
associated parameter $f$ are therefore sensitive to the fit range and
any variation in the parametrisation. For example, we also considered
allowing a term proportional to $|\rvec|^{-2}$ in an attempt to
describe better the short distance potential.  However, the coarseness
of the lattice and crudeness of the parametrisation prevented reliable
fits. In the continuum limit one would expect the short range
potential to behave as
\bea
V(\rvec)=-\frac{4}{3}{{\alpha_s(\mu)}\over{|\rvec|}}  \, .
\label{eq:Vrcont}
\eea
where $\mu$ is some scale. Lowest-order perturbation theory then
suggests an enhancement of some 14\%{} in $\alpha_s$ arising from the
change in the factor $33-2N_f$ when unquenching the theory (at fixed
scale).  Using the above parametrisation (\ref{eq:Vr}) we can see if
such an effect is reflected in a corresponding increase of 18\%{} in
the parameter $e$.

We have performed correlated fits to the potential with a constant
choice of parametrisation and fit ranges. Some reasonable variation in
the latter was then used to give an estimate of systematic errors. The
fits for the central values of parameters included all data from
Table~\ref{tb:Vr} satisfying $\sqrt{2}\leq |\rvec|/a \leq 9$.  The
statistical errors were produced via an overall bootstrap of the full
analysis (with 500 bootstrap samples). The results are included in
Table~\ref{tb:r0}.  The coupling parameter $e$ does seem to show an
increase due to unquenching. For the matched ensembles the increase is
$18\err{13}{10}\%$ in going from quenched to $\ksea=.13500$.

Similar findings in the case of two flavours of Wilson fermions have
been reported by the SESAM-T$\chi$L collaboration~\cite{Bali:2000vr} where
an increase of $16-33\%$ was found.

For comparison with other scale determinations, we have included the
fit parameter $\sqrt{\sigma}$ expressed in units of MeV as deduced
from $r_0=0.49$ fm. We repeat the caveat offered above that the
parameter $\sqrt{\sigma}$ reflects the medium-range shape of the
potential and does not represent a definitive determination of the
asymptotic string tension.  Phenomenological models for the hadronic
string suggest a value of around $440$ MeV. The energy scale
determination based on $r_0/a$ is therefore some 6 to 7\%{} higher
than that based on the string tension.  In the next section we compare
the above scale determination with values deduced from the vector
meson mass.

Recently the MILC collaboration~\cite{Bernard:2000gd,Bernard:2001av}
has presented results of a comparison of the quenched static potential
with that due to three flavours of staggered fermions. As in the
present analysis, the authors have noted the strong influence of the
dynamical quarks on the effective lattice spacing and have compared
the shapes of the potential measured on matched ensembles.
%}}}

%
% hadron spectrum
%

\section{Light hadron spectrum}
\label{sec:spectrum}

%{{{ spectrum intro

Throughout this section one of our main aims will be to uncover
any unquenching effects in the light hadron spectrum.
Because we have a {\em matched} data set, any differences can more
directly be attributed to unquenching effects. However, 
the task of identifying differences is likely to be
hard for those quantities which are primarily sensitive
to physics at the same scale as that used to define the matching
trajectory in the $(\beta,\ksea)$ parameter space
($r_0$ in this case). This is expected to be the
case for the hadron spectrum considered here where the
quark masses are still relatively heavy.

Two-point hadronic correlation functions were produced for each of the
datasets appearing in Table \ref{tb:config_sum}.  The interpolating
operators for pseudoscalar, vector, nucleon and delta channels were
those described in~\cite{Allton:1994ps}.  Mesonic correlators were
constructed using both degenerate and non-degenerate valence quarks,
whereas only degenerate valence quarks were used for the baryonic
correlators.

The hadronic masses are presented in Tables
\ref{tb:ps}-\ref{tb:delta}. These are expressed in both lattice units
($\hat{M}\equiv Ma$), and in the dimensionless form $r_0 M$.  Note
that the errors displayed are statistical only.  We estimate that the
systematic errors arising from different choices of fitting procedure
are similar in size to the statistical errors.

In the following, we review the main fitting procedures 
which were used to obtain the light hadron spectrum results.
Further details of the fitting procedure can be found in
\cite{Garden:thesis}.

%}}}

%{{{ fitting procedure

\subsection{Fitting procedure}

We used the fuzzing procedures of~\cite{Lacock:1995qx} to generate
correlators of the type LL, FL and FF where F denotes fuzzed, and L
local operators.  Conforming to our usual convention, FL means fuzzed
at the source and local at the sink.  The fuzzing radius was set to
$R_{\rm fuzz} = 2$.

Effective mass plots for the three types of fuzzed correlators (LL,FL
and FF) are shown in Fig. \ref{fig:effm} for the $\beta = 5.2$, $\ksea
= \kval = 0.13500$ data set. Note that all the effective mass plots
approach their asymptote from above. 
The FF correlator exhibits the fastest approach.  
This behaviour is universal throughout all the datasets.
%%An optimised value of $R_{\rm fuzz}$ 
%%would result in still faster
%%isolation of the ground state but, since the plateaux in the current
%%data are sufficient in length to obtain ground state parameters, no
%%further optimisation over $R_{\rm fuzz}$ was performed.
For technical reasons, the fuzzing procedures used in practice 
for the hadron correlators introduced some unbiassed stochastic noise.
We have checked that this has indeed had no significant effect on the
hadronic quantities presented here but has resulted in increased error
estimates at the level of less than $10\%$  for the pion and less than $20\%$
for the nucleon.

Correlated fits were used throughout the fitting analysis of the
correlation functions and the eigenvalue smoothing technique
of~\cite{Michael:1995sz} was employed.  Ensembles of 500 bootstrap
samples were used to estimate the errors \cite{siam}.

We performed a {\em factorising fit} which we now describe for the
baryonic case.  The three fuzzed correlators LL, FL and FF are fitted
together, where the fitting function used for, say, the FL channel is
\[
Z_0^L Z_0^F \e^{-m_0\,t} + Z_1^L Z_1^F \e^{-m_1\,t},
\]
and the LL and FF fitting functions are similarly defined (see
e.g. \cite{Duncan:1995uq}).
Note that both the coefficients, $Z_{0,1}$ and the masses $m_{0,1}$
are common to all the channels, and that the $\chi^2$ comprises
the individual $\chi^2$ of the three channels and includes the correlation
between different times and channels.

For the mesonic case, we modify the above as usual by including the
backward-propagating state, i.e. $\e^{-mt} \rightarrow \e^{-m(T-t)}$
where $T$ is the temporal extent of the lattice.

Within these three different fitting types, a \emph{sliding window}
analysis was used to determine the optimal fitting range ($\Tmin -
\Tmax$) \cite{Bowler:1999ae}.  In this analysis, fits for various
$\Tmin$ were obtained with $\Tmax$ fixed generally to 15. Stability
requirements in the baryonic sector forced $\Tmax=14$ in some cases.
The masses so obtained are displayed in Tables
\ref{tb:ps}-\ref{tb:delta}.

%}}}

%{{{ PCAC Mass

\subsection{PCAC Mass}

The PCAC mass can be defined using the relation
\[
\partial_\mu A_\mu(x) = 2\, m_{\rm PCAC} P(x),
\]
where $P(x)$ and $A(x)$ are pseudoscalar and axial current 
densities.
On the lattice, the following expression can be used to obtain an
estimate of $m_{\rm PCAC}$~\cite{Luscher:1996sc}.
\bea\nonumber
m_{\rm PCAC} &=& \big< \frac{\tilde{\partial}_4 C_{A_4 P^\dagger}(\vec{0},t)
 + a c_A \partial^\ast_4 \partial_4 C_{P   P^\dagger}(\vec{0},t)}
{2 C_{PP^\dagger}(\vec{0},t)} \big> \\
&=& \langle r(t)\rangle + c_A \langle s(t)\rangle
\label{eq:mpcac}
\eea
where $\tilde{\partial}_4$ is the temporal lattice
derivative averaged over the forward, $\partial$, and backward,
$\partial^\ast$, directions, and $\langle \ldots \rangle$ represents averaging
over times, $t$, where the asymptotic state dominates.
The correlators $C$ are defined in~\cite{Allton:1994ps}.
The value of the coefficient used is
\be
c_A = - 0.00756 g_0^2,
\label{eq:ca}
\ee
with $g_0^2 = 6/\beta$ (the bare coupling).
This is the one-loop, dynamical value \cite{Luscher:1996vw},
and hence eq.(\ref{eq:mpcac}) suffers from ${\cal O}(a)$ errors.
Table \ref{tb:mpcac} shows the results for $m_{\rm PCAC}$ for all the
datasets with $c_A$ defined as in eq.(\ref{eq:ca}).

There has been some recent debate in the literature regarding the most
suitable non-perturbatively improved $c_A$ value (see e.g.
\cite{Collins:2000jg,Bhattacharya:1999uq,Bhattacharya:2000pn}) and a
reliable value may not yet have been determined.  In the absence of a
non-perturbatively improved value of $c_A$ (for $N_f=2)$, we choose to
display also in Table \ref{tb:rs} the values for $<r(t)>$ and
$<s(t)>$. With these numbers the reader can readily obtain the values
for $m_{\rm PCAC}$ with any choice of $c_A$.

%}}}

%{{{ Mesonic Sector and the $\mathbf{J}$ parameter

\subsection{The J parameter}
\label{sec:J}

In Figures \ref{fig:mesons} and \ref{fig:hyperfine} the vector meson
masses and hyperfine splittings are plotted against
the corresponding pseudoscalar masses for all the datasets.  It is
difficult to identify an unquenching signal from these plots - the data
seem to overlay each other.  Note that in \cite{Allton:1998gi}, it was
reported that there was a tendency for the vector mass to {\em
 increase} as the sea quark mass {\em decreases} (for fixed
pseudoscalar mass). The observations for the present {\em matched}
dataset imply that this may have been due to either an ${\cal O}(a)$
effect (since the dataset in \cite{Allton:1998gi} was not fully
improved at this level) or a finite volume effect.  The conclusion
therefore is that it is important to run at a fixed $a$ in order to
disentangle unquenching effects from lattice artifacts or finite
volume effects.

A possible explanation as to why there is no signal of unquenching in
our meson spectrum is the following. Our matched ensembles are defined
to have a common $r_0$ value, so any physical quantity that is
sensitive to this distance scale (and the static quark potential
itself) will also, by definition, be matched. Our mesons, because they
are composed of relatively heavy quarks, are examples of such
quantities, and this is a possible reason why there is no significant
evidence of unquenching effects in the meson spectrum.

When comparing the experimental data points with the lattice data 
in Figures \ref{fig:mesons} and \ref{fig:hyperfine} 
we note that the lattice
data are high. This could be due to an incorrect value of $r_0$ being used
($r_0 = 0.49$ fm) and that the true value of $r_0$ is somewhat higher.
This possibility is discussed again in the next section.

A further point regarding hyperfine splitting in Figure \ref{fig:hyperfine}
is that the lattice data for the {\em matched} ensembles tends to
flatten as the sea quark mass decreases. (The quenched data has a distinctly
negative slope, whereas the $\ksea=0.1350$ data is flat.) Thus the
lattice data is tending towards the same behaviour as the experimental
data which lies on a line with {\em positive} slope (independent of
the value used for $r_0$).
This behaviour is apparently spoiled by the unmatched run with $\ksea=0.1355$
(see Figure \ref{fig:hyperfine}.) which has a clear {\em negative} slope.
However the $\ksea=0.1355$ data does not satisfy the finite volume bound
of \cite{Allton:1998gi} (see Sec.\ref{sec:edinplot}). One would expect that
these finite volume effects would squeeze the vector meson state more than
the pseudoscalar state (the $\rho$ is an extended object). Furthermore,
as the valence quark mass was decreased, the more the vector mass would
be raised by finite volume systematics. These considerations match with the
observed behaviour of the $\ksea=0.1355$ data in Figure \ref{fig:hyperfine}.
The JLQCD collaboration~\cite{Aoki:2000yi} 
has recently reported on a finite-volume
analysis with the same action as used in this work.
For $\beta=5.2, \ksea = 0.1350$, they found no evidence of finite-volume
effects in their $16^3$ data for either the pseudoscalar or vector meson. 
It would be interesting to extend this analysis to their
$\beta=5.2, \ksea = 0.1355$ dataset.

The $J$-parameter is defined~\cite{Lacock:1995tq} as 
\be
J = M_V \frac{dM_V}{dM_{PS}^2} \bigg|_{K,K^\ast}.
\label{eq:J}
\ee

In the context of dynamical fermion simulations, this parameter can be
calculated in two ways. The first is to define a partially quenched
$J$ for each value of the sea quark mass. In this case, the derivative
in (\ref{eq:J}) is with respect to variations in the valence quark
mass (with the sea quark mass fixed). The second approach is to define
$J$ along what we will term the `unitary' trajectory,
i.e. along $\ksea = \kval$.
In Table \ref{tb:J}, the results from both methods are given.
These values of $J$ are around 25\% lower than
the experimental value $J_{expt} = 0.48(2)$\footnote{Note, however,
that the experimental value of $J$
does depend on assumptions regarding the mixing of the 
strange and non-strange quark states.}

Finally we note that the physical value of $J$ (i.e. that which
most closely follows the procedure used to determine the
experimental value of $J_{expt} = 0.48(2)$),
should be obtained from extrapolating the results from the first approach
to the physical sea quark masses. We call this the third approach.
In order to perform this extrapolation, we extrapolate the three matched
dynamical $J$ values
obtained from the first approach linearly in $(M_{PS}^{\rm unitary})^2$
to $(M_{PS}^{\rm unitary})^2=0$.
$M_{PS}^{\rm unitary}$ is the pseudoscalar meson mass at the unitary
point (i.e. where the valence and sea quark masses are all degenerate).
The value for $J$ from the third approach is presented in Table \ref{tb:J}
and we note that it is approaching the experimental value for $J$.

The results from all three approaches are plotted
in Figure \ref{fig:J}, together with the experimental result.
There is some promising evidence that the lattice estimate of $J$ increases
towards the experimental point as the sea quark mass decreases (see
the $J$ value from approaches 1 and 3).
This effect will move the lattice estimates of the $J$
parameter towards the experimental value as simulations are performed at
more physical values of quark mass.

Recently there has been a proposed ansatz for the functional form of
$M_V$ as a function of $M_{PS}^2$ \cite{Leinweber:2001ac}.
However, all our data have $M_{PS}/M_V \gtap 0.6$, and for this region, 
the ansatz of \cite{Leinweber:2001ac} is linear to a good approximation.
Therefore we choose to interpolate our data with a simple linear function
and await more chiral data before using the ansatz of \cite{Leinweber:2001ac}.

Two groups have recently reported results on the $J$ parameter from
dynamical simulations. The CP-PACS collaboration results at $a \approx
0.11$ fm, found $J^{Dynamical} > J^{Quenched}$ using a clover action
\cite{AliKhan:2001tx}.  Furthermore they found that this discrepancy
increased as the continuum limit was taken.  A similar result was
found by the MILC collaboration who used an improved staggered action with $a
\approx 0.13$ fm \cite{Bernard:2001av}. Both these groups' results
match those found in this work.

%}}}

%{{{ Lattice Spacing Determination

\subsection{Lattice spacing}
\label{sec:lspac}
In Section \ref{sec:pot} the lattice spacing, $a$, was determined from
the intermediate range properties of the static quark potential.  In
this subsection, we present a complementary determination of $a$ from
the meson spectrum.

A common method of determining $a$ from the meson spectrum uses the
$\rho$ mass. However, this requires the chiral extrapolation of the
vector meson mass down to (almost) the chiral limit. This
extrapolation is often performed using a linear function. However, as
was discussed in the previous subsection, a linear chiral
extrapolation may not be appropriate for $M_V \ltap 0.8$ GeV.  An
alternative method of extracting the lattice spacing using the vector
meson mass at the {\em simulated} data points (i.e.  without any
chiral extrapolation) was given in \cite{Allton:1997yv}.  Using this
method, we obtain the lattice spacing values as shown in Table
\ref{tb:aJ}. Note that these are in general 10-15\% larger than the
values from Section \ref{sec:pot} where the lattice spacing was
determined from $r_0$. A possible explanation for this discrepancy is
that the potential and mesonic spectrum are contaminated with
different ${\cal O}(a^2)$ errors, or that the value $r_0 = 0.49$ fm
is 10-15\% too small, and that the true value is $r_0 \approx
0.55 $fm.

It is interesting to study the lattice spacing determinations in more
detail since they are a measure of unquenching effects in dynamical
simulations. Specifically, it is often assumed that the reason the
various quenched determinations of $a$ from e.g. the meson spectrum
differ from that of $r_0$ or the string tension is due to dynamical
quark effects. An obvious quantity to monitor the merging of the
various $a$ determinations can be defined:
\be
\delta_{i,j}(\beta,\hat{m}_{\rm sea}) = 
1 - \frac{a_i(\beta,\hat{m}_{\rm sea})}{a_j(\beta,\hat{m}_{\rm sea})},
\label{eq:delta_ij}
\ee
where $a_i$ is the lattice spacing determined from the physical quantity
$i = \{ M_\rho, M_K, f_\pi \ldots \}$.
Obviously, if $\delta_{i,j} = 0$ then the lattice prediction for
quantity $i$ using the scale determined from $j$ (or vice
versa) is in exact agreement with experiment.

Since our simulations are improved to ${\cal O}(a^2)$ we expect that
$\delta \rightarrow {\cal O}(a^2)$ as $m_{\rm sea=val} \rightarrow
m_l$ (where $m_l$ is the average $u,d$ quark mass).  Thus a plot of
$\delta$ against $(\hat{M}_{PS}^{\rm unitary})^{-2}$ would be
insightful, where $\hat{M}^{\rm unitary}_{PS}$ is the pseudoscalar
mass at the unitary point i.e. for degenerate valence and sea quarks
(so $\hat{M}^{\rm unitary}_{PS} = \infty$ for the quenched data).
Here, we work with $(\hat{M}^{\rm unitary}_{PS})^{-2}$ rather than
$1/\hat{m}_{\rm sea}$ for the $x-$co-ordinate since it is equivalent to, but
easier to define than $1/\hat{m}_{\rm sea}$.  It is important to note
that the $x-$co-ordinate in this plot is the `control parameter' for the
study of unquenching effects. i.e. when we vary this parameter from
its quenched value towards its experimental value, we hope to see the
data plotted in the $y-$co-ordinate move towards its appropriate
experimental value. Thus it is easier to interpret unquenching effects
directly from this plot than from, e.g., plots of $\hat{M}_V$ against
$\hat{M}_{PS}^2$ for various $\hat{m}_{\rm sea}$.

In Figure \ref{fig:delta}, $\delta_{i,j}$ is plotted against
$(\hat{M}_{PS}^{\rm unitary})^{-2}$ for the matched datasets.
In this plot we have fixed $j = r_0$ and
the various physical quantities $i$ are $\sqrt\sigma$ and the hadronic
mass pairs $(M_K*,M_K)$ \& $(M_\rho,M_\pi)$.
The method that was used to determine the scale $a_i$ from these mass
pairs is that of \cite{Allton:1997yv}.
It is worth noting that the experimental point on this same plot would occur
at an $x-$co-ordinate of $(\hat{M}_\pi)^{-2} \approx 200$.

Figure \ref{fig:delta} does not show signs
of unquenching for quantities involving the hadronic spectrum
i.e. the mass pairs $(M_K*,M_K)$ \& $(M_\rho,M_\pi)$.
(Future work will study $\delta_i$ for the matrix
element quantities $i = f_\pi$ and $f_K$.)
However, there is evidence of unquenching effects when comparing the
scale from $r_0$ with that from $\sqrt\sigma$. The quenched value
of $\delta_{\sqrt\sigma}$ is distinct from the dynamical values, though
we note that the method used to obtain $\sigma$ was optimised for the
extraction of $r_0$ rather than $\sigma$ itself (see Sec.\ref{sec:r0}).

One may wonder if the the $\delta$ values may have been distorted by
not choosing the simulation parameters $(\beta,\hat{m}_{\rm sea})$
exactly on the matched trajectory.
In order to obtain a rough estimate of the effect of a mis-matched value
of $\beta$, we use the
renormalisation group inspired ansatz for $a_i$ \cite{Allton:1996kr,Allton:1997dn}:
\begin{equation}
a_i(g_0^2) = \Lambda^{-1} f_{PT}(g_0^2)
\times  \left[ 1 + X_i f_{PT}(g_0^2)^{n_i} \right],
\label{eq:ldpt_fit}
\ee
where $f_{PT}(g^2)$ is the usual asymptotic scaling function obtained
from integrating the $\beta$-function of QCD and $X_i$ is the coefficient
of the ${\cal O}(a^n)$ lattice systematic.
The functional form for $a(g_0^2)$ was originally applied for the quenched
theory, but let us assume that it can also be applied in the unquenched case.
Using Eq.(\ref{eq:ldpt_fit}), we see that a mismatch in $\beta$ of
$\Delta\beta$ would lead to a relative error in $\delta$ of
\begin{equation}
\frac{\delta(\beta+\Delta\beta) - \delta(\beta)}{\delta(\beta)}
\approx -3 \Delta\beta.
\label{eq:deltabeta}
\end{equation}
This shows that even an error in $\beta$ of as much as
$\Delta\beta \approx 0.01$
introduces a relative error in $\delta(\beta)$ of only 3\%,
ruling out any possible mismatching in $\beta$ as leading to
a significant distortion in $\delta$.

%}}}

%{{{ Edinburgh Plot

\subsection{Edinburgh Plot}
\label{sec:edinplot}

In Table \ref{tb:rat_mass} the ratios $M_{PS} / M_{V}$ are displayed
for the case \ksea = \kval.  The average $u$ and $d$ quark mass is
fixed by requiring $M_{PS} / M_{V}= 0.18$.  As can been seen, the
simulations are at much larger dynamical quark masses. Figure
\ref{fig:edin} shows the \lq Edinburgh plot\rq{} ($M_N/M_V$ v.s.
$M_{PS}/M_V$) for all the data sets.  There is no significant
variation within the dynamical data as the sea quark mass is changed
but the dynamical data does tend to lie above the (matched) quenched
data.  This latter feature may be indicative of finite volume effects
since these are expected to be larger in full QCD compared to the
quenched case \cite{Ukawa:1993hz}.  In \cite{Allton:1998gi} an analysis
of dynamical finite volume effects concluded that they were
statistically insignificant for spatial extents of $L \gtap 1.6$ fm
and sea quark masses corresponding to $M_{PS} / M_V \gtap 0.67$ with
around 100 configurations.  This bound is satisfied for the matched
ensembles, but not for the $\ksea = 0.1355$ case where $L = 1.60$ fm
and $M_{PS}/M_V = 0.58$.

Note also that for heavy valence quark masses, the dynamical data lies
close to the phenomenological curve~\cite{Ono:1978}, 
whereas it tends to drift
higher than the curve for small valence quark masses. The (matched)
quenched data agrees well with the curve. 

Dynamical results for baryons have recently been reported by two
groups. CP-PACS (using a clover action) find good agreement with
experiment for strange baryons, but their light baryons (in the
continuum limit) are around 10\% higher than experiment (see sec.VC in
\cite{AliKhan:2001tx}). They discuss the possibility that this is caused
by finite volume effects.
The MILC collaboration (using an improved staggered action) find their
dynamical and quenched Edinburgh plots overlay each other
\cite{Bernard:2001av}.

%}}}

%{{{ Chiral Extrapolations

\subsection{Chiral Extrapolations}

There are a number of different \lq chiral extrapolations\rq{} that
one can perform in the case of dynamical fermions where there is a
two-dimensional quark mass parameter space, $(m_{\rm sea},m_{\rm
  val})$.  We describe three such extrapolations of the data.  The
first uses a {\em partially quenched} analysis where each of the
$m_{\rm sea}$ datasets is extrapolated entirely separately.  The
second uses only the {\em unitary} sub-set with $m_{\rm sea} \equiv
m_{\rm val}$.  The third does a {\em combined} fit of all the matched
data using a fitting ansatz to model the variation between the
different $m_{\rm sea}$ values.

Note that there have been recent proposals for the functional form of
$M_N$ and $M_V$ as a function of $M_{PS}^2$
which go beyond the usual chiral linear ansatz normally used in extrapolations
of lattice data \cite{Leinweber:1999ig,Leinweber:2001ac}.
However, as reported in sec.\ref{sec:J}, the non-linearity of these
functional forms becomes relevant only for lattice data lighter than in
our simulations and therefore we choose to use naive linear chiral
extrapolations in the following.

%{{{ Partially quenched Chiral Extrapolations

\subsubsection{Partially Quenched Chiral Extrapolations}

A partially quenched chiral extrapolation was performed for the hadronic masses
$\hat{M} = \hat{M}_V, \hat{M}_N$ and $\hat{M}_\Delta$ against $\hat{M}_{PS}^2$,
i.e. the following ansatz
was used,
\be
\hat{M}(\beta,\ksea;\kval) = A + B \hat{M}_{PS}(\beta,\ksea;\kval)^2.
\label{eq:pseudoq}
\ee
We have introduced the following nomenclature.
In $\hat{M}(\beta,\ksea;\kval)$, the first two arguments refer to the
sea parameters: the gauge coupling $\beta$ and the sea quark mass $\ksea$.
The third argument refers to the valence quark mass $\kval$.
The results for these partially quenched extrapolations appear in Table \ref{tb:pseudoq}.
Note that there is no convincing sign of unquenching effects in that
the $A$ and $B$ values for the matched datasets tend to overlay
eachother, and there is no clear trend for these values as
a function of $m_{sea}$.

Although we choose to extrapolate with respect to
$\hat{M}_{PS}(\beta,\ksea;\kval)^2$, we also show,
for completeness, the values of $\kcrit$ in Table \ref{tb:kcrit}.
These were obtained from the
usual fit of $\hat{M}_{PS}(\beta,\ksea;\kval)^2$ versus $1/\kval -
1/\kcrit$.

%}}}

%{{{ Unitary Chiral Extrapolations

\subsubsection{Unitary Chiral Extrapolations}

An extrapolation of the hadronic masses
$\hat{M} = \hat{M}_V$, $\hat{M}_N$ and $\hat{M}_\Delta$ against $\hat{M}_{PS}^2$
was performed for the unitary subset of data i.e. the following ansatz
was used,
\be
\hat{M}(\beta,\ksea;\ksea) = A^{\rm unitary} + B^{\rm unitary} \hat{M}_{PS}(\beta,\ksea;\ksea)^2,
\label{eq:unitary}
\ee
Note that only the matched, dynamical datasets were included in these fits.
The results appear in Table \ref{tb:unitary}.

%}}}

%{{{ Combined Chiral Fits

\subsubsection{Combined Chiral Fits}

It is instructive to perform a combined chiral fit to
the entire matched dataset.
In order to achieve this we consider the following fitting ansatz
for the fitting of the hadronic mass $\hat{M}$, where
\bea\nonumber
\hat{M}(\beta,\ksea;\kval)
&=& A^{combined} + B^{combined} \hat{M}_{PS}(\beta,\ksea;\kval)^2 \\\nonumber
&=& A_0 + A_1 \hat{M}_{PS}(\beta,\ksea;\ksea)^{-2} \\
&+& \left[ B_0 + B_1 \hat{M}_{PS}(\beta,\ksea;\ksea)^{-2} \right] 
\hat{M}_{PS}(\beta,\ksea;\kval)^2.
\label{eq:combined}
\eea
One advantage of such a fitting procedure is that in total, to fit the
entire matched dataset, there are fewer fitting parameters than are
required in the partially quenched analysis.  The functional form in
eq.(\ref{eq:combined}) is the simplest functional form which allows
for a variation of $A$ and $B$ with the sea quark mass, and which is
finite for all the datasets studied.  (Note that
$\hat{M}_{PS}(\beta,\ksea;\ksea) \equiv \infty$ for the quenched
data.)  The other advantage is that the parameters $A_1$ and $B_1$ are
a direct measure of unquenching effects.

The results for the fitting parameters $A_{0,1}$ and $B_{0,1}$ are
displayed in Table \ref{tb:combined} for the hadronic masses
$\hat{M}_V, \hat{M}_N$ and $\hat{M}_\Delta$.  The parameters $A_1$ and
$B_1$ for all the hadrons are compatible with zero at the $2\sigma$
level, underlining again the fact that we have not unambiguously
uncovered unquenching effects in the meson and baryon spectra.

%}}}

%}}}

%{{{ SECOND PART

% 
% glueballs 
%

\subsection{Glueballs and torelons}
\label{sec:gball}

Experiment has not so far detected glueball states unambiguously in
the light hadron spectrum.  This failure is usually believed to be a
consequence of mixing between the light glueballs and $q\bar{q}$
states (``quarkonia'') with the same quantum numbers and similar
masses. We lack, however, a clear understanding of the mixing
matrix elements that lead to the strong interaction eigenstates that
would be observed, and thus phenomenological attempts to describe the
content (gluonic or quarkonium) of the scalar sector glueball
candidates have led to widely differing results
\cite{Close:2001ga,Lee:1999kv}.

Lattice QCD can in principle predict these mixing parameters, and in the
quenched approximation precise values are known for the continuum
gluodynamics (quenched QCD) glueball masses (see
%
%\cite{teper98,toussaint00} 
\cite{Teper:1998kw,Toussaint:1999kh} 
for reviews). Attempts to measure the mixing matrix have been
made (see
%
%\cite{toussaint00}
\cite{Toussaint:1999kh}
for a review of quenched measurements, and 
%
%\cite{michael00}
\cite{Michael:1999rs}
for first determinations in the presence of sea quarks) and are in
progress using the current UKQCD field configurations
\cite{michael_prog}.
Simultaneously, the validity of such a simple mixing scenario can also
be addressed
\cite{ukqcd_glue_prog}. 

Quenched glueball calculations require large ensembles and, until
recently, it had been assumed that a similar level of statistical
noise would preclude accurate measurements in simulations with
dynamical fermions. We find, however, and in common with other recent
studies
%
%\cite{bali00},
\cite{Bali:2000vr}
that statistical errors are, somewhat surprisingly, reduced in
dynamical simulation estimates of glueball masses at present parameter
values, at least compared to similarly sized quenched ensembles.

Before continuing with a discussion of our calculations
we need to be a little more specific about what we mean
when we talk of ``glueballs'' in QCD. The point is
that the presence of
quarks will change the vacuum and there is no fundamental
reason to think that the mass spectrum of QCD can be 
approximately described as consisting of the glueballs 
of the pure gauge theory, the usual quarkonia and, where these 
are close in mass, mixtures of the two. There is, however,
a collection of phenomena --- the OZI rule,
small sea quark effects etc. --- that creates a reasonable
prejudice that this might be so. This question will be
examined more explicitly elsewhere
\cite{ukqcd_glue_prog}. 
Here we shall follow the usual view and assume 
it to be so. In that case we expect that if there are no
nearby quarkonia then the states most readily visible using 
purely gluonic operators similar to those used in pure gauge
theories will be almost entirely glueball-like.
This is (probably) the case for the scalar ``glueball'' state 
we discuss herein. The fact that the overlap of this
state onto these purely gluonic operators is similar
to that in the pure gauge theory reinforces our prejudice.
Thus we will refer to this state as the scalar glueball
during the remainder of our analysis.

If we then assume that the glueball spectrum of the dynamical theory
is not radically different to that of the pure gauge gluodynamics, we
expect the lightest states to be the scalar and tensor ground states.
In terms of the reduced symmetries of the space-time lattice, these
correspond to the $A_1^{++}$ and $T_2^{++}$ representations of the
appropriate cubic group. In the continuum where full rotational
symmetry is restored, these match onto the $J^{PC} = 0^{++}, \,
2^{++}$ states. Given the size of our ensembles, we find it difficult
to resolve lattice masses much beyond $\hat{M}_G \sim 1.2$. In
gluodynamics, the heavier tensor state has a (continuum extrapolated)
mass in units of the Sommer scale around $r_0 M_G \simeq 6$. The
$\hat{r}_0$ values tabulated for our ensembles in Table~\ref{tb:r0}
thus suggest that the scalar and tensor are the states we are most
likely to be able to study.

Using a full arsenal of noise reduction techniques it is now possible to
make good estimates of the masses of these lightest glueball masses
using existing ensembles. In this section we present, as an
example, the scalar and tensor states extracted from one ensemble,
that at $(\beta,\kappa) = (5.20,0.13550)$.  Full results for all
couplings, and giving greater details of methodology, will be reported
in
\cite{ukqcd_glue_prog}. 
Preliminary results have appeared in
\cite{Irving:2000hs}.

Measurements were made after every tenth HMC trajectory giving an
ensemble of 830 configurations, which may not be uncorrelated. A
jack-knife error analysis was performed using ten bins, each 830
trajectories in size, which were much larger than the autocorrelation
times of the observables.  This ensured statistically uncorrelated
averages for neighbouring bins.

To reduce statistical errors on mass estimates, operators should have
a good overlap onto the ground state excitation with the specified
quantum numbers. This was achieved in two ways.

Each operator is based on a traced, closed contour of gauge links,
which is gauge invariant. We may improve the overlap of these
operators onto the ground state excitations by ``smearing'' and
``blocking'' the links. The former is computationally cheap, but the
latter has the advantage of doubling the spatial extent of the
operator with each iteration. This proves especially useful for
measuring wavefunctions that are not spherically symmetric, such as the
tensor. The details of this procedure will be discussed further in
\cite{ukqcd_glue_prog}.

A suite of four glueball operators was constructed in each time-slice
of the gauge field configurations by summing similarly improved
contours in the appropriate symmetry combinations
%
%\cite{michael90}.
\cite{Michael:1990ry}.
Overall this gave twenty-eight operators per symmetry channel. These
were cross correlated and a L\"uscher--Wolff variational analysis
%
%\cite{luescher90}
\cite{Luscher:1990ck}
(for details of the exact procedure see Section~3.2 of
%
%\cite{hart00b})
\cite{Hart:2000ha})
used to extract the ground states for each of the lowest momentum
combinations of the operators (labelled as $P \cdot P=0,1\ldots$ where
$P\equiv\hat{p}=pa$). All scalar operators ($A_1^{++}$), for example,
were found to have a good overlap (typically greater than $0.7$) onto
the ground state.  The robustness of the variational analysis was
checked by examining the behaviour of individual correlation
functions, and of sub--sets of the full operator basis. In each case
the mass estimates were found to be consistent as expected given the
good overlap of all operators onto the ground state.

From correlation functions we may define an effective energy as a
function of the Euclidean timelike separation $t$ (in lattice
units) of the creation and
annihilation operators:
\be
\hat{E}_{\rm eff}(t) \equiv - \log \frac
{\langle {\cal O}^\dagger(t+1) {\cal O}(0) \rangle}
{\langle {\cal O}^\dagger(t) {\cal O}(0) \rangle}
\ee
The effective energies of the non--zero momentum states were converted
to effective masses assuming the lattice dispersion relation
\be 
\hat{E}(P)^2 = \hat{M}^2 + \sum_{\mu=1}^3 
\sin^2 \left( \frac{2 \pi P_\mu}{L} \right) 
\ee
The signal from the $P \cdot P=1$ channel was found to be particularly
useful. The mass of the ground state excitation were still small
enough for reliable effective energy plateaux to be observed, and
statistical noise was observed to be only of a similar magnitude to
the $P \cdot P=0$ channel. For $P \cdot P=2$, however, the energies of
the states were too large to be confidently assessed. Where they could
be extracted, they showed effective mass plateaux consistent with
lower momentum channels. Since they did not improve the quality of the
fits, however, they were not included.  Correlated and uncorrelated
plateau fits were then carried out using $P \cdot P=0,1$ together. As
the former fits differed only within errors, for robustness we quote
uncorrelated results in this summary.

In Fig.~\ref{fig_glue_scal} we plot the effective masses for the
various momentum channels of the scalar glueball. A clear plateau is
seen in each of the momentum channels. Since these plateaux are
compatible, indicating a restoration of the continuum Lorentz
symmetry, we can combine the lowest momentum channels to estimate the
pure scalar glueball mass $\hat{M}_G = 0.628 \, (30)$ in lattice
units, or $r_0 M_G = 3.17 \, (15)$ in units of the Sommer scale.  We
note here that the interpolated quenched glueball mass at this lattice
spacing is $r_0 M_G = 3.79 \, (16)$
\cite{ukqcd_glue_prog},
which is significantly above the scalar mass measured here. There
would thus appear to be strong evidence for a quenching effect in the
scalar glueball channel of QCD. We should temper this statement
slightly, as there are other possible sources of suppression of the
scalar glueball mass. Firstly there are finite volume effects which
are known to suppress the scalar glueball mass. In quenched QCD the
principle source of this suppression is the mixing of the glueball
with torelon pair states, e.g.
%
%\cite{teper98a},
\cite{Teper:1998te},
but we shall demonstrate below that in the present case our lattices
are large enough for any such effects to be very small.

More seriously, we do not know the size of this effect in the
continuum limit. In the quenched theory there are known to be
large scaling violations in the $A_1^{++}$ channel for the Wilson
action
%
%\cite{morning99}
\cite{Morningstar:1999rf}
with the ``scalar dip'' tending to suppress the mass below the
continuum value even at relatively small lattice spacings. Without
a continuum limit extrapolation of the glueball mass, we cannot here
speculate as to the size of the corresponding effect in the presence
of dynamical fermions, but preliminary work suggests the scalar dip
may indeed be enhanced in the ensemble considered here
\cite{ukqcd_dip_prog}.

A similar analysis yields a tensor mass estimate of $\hat{M}_G = 1.28
\, (9)$ in lattice units, or $r_0 M_G = 6.43 \, (42)$ in units of the
Sommer scale. This is compatible with the interpolated mass in the
pure glue theory $r_0 M_G = 5.91 \, (23)$.

Colour flux tubes, analogous to that between a static quark and
anti-quark pair, but without source or sink can exist on a periodic
volume. Rather, the flux tube closes on itself through a spatial
boundary (assuming it to be in the `confined' phase), forming what is
usually termed a torelon. To a first approximation the mass of the
lightest such state equals the spatial extent of the lattice
multiplied by the energy per unit length of the flux tube (the string
tension). In the infinite volume limit such states become very massive
and decouple from the observed spectrum.

The vacuum expectation value (VEV) of the Polyakov operator that
couples to such a torelon loop is zero in the confined phase of
gluodynamics, as the loop cannot be broken when no sources in the
fundamental representation exist. Thus, only a combination of at least
two torelons with the appropriate symmetries can couple to the
particle states in the theory.  On lattices small enough that the mass
of the lightest torelon pair is comparable to the scalar glueball mass
we will see significant finite volume effects.

When light dynamical quarks are present, the torelon becomes unstable
to decay.  In this case, the Polyakov loop operator gains a non--zero
expectation value.  This is an effect analogous to the string breaking
seen in the static quark potential measured using Wilson loops and is
another explicit signal for the presence of light dynamical quarks in
these simulations. In addition, it becomes possible for torelon states
to mix with glueballs. Such states are, of course, lighter than the
pairs of torelons that mix in the quenched theory, and so we
might expect to see finite volume effects on larger lattices
in the presence of dynamical quarks.

The Polyakov loop operator is defined as the traced product of links
in a line through the periodic spatial boundary:
\be
p_\mu(n) = \Tr \prod_{k=1}^L U_\mu(n+k\hat{\mu})\, ,
\ee
for $\mu = 1...3$.  In order to improve statistics we create a basis
of operators using improved spatial links as before. 

In Fig.~\ref{fig_poly_vevs} we plot the vacuum expectation value of
the $P \cdot P=0,1,2$ Polyakov loop operators. From momentum
conservation we expect the VEVs of the non--zero momentum operators to
be zero.  This is seen to be satisfied within less than two standard
deviations in all cases, indicating that the statistical errors are
under control.  It also adds significance to the fact that the $P
\cdot P=0$ operators have a vacuum expectation value that deviates
substantially from zero. This is clear evidence of flux tube breaking
by dynamical quark pair production.

Fitting effective masses from $P \cdot P=0,1$ after a L\"uscher--Wolff
analysis as before, we estimate the torelon mass as $\hat{M}_P = 0.77
\, (5)$. Including the leading order universal string correction
%
%\cite{forcrand85},
\cite{deForcrand:1985cz},
we expect the loop mass to vary with the lattice size in $D$
dimensions as
\be
\hat{M}_P = \sigma L - \frac{\pi(D-2)}{6 L}.
\ee
From this we estimate the string tension to be $\hat{\sigma} = 0.052
\, (3)$ or, using the Sommer scale to set physical units,
$\sqrt{\sigma} = 462 \, (13) \, {\rm MeV}$ in good agreement with the
value quoted in Table~\ref{tb:r0}.

The mass of the lightest torelon pair, around twice the torelon mass,
is thus clearly too heavy to induce finite volume effects.  Likewise,
finite volume effects from meson exchange through the boundary should
be small, although we do not consider this process here. The mass of
the torelon, on the other hand, is not much larger than that of the
scalar glueball, and there is a possibility of mixing occurring
between the two which would lead to a finite volume contamination. We
thus perform a variational analysis where we cross correlate a basis
of eight of the ``best'' scalar glueball operators with the two
``best'' torelon operators. We find the matrix to be block diagonal
within errors, and the two lowest eigenstates match closely the
original glueball and torelon in mass and operator overlap. Thus this
finite volume contamination is negligible, something which could have
been anticipated from the small size of the Polyakov line VEV.

In summary, we have presented measurements of the scalar and tensor
glueball and torelon masses on an ensemble of configurations at
$(\beta,\kappa) = (5.20,0.13550)$. We find clear signals for the
presence of light sea quarks, both in a scalar glueball mass that is
significantly suppressed below the quenched value at a comparable
lattice spacing, and in the breaking of the confining flux tube as
demonstrated by a non--zero expectation value for the spatial Polyakov
loop operator. Although non--zero, the smallness of these VEVs
together with the fact that the torelon and torelon pair masses are
significantly larger than the scalar glueball mass lead us to believe
that the suppression of the scalar glueball mass is not a finite
volume effect, a conclusion which is reinforced by an explicit mixing
analysis. The dependence of these effects on the sea quark mass, and
whether this effect persists in the continuum limit is not, however,
resolved here.

%
%   topology
%

\section{The topological susceptibility and \boldmath{$\eff_\pi$}.}
\label{sec:topol}

The ability to access the non--perturbative sectors, and to vary
parameters fixed in Nature has made lattice Monte Carlo simulation a
valuable tool for investigating the r\^{o}le of topological excitations
in QCD and related theories, and it is these that we now consider.

In quenched lattice calculations, the continuum topological
susceptibility now appears to be relatively free of the systematic
errors arising from the discretisation, the finite volumes and the
various measurement algorithms employed. Attempts to measure the
microscopic topological structure of the vacuum are also well advanced
(for a recent review, see
\cite{Teper:1999wp}).
The inclusion of sea quarks in lattice simulations, even at the
relatively large quark masses currently employed, is numerically
extremely expensive, and to avoid significant finite volume
contamination of the results, the lattice must be relatively coarse,
with a spacing $a \simeq 0.1 \ \fm$ as in this study. Compared to
quenched lattice studies at least, this is a significant fraction of
the mean instanton radius, and has so far precluded a robust, detailed
study of the local topological features of the vacuum in the presence
of sea quarks. The topological susceptibility, on the other hand, may
be calculated with some confidence and provides one of the first
opportunities to test some of the more interesting predictions for
QCD.  Indeed, it is in these measurements that we find some of the
most striking evidence for the effects of sea quarks (or,
alternatively, for a strong quenching effect) in the lattice
simulations described in this paper.

We find clear evidence for the expected suppression of the topological
susceptibility in the chiral limit, despite our relatively large quark
masses. From this behaviour we can directly estimate the pion decay
constant without needing to know the lattice operator renormalisation
factors that arise in more conventional calculations.

These results were presented at the IOP2000 
\cite{Hart:2000wr},
the Confinement IV
\cite{hart_wien}
and, in a much more preliminary form, the Lattice '99
\cite{Hart:1999hy}
conferences. Since then, we have increased the size of several
ensembles and included a new parameter set. We also have more accurate
results from the quenched theory with which to compare. 
Related results have been presented by the CP-PACS collaboration
\cite{AliKhan:1999zi,Durr:2001ty,AliKhan:2001ym},
the Pisa group
\cite{Alles:1999kf,Alles:2000cg},
the SESAM--T$\chi$L collaboration
\cite{Bali:2001gk}
and the Boulder group
\cite{Hasenfratz:2001wd}.
A detailed analysis of our data set, and its relation to these other
studies will be given in
\cite{top_prog}.

The topological charge is 
\be
Q = \frac{1}{32\pi^2} \int d^4x 
\frac{1}{2} \varepsilon_{\mu \nu \sigma \tau}
F^a_{\mu \nu}(x) F^a_{\sigma \tau}(x).
\ee
The topological susceptibility is the squared expectation value of the
topological charge, normalised by the volume
\be
\chi = \frac{\langle Q^2 \rangle}{V}.
\ee
Sea quarks induce an instanton--anti-instanton attraction which
in the chiral limit becomes stronger, suppressing $Q$ and $\chi$
\cite{DiVecchia:1980ve}
%\cite{vecchia80}
%
\be
\chi = \Sigma \left( \frac{1}{m_{u}} + \frac{1}{m_{d}} 
 \right)^{-1},
\ee
where
\be
\Sigma = - \lim_{m_q \to 0} \lim_{V \to \infty} 
\langle 0 | \bar{\psi} \psi | 0 \rangle
\ee
is the chiral condensate
\cite{Leutwyler:1992yt}. 
%\cite{leutwyler92} 
%
We assume
$\langle 0 | \bar{\psi} \psi | 0 \rangle =
\langle 0 | \bar{u} u | 0 \rangle =
\langle 0 | \bar{d} d | 0 \rangle $
and neglect contributions of heavier quarks.
The  Gell-Mann--Oakes--Renner relation,
\be
f_\pi^2 M_\pi^2 = 2(m_{u} + m_{d}) \Sigma + {\cal O}(m_q^2)
\ee
implies
\be
\chi = \frac{f_\pi^2 M_\pi^2}{4 N_f} + {\cal O}(M_\pi^4)
\label{eqn_chi_pi2}
\ee
for $N_f$ degenerate light flavours, in a convention where the
experimental value of the pion decay constant $f_\pi \simeq 132 \MeV$
\footnote{ 
N.B. there is a common alternative convention, used in
earlier presentations of this data
\cite{Hart:2000wr,hart_wien},
where a factor of 2 is absorbed into $f_\pi^2$ in (\ref{eqn_chi_pi2}),
and where $f_\pi$ is a factor of $\sqrt{2}$ smaller, around $93\MeV$.}.
Equation~(\ref{eqn_chi_pi2}) holds in the limit $f_\pi^2 M_\pi^2 V \gg1$, which is satisfied by all our lattices.  The higher order terms
ensure that $\chi \rightarrow \chiqu$, the quenched value, as
$m_q,M_\pi \to \infty$.  We find, however, that our measured values
are not very much smaller than $\chiqu$, so we must consider two
possibilities.

Firstly, there are phenomenological reasons
\cite{tHooft:1974hx,Witten:1979kh}
%\cite{thooft74,witten79}
%
for believing that QCD is `close' to $N_c = \infty$, and in the case
of gluodynamics even SU(2) is demonstrably close to SU($\infty$)
\cite{Teper:1998te,Teper:1998kw,Lucini:2001ej}.
%\cite{teper98,teper99,Lucini:2001ej}
%
Fermion effects are non-leading in $N_c$, so we expect $\chi \to
\chiqu$ for any fixed value of $m_q$ as the number of colours $N_c
\rightarrow \infty$. For small $m_q$ we expect 
\be
\chi = \frac{\chinfty M_\pi^2}
{\frac{4 N_f \chinfty}{f_\infty^2} + M_\pi^2},
\label{eqn_nlge_form}
\ee
with $\chinfty$, $f_\infty$ the quantities at leading order in $N_c$
\cite{Leutwyler:1992yt}.
%\cite{leutwyler92} 
%
Alternatively, our $m_q \simeq m_{\rm strange}$ and perhaps
higher order terms are important. In the
absence of a QCD prediction, 
\be
\chi  =  
\frac{f_{\pi}^2}{2 \pi N_f}  M_{\pi}^2
\arctan \left(\frac{2 \pi N_f}{f_{\pi}^2} \chiqu
\frac{1}{M_{\pi}^2} \right)
\label{eqn_mlge_form} 
\ee
interpolates between (\ref{eqn_chi_pi2}) and the quenched 
limit.\footnote{Note that, in describing chiral extrapolations, 
we adopt the common convention of
using $\pi$ to label quantities associated with the 
pseudoscalar channel irrespective 
of the quark mass}.
Measurements of $\chi$ were made on a number of ensembles of $N_f=2$
lattice field configurations. We
reiterate here that these ensembles have two notable features. The
improvement is fully non--perturbative, with discretisation errors
being quadratic rather than linear in the lattice spacing. Second, the
couplings are chosen to maintain an approximately constant lattice
spacing (as defined by the Sommer scale, $r_0=0.49 \ \fm$~%
\cite{Sommer:1994ce}) 
%\cite{sommer94}) 
%
as the quark mass is varied. This is important, as the susceptibility
in gluodynamics varies considerably with the lattice spacing
\cite{Teper:1998kw,Lucini:2001ej},
%\cite{teper99,Lucini:2001ej},
%
in competition with the variation with $m_q$.  The topological
susceptibility is measured from the gauge fields after cooling to
remove the UV noise. Further details of the procedure may be found in
\cite{Hart:2000wr,top_prog}.

We plot data for the ensembles presented in this paper in
Figs.~\ref{fig_r04} and~\ref{fig_r02}, as well as for preliminary
results for two further data sets at $(\beta,\kappa) = (5.20,0.13565)$
and $(5.25,0.13520)$. Also shown, as a band, is the interpolated
$\chiqu$ at an equivalent lattice spacing. Owing to the systematic
differences in the methods for determining $\hat{r}_0$ (which can
amount to a 20\% difference in $\hat{r}_0^4$),
the value chosen is for the quenched coupling $\beta=5.93$, taken from
\cite{Lucini:2001ej},
where we have an estimate of $\hat{r}_0$ determined in a consistent
manner. The variation in the equivalent quenched susceptibility over
the range in $\hat{r}_0$ spanned by our data is much smaller
than the error on the $\beta=5.93$ point shown, a useful consequence
of the matching programme.

The behaviour of $\hat{r}_0^4 \hat{\chi}$ with $(\hat{r_0}
\hat{M}_\pi)^2$ is qualitatively as expected and, more quantitatively,
we attempt fits motivated by
(\ref{eqn_chi_pi2}),~(\ref{eqn_nlge_form})
and~(\ref{eqn_mlge_form}). The leading order chiral behaviour will be
\be
\frac{\hat{r_0}^2 \hat{\chi}}{M_\pi^2}=
c_0, 
\label{eqn_fit_fl}
\ee
with the first correction term generically being
\be
\frac{\hat{r_0}^2 \hat{\chi}}{M_\pi^2}=
c_0 + c_1 (\hat{r_0} \hat{M}_\pi)^2.
\label{eqn_fit_fl_lin}
\ee
Attempting to include data further from the chiral limit, large-$N_c$
theory suggests a functional form
\be
\frac{\hat{r_0}^2 \hat{\chi}}{M_\pi^2}=
\frac{c_0 c_3}
{c_3 + c_0 (\hat{r_0} \hat{M}_\pi)^2},
\label{eqn_fit_nlge}
\ee
whilst a more general interpolation is provided by
\be 
\frac{\hat{r_0}^2 \hat{\chi}}{M_\pi^2}= \frac{2c_0}{\pi} {\rm arctan}
\left( \frac{\pi c_3}{2 c_0 (\hat{r_0} \hat{M}_\pi)^2} \right).
\label{eqn_fit_atan} 
\ee
In each case the intercept is related to the decay constant by $c_0 =
(\hat{r}_0 \hat{f}_\pi)^2/8$. The corresponding fits are shown in
Figs.~\ref{fig_r04} and~\ref{fig_r02}. The extent of the curves
indicates which points were included in fit. We include progressively
less chiral points until the $\chi^2/{\rm dof}$ of the fit becomes
unacceptably bad. We note the wide range fitted simply by including an
$M_\pi^4$ term, and the consistency of our data with large-$N_c$
predictions. The stability and similarity of the fits motivates us to
use $c_0$ from (\ref{eqn_fit_fl_lin}) to estimate $f_\pi = 149 \ \pm 8
\ ^{+25}_{-14} \ \MeV$ at a lattice spacing $a \simeq 0.1 \ \fm$, with
variation between other fits providing the second, systematic error,
and in good agreement with the experimental value around $132 \ \MeV$.

%
% conclusions
%

\section{Conclusions}
\label{sec:conc}

Two particular features distinguish this work from
previous published reports on lattice simulations of
QCD with dynamical fermions. It represents the first 
presentation of a wide range of 
results using the fully non-perturbativley improved
Wilson action. It also demonstrates the value of a new strategy
of using so-called `matched ensembles' which
allows a more controlled study of unquenching effects 
than would otherwise be possible at finite lattice spacing.

We have presented detailed measurements of 
the static inter-quark potential, 
light hadron spectrum, 
scalar and tensor glueballs,
torelon states
and the topological charge and susceptibility.

From the analysis of these quantities, 
we have presented significant evidence of
effects attributable to dynamical effects (two flavours
of light quarks) on
\begin{itemize}
\item 
the static inter-quark potential, particularly at 
short range (section~\ref{sec:chscreen});
\item 
the topological susceptibility (section~\ref{sec:topol});
\end{itemize}

We have also seen some evidence of dynamical quark effects in
\begin{itemize}
\item 
the effective string tension (section~\ref{sec:lspac});
\item 
the nucleon mass (section~\ref{sec:edinplot});
\item 
the scalar glueball mass (section~\ref{sec:gball});
\end{itemize}

For the present range of light quark masses ($M_\pi/M_\rho\gtap 0.58$)
there is no convincing evidence of effects on the light meson
spectrum. Nor do we see evidence of string breaking, save indirectly
in the small, but non--zero, VEV of the winding gluonic flux tube
(torelon) operator.

Further analyses of these ensembles and complementary ones
being produced by the QCDSF
collaboration~\cite{Pleiter:2000qd,Booth:2001qp}
are underway.

\section*{Acknowledgements}

We acknowledge the support of the U.K. Particle Physics and Astronomy
Research Council under grants GR/L22744, GR/L29927, GR/L56374,
PPA/G/O/1998/00621 and PPA/G/O/1998/00518.
A.H. wishes to thank the Aspen Center for
Physics for its hospitality during part of this work.  H.W.
acknowledges the support of PPARC through the award of an Advanced
Fellowship. C.R.A. wishes to thank R.G. Edwards for useful conversations.

%
% REFS
%
% Before submission, comment out the BibTeX lines 
%
%\bibliographystyle{h-elsevier2}
%\bibliographystyle{h-physrev3}
%\bibliography{csw202paper}
%
% and insert csw202paper.bbl directly here
%
%

\clearpage

%
% TABLES
%
%
%----------------------------------------------------------------------
\begin{table}
\begin{center}
\begin{tabular}{llrllllll}
$\beta$ & $\csw$ & \#conf. & $\ksea$ &
               \multicolumn{5}{c}{$\kval$} \\
\hline
 5.20 & 2.0171 &  244 & 0.13565 & 0.13565 & & & &
\\
 5.20 & 2.0171 &  832 & 0.1355 & 0.1355 & 0.1350 & 0.1345 & 0.1340 &
\\
\hline
 5.20 & 2.0171 &  600 & 0.1350 & 0.1350 & 0.1345 & 0.1340 & 0.1335 &
\\
 5.26 & 1.9497 &  404 & 0.1345 & 0.1350 & 0.1345 & 0.1340 & 0.1335 &
\\
 5.29 & 1.9192 &  404 & 0.1340 & 0.1350 & 0.1345 & 0.1340 & 0.1335 &
\\
\hline 
 5.93 & 1.82   &  623 & 0 & 0.1339 & 0.1337 & 0.1334 & 0.1332 &
0.1327 \\
\end{tabular}
\end{center}
\caption{Summary of simulation parameters and statistics used in the
computation of the static potential and light hadron spectrum.}
\label{tb:config_sum}
\end{table}

%----------------------------------------------------------------------
\begin{table}[tp]
\begin{center}
\begin{tabular}{llllll}
$L^3\cdot T$ &$\beta$ &$\csw$  &$\ksea$ & \#traj. & $\tint$ \\
\hline
$16^3\cdot32$   &5.20 &2.0171 & 0.13565 & 2400 & 13(5)       \\
                &5.20 &2.0171 & 0.13550 & 8000 & 14(1)       \\
                &5.20 &2.0171 & 0.13500 & 6000 & 16(3)       \\
                &5.26 &1.9497 & 0.13450 & 6000 & 18(3)       \\
                &5.29 &1.9192 & 0.13400 & 5000 & 25(7)       \\
\hline
$16^3\cdot24$   &5.20 &1.76   & 0.1390 &  3800 & 37(3)    \\
                &5.20 &1.76   & 0.1395 &  3200 & 27(18)       \\
                &5.20 &1.76   & 0.1398 &  3000 & 32(8)\\
\end{tabular}
\end{center}
\caption{Comparison of integrated autocorrelation times $\tint$
for the average plaquette measured in the present simulations
with those in previous simulations at 
$\beta=5.20$, $\csw=1.76$.}
\label{tb:autocor}
\end{table}
%----------------------------------------------------------------------

\begin{table}
\begin{center}
\begin{tabular}{cr@{.}lr@{.}l}
$(\beta,\kappa)$ & 
\multicolumn{2}{c}{$L/r_0$} &
\multicolumn{2}{c}{$L M_\pi$} \\
\hline
(5.20, 0.13565) & 3&07 (3) & 4&18 (5) \\
(5.20, 0.13550) & 3&17 (3) & 4&70 (6) \\
(5.20, 0.13500) & 3&37 (3) & 6&48 (8) \\
(5.26, 0.13450) & 3&40 (4) & 8&14 (3) \\
(5.29, 0.13400) & 3&32 (3) & 9&23 (4) \\
\end{tabular}
\caption{ \label{tab_fin_vol}
  { Measures of finite volume effects in simulations.}}
\end{center}
\end{table}
%----------------------------------------------------------------------

\begin{table}[ht]
\begin{center}
\begin{tabular}{cllllll}
\multicolumn{7}{c}{$(\beta,\ksea)$} \\
\hline
$\rvec$  &$(5.20,.1350)$ &$(5.26,.1345)$ &$(5.29,.1340)$  &$(5.93,0)$ &$(5.2,.1355)$ &$(5.2,.13565)$\\
\hline
$(1,0,0)$ &$0.4823(02)$ &$0.4739(04)$   &$0.4707(03)$   &$0.4259(01)$   &$0.4762(02)$  &$0.4749(02)$ \\
$(2,0,0)$ &$0.6970(08)$ &$0.6839(11)$   &$0.6782(10)$   &$0.6268(03)$   &$0.6832(08)$  &$0.6794(06)$ \\
$(3,0,0)$ &$0.8253(17)$ &$0.8100(15)$   &$0.8027(17)$   &$0.7439(05)$   &$0.7999(14)$  &$0.7954(12)$ \\
$(4,0,0)$ &$0.9193(22)$ &$0.9001(27)$   &$0.8920(28)$   &$0.8307(06)$   &$0.8839(18)$  &$0.8745(14)$ \\
$(5,0,0)$ &$0.9945(30)$ &$0.9777(36)$   &$0.9654(36)$   &$0.9070(07)$   &$0.9504(28)$  &$0.939(02)$ \\
$(6,0,0)$ &$1.0628(43)$ &$1.042(06)$   &$1.0342(43)$   &$0.9780(09)$   &$1.0168(29)$   &$1.002(02)$ \\
$(7,0,0)$ &$1.130(06)$ &$1.105(06)$   &$1.098(07)$   &$1.0484(13)$   &$1.0828(39)$   &$1.061(04)$ \\
$(8,0,0)$ &$1.183(08)$ &$1.175(09)$   &$1.170(11)$   &$1.1117(16)$   &$1.135(05)$   &$1.114(04)$ \\
$(9,0,0)$ &$1.262(11)$ &$1.244(11)$   &$1.244(11)$   &$1.1802(26)$   &$1.186(07)$   &$1.165(05)$ \\
$(10,0,0) $&$1.321(17)$ &$1.285(21)$   &$1.310(15)$  &$1.243(4)$   &$1.246(08)$  &$1.221(07)$ \\
$(11,0,0) $&$1.398(21)$ &$1.414(23)$   &$1.367(16)$  &$1.301(5)$   &$1.298(10)$  &$1.277(11)$ \\
$(12,0,0) $&$1.467(24)$ &-         &  -              &$1.365(8)$   &$1.330(17)$   &$1.287(25)$ \\
\hline
$(1,1,0)$ &$0.6276(05)$ &$0.6156(06)$   &$0.6103(07)$   &$0.5514(2)$   &$0.6173(05)$   &$0.6140(04)$ \\
$(2,1,0)$ &$0.7495(09)$ &$0.7315(13)$   &$0.7288(11)$   &$0.6671(4)$   &$0.7310(09)$   &$0.7262(07)$ \\
$(2,2,0)$ &$0.8163(14)$ &$0.8001(17)$   &$0.7940(15)$   &$0.7319(5)$   &$0.7944(10)$   &$0.7884(10)$ \\
$(3,1,0)$ &$0.8483(15)$ &$0.8296(16)$   &$0.8226(16)$   &$0.7616(6)$   &$0.8215(15)$   &$0.8138(11)$ \\
$(3,2,0)$ &$0.8873(18)$ &$0.8687(27)$   &$0.8636(23)$   &$0.8009(7)$   &$0.8599(15)$   &$0.8497(14)$ \\
$(3,3,0)$ &$0.9387(24)$ &$0.9235(26)$   &$0.9122(22)$   &$0.8517(9)$   &$0.9051(17)$   &$0.8939(18)$ \\
\hline
\# conf. & $150$        &$101$          &$101$          &$623$          &$208$          &244\\
traj. spac. & $40$  &$40$   &$40$   &-   &$40$   &10$\times 2$  \\
\end{tabular}
\end{center}
\caption{The static potential $V(\rvec)$
in lattice units. For the preliminary data at $\ksea=0.13565$ the
configurations
were measured every 10 trajectories and analysed in bins of 2.}
\label{tb:Vr}
\end{table}
%----------------------------------------------------------------------

%
\begin{table}[th]
\begin{center}
\begin{tabular}{lllll}
$(\beta,\ksea)$         &$r_0/a$                &$a [fm]$       &$e$    &$\sqrt{\sigma}[MeV]$ \\
\hline
$(5.2,.13565)$          &$5.21(05)(+0-8)$ &$0.0941(8)(+13-0)$ &$0.315(7)(+18-11)$ &$465(1)(+19-3)$ \\
$(5.2,.13550)$          &$5.041(40)(+0-10)$ &$0.0972(8)(+7-0)$ &$0.307(6)(+17-1)$ &$467(1)(+17-3)$ \\
\hline
$(5.20,.1350)$  &$4.754(40)(+2-90)$ &$0.1031(09)(+20-1)$ &$0.326(07)(+32-12)$ &$463(2)(+2-6)$ \\
$(5.26,.1345)$  &$4.708(52)(+45-50)$ &$0.1041(12)(+11-10)$ &$0.298(09)(+100-8)$ &$468(2)(+2-18)$ \\
$(5.29,.1340)$  &$4.813(45)(+35-84)$ &$0.1018(10)(+20-7)$ &$0.310(10)(+0-61)$ &$466(2)(+10-0)$ \\
$(5.93,0)$              &$4.714(13)(+0-18)$ &$0.1040(03)(+4-0)$ &$0.276(03)(+17-2)$ &$471(1)(+21-3)$ \\
\end{tabular}
\end{center}
\caption{Sommer scale $r_0$ and other parameters deduced from the lattice potential.} 
\label{tb:r0}
\end{table}
%----------------------------------------------------------------------

%% /research/python/staff/pyca/ukqcd/csw202/tex/M_p5.px
%% Tue May 22 16:23:03 BST 2001
\begin{table}[*htbp]
\begin{center}
 \begin{tabular}{cccccccc}
  & $\beta$ & $\ksea$ & $\kvala$ & $\kvalb$ & $r_0 M_{PS}$ & $a M_{PS}$ & \\
 &&&&&&&\\
\hline
 &&&&&&&\\
  & 5.2000 & 0.1355 & 0.1340 & 0.1340 & 2.39\err{ 3}{ 2} & 0.473\err{ 2}{ 2} & \\
  & 5.2000 & 0.1355 & 0.1345 & 0.1340 & 2.25\err{ 3}{ 2} & 0.447\err{ 2}{ 2} & \\
  & 5.2000 & 0.1355 & 0.1345 & 0.1345 & 2.12\err{ 3}{ 2} & 0.420\err{ 2}{ 2} & \\
  & 5.2000 & 0.1355 & 0.1350 & 0.1340 & 2.12\err{ 3}{ 2} & 0.420\err{ 2}{ 2} & \\
  & 5.2000 & 0.1355 & 0.1350 & 0.1345 & 1.97\err{ 3}{ 2} & 0.391\err{ 3}{ 2} & \\
  & 5.2000 & 0.1355 & 0.1350 & 0.1350 & 1.82\err{3}{1} & 0.362\err{ 3}{ 3} & \\
  & 5.2000 & 0.1355 & 0.1355 & 0.1340 & 1.98\err{3}{1} & 0.392\err{ 3}{ 2} & \\
  & 5.2000 & 0.1355 & 0.1355 & 0.1345 & 1.82\err{3}{1} & 0.362\err{ 3}{ 3} & \\
  & 5.2000 & 0.1355 & 0.1355 & 0.1350 & 1.66\err{3}{1} & 0.329\err{ 3}{ 3} & \\
  & 5.2000 & 0.1355 & 0.1355 & 0.1355 & 1.48\err{ 3}{ 2} & 0.294\err{ 4}{ 3} & \\
 &&&&&&&\\
\hline
 &&&&&&&\\
  & 5.2000 & 0.1350 & 0.1335 & 0.1335 & 2.68\err{ 2}{ 3} & 0.563\err{ 3}{ 3} & \\
  & 5.2000 & 0.1350 & 0.1340 & 0.1335 & 2.56\err{ 2}{ 3} & 0.539\err{ 3}{ 4} & \\
  & 5.2000 & 0.1350 & 0.1340 & 0.1340 & 2.45\err{ 2}{ 3} & 0.514\err{ 3}{ 4} & \\
  & 5.2000 & 0.1350 & 0.1345 & 0.1335 & 2.44\err{ 2}{ 3} & 0.514\err{ 3}{ 4} & \\
  & 5.2000 & 0.1350 & 0.1345 & 0.1340 & 2.32\err{ 2}{ 3} & 0.489\err{ 3}{ 4} & \\
  & 5.2000 & 0.1350 & 0.1345 & 0.1345 & 2.20\err{ 2}{ 3} & 0.462\err{ 4}{ 5} & \\
  & 5.2000 & 0.1350 & 0.1350 & 0.1335 & 2.32\err{ 2}{ 3} & 0.488\err{ 3}{ 4} & \\
  & 5.2000 & 0.1350 & 0.1350 & 0.1340 & 2.20\err{ 2}{ 3} & 0.462\err{ 4}{ 5} & \\
  & 5.2000 & 0.1350 & 0.1350 & 0.1345 & 2.06\err{ 2}{ 3} & 0.434\err{ 4}{ 5} & \\
  & 5.2000 & 0.1350 & 0.1350 & 0.1350 & 1.93\err{ 2}{ 3} & 0.405\err{ 4}{ 5} & \\
 &&&&&&&\\
\hline
 &&&&&&&\\
  & 5.2600 & 0.1345 & 0.1335 & 0.1335 & 2.85\err{ 2}{ 4} & 0.603\err{ 2}{ 2} & \\
  & 5.2600 & 0.1345 & 0.1340 & 0.1335 & 2.74\err{ 2}{ 4} & 0.580\err{ 2}{ 2} & \\
  & 5.2600 & 0.1345 & 0.1340 & 0.1340 & 2.63\err{ 2}{ 4} & 0.557\err{ 2}{ 2} & \\
  & 5.2600 & 0.1345 & 0.1345 & 0.1335 & 2.63\err{ 2}{ 4} & 0.557\err{ 2}{ 2} & \\
  & 5.2600 & 0.1345 & 0.1345 & 0.1340 & 2.52\err{ 2}{ 4} & 0.533\err{ 2}{ 2} & \\
  & 5.2600 & 0.1345 & 0.1345 & 0.1345 & 2.41\err{ 2}{ 4} & 0.509\err{ 2}{ 2} & \\
  & 5.2600 & 0.1345 & 0.1350 & 0.1335 & 2.52\err{ 2}{ 4} & 0.533\err{ 2}{ 2} & \\
  & 5.2600 & 0.1345 & 0.1350 & 0.1340 & 2.41\err{ 2}{ 4} & 0.509\err{ 2}{ 2} & \\
  & 5.2600 & 0.1345 & 0.1350 & 0.1345 & 2.29\err{ 2}{ 4} & 0.484\err{ 2}{ 2} & \\
  & 5.2600 & 0.1345 & 0.1350 & 0.1350 & 2.16\err{ 2}{ 3} & 0.458\err{ 2}{ 2} & \\
 &&&&&&&\\
\hline
 &&&&&&&\\
  & 5.2900 & 0.1340 & 0.1335 & 0.1335 & 2.99\err{ 2}{ 4} & 0.621\err{ 2}{ 2} & \\
  & 5.2900 & 0.1340 & 0.1340 & 0.1335 & 2.88\err{ 2}{ 4} & 0.599\err{ 2}{ 2} & \\
  & 5.2900 & 0.1340 & 0.1340 & 0.1340 & 2.78\err{ 2}{ 4} & 0.577\err{ 2}{ 2} & \\
  & 5.2900 & 0.1340 & 0.1345 & 0.1335 & 2.78\err{ 2}{ 4} & 0.577\err{ 2}{ 2} & \\
  & 5.2900 & 0.1340 & 0.1345 & 0.1340 & 2.67\err{ 2}{ 4} & 0.554\err{ 2}{ 2} & \\
  & 5.2900 & 0.1340 & 0.1345 & 0.1345 & 2.55\err{ 2}{ 4} & 0.530\err{ 2}{ 3} & \\
  & 5.2900 & 0.1340 & 0.1350 & 0.1335 & 2.67\err{ 2}{ 4} & 0.554\err{ 2}{ 2} & \\
  & 5.2900 & 0.1340 & 0.1350 & 0.1340 & 2.55\err{ 2}{ 4} & 0.530\err{ 2}{ 3} & \\
  & 5.2900 & 0.1340 & 0.1350 & 0.1345 & 2.43\err{ 2}{ 4} & 0.506\err{ 2}{ 3} & \\
  & 5.2900 & 0.1340 & 0.1350 & 0.1350 & 2.31\err{ 2}{ 3} & 0.480\err{ 3}{ 3} & \\
 &&&&&&&\\
\hline
 &&&&&&&\\
  & 5.9300 & 0.0000 & 0.1327 & 0.1327 & 2.334\err{ 6}{10} & 0.495\err{1}{1} & \\
  & 5.9300 & 0.0000 & 0.1332 & 0.1327 & 2.211\err{ 6}{ 9} & 0.469\err{1}{1} & \\
  & 5.9300 & 0.0000 & 0.1332 & 0.1332 & 2.081\err{ 6}{ 9} & 0.442\err{1}{1} & \\
  & 5.9300 & 0.0000 & 0.1334 & 0.1327 & 2.159\err{ 6}{ 9} & 0.458\err{1}{1} & \\
  & 5.9300 & 0.0000 & 0.1334 & 0.1332 & 2.028\err{ 6}{ 9} & 0.430\err{1}{1} & \\
  & 5.9300 & 0.0000 & 0.1334 & 0.1334 & 1.973\err{ 6}{ 9} & 0.419\err{1}{1} & \\
  & 5.9300 & 0.0000 & 0.1337 & 0.1337 & 1.800\err{ 6}{ 9} & 0.382\err{1}{1} & \\
  & 5.9300 & 0.0000 & 0.1339 & 0.1337 & 1.739\err{ 6}{ 9} & 0.369\err{1}{1} & \\
  & 5.9300 & 0.0000 & 0.1339 & 0.1339 & 1.676\err{ 6}{ 9} & 0.356\err{1}{1} & \\
\end{tabular}
\end{center}
\caption{Pseudoscalar meson masses for all data sets.}
\protect\label{tb:ps}
\end{table}

%}}}

%{{{    table {tb:vector}
%% /research/python/staff/pyca/ukqcd/csw202/tex/M_Vi.px
%% Tue May 22 16:23:40 BST 2001
\begin{table}[*htbp]
\begin{center}
 \begin{tabular}{cccccccc}
  & $\beta$ & $\ksea$ & $\kvala$ & $\kvalb$ & $r_0 M_{V}$ & $a M_V$ & \\
 &&&&&&&\\
\hline
 &&&&&&&\\
  & 5.2000 & 0.1355 & 0.1340 & 0.1340 & 3.01\err{ 5}{ 2} & 0.596\err{ 6}{ 5} & \\
  & 5.2000 & 0.1355 & 0.1345 & 0.1340 & 2.92\err{ 5}{ 3} & 0.578\err{ 6}{ 6} & \\
  & 5.2000 & 0.1355 & 0.1345 & 0.1345 & 2.82\err{ 5}{ 3} & 0.560\err{ 7}{ 6} & \\
  & 5.2000 & 0.1355 & 0.1350 & 0.1340 & 2.84\err{ 5}{ 3} & 0.563\err{ 7}{ 6} & \\
  & 5.2000 & 0.1355 & 0.1350 & 0.1345 & 2.75\err{ 6}{ 3} & 0.546\err{ 8}{ 7} & \\
  & 5.2000 & 0.1355 & 0.1350 & 0.1350 & 2.68\err{ 7}{ 4} & 0.531\err{10}{ 8} & \\
  & 5.2000 & 0.1355 & 0.1355 & 0.1340 & 2.79\err{ 6}{ 4} & 0.553\err{10}{ 8} & \\
  & 5.2000 & 0.1355 & 0.1355 & 0.1345 & 2.71\err{ 7}{ 4} & 0.537\err{11}{ 9} & \\
  & 5.2000 & 0.1355 & 0.1355 & 0.1350 & 2.63\err{ 8}{ 5} & 0.522\err{13}{ 9} & \\
  & 5.2000 & 0.1355 & 0.1355 & 0.1355 & 2.56\err{10}{ 4} & 0.508\err{18}{10} & \\
 &&&&&&&\\
\hline
 &&&&&&&\\
  & 5.2000 & 0.1350 & 0.1335 & 0.1335 & 3.31\err{ 3}{ 4} & 0.695\err{ 4}{ 4} & \\
  & 5.2000 & 0.1350 & 0.1340 & 0.1335 & 3.22\err{ 3}{ 4} & 0.677\err{ 4}{ 5} & \\
  & 5.2000 & 0.1350 & 0.1340 & 0.1340 & 3.13\err{ 3}{ 4} & 0.658\err{ 5}{ 5} & \\
  & 5.2000 & 0.1350 & 0.1345 & 0.1335 & 3.13\err{ 3}{ 4} & 0.658\err{ 5}{ 5} & \\
  & 5.2000 & 0.1350 & 0.1345 & 0.1340 & 3.04\err{ 3}{ 4} & 0.638\err{ 5}{ 6} & \\
  & 5.2000 & 0.1350 & 0.1345 & 0.1345 & 2.94\err{ 3}{ 4} & 0.619\err{ 6}{ 7} & \\
  & 5.2000 & 0.1350 & 0.1350 & 0.1335 & 3.03\err{ 3}{ 4} & 0.638\err{ 5}{ 5} & \\
  & 5.2000 & 0.1350 & 0.1350 & 0.1340 & 2.94\err{ 3}{ 4} & 0.618\err{ 6}{ 6} & \\
  & 5.2000 & 0.1350 & 0.1350 & 0.1345 & 2.85\err{ 3}{ 5} & 0.599\err{ 6}{ 7} & \\
  & 5.2000 & 0.1350 & 0.1350 & 0.1350 & 2.75\err{ 4}{ 5} & 0.579\err{ 7}{ 9} & \\
 &&&&&&&\\
\hline
 &&&&&&&\\
  & 5.2600 & 0.1345 & 0.1335 & 0.1335 & 3.41\err{ 3}{ 5} & 0.721\err{ 4}{ 4} & \\
  & 5.2600 & 0.1345 & 0.1340 & 0.1335 & 3.32\err{ 3}{ 5} & 0.703\err{ 4}{ 4} & \\
  & 5.2600 & 0.1345 & 0.1340 & 0.1340 & 3.24\err{ 3}{ 5} & 0.685\err{ 4}{ 4} & \\
  & 5.2600 & 0.1345 & 0.1345 & 0.1335 & 3.24\err{ 3}{ 5} & 0.685\err{ 4}{ 4} & \\
  & 5.2600 & 0.1345 & 0.1345 & 0.1340 & 3.16\err{ 3}{ 5} & 0.668\err{ 4}{ 4} & \\
  & 5.2600 & 0.1345 & 0.1345 & 0.1345 & 3.07\err{ 3}{ 5} & 0.650\err{ 4}{ 4} & \\
  & 5.2600 & 0.1345 & 0.1350 & 0.1335 & 3.16\err{ 3}{ 5} & 0.668\err{ 4}{ 5} & \\
  & 5.2600 & 0.1345 & 0.1350 & 0.1340 & 3.08\err{ 3}{ 5} & 0.651\err{ 4}{ 5} & \\
  & 5.2600 & 0.1345 & 0.1350 & 0.1345 & 2.99\err{ 3}{ 5} & 0.633\err{ 5}{ 5} & \\
  & 5.2600 & 0.1345 & 0.1350 & 0.1350 & 2.91\err{ 3}{ 5} & 0.614\err{ 5}{ 5} & \\
 &&&&&&&\\
\hline
 &&&&&&&\\
  & 5.2900 & 0.1340 & 0.1335 & 0.1335 & 3.49\err{ 3}{ 6} & 0.725\err{ 5}{ 5} & \\
  & 5.2900 & 0.1340 & 0.1340 & 0.1335 & 3.41\err{ 3}{ 6} & 0.708\err{ 6}{ 6} & \\
  & 5.2900 & 0.1340 & 0.1340 & 0.1340 & 3.32\err{ 3}{ 6} & 0.691\err{ 6}{ 6} & \\
  & 5.2900 & 0.1340 & 0.1345 & 0.1335 & 3.32\err{ 3}{ 6} & 0.691\err{ 6}{ 6} & \\
  & 5.2900 & 0.1340 & 0.1345 & 0.1340 & 3.24\err{ 4}{ 6} & 0.674\err{ 6}{ 6} & \\
  & 5.2900 & 0.1340 & 0.1345 & 0.1345 & 3.16\err{ 4}{ 6} & 0.656\err{ 7}{ 7} & \\
  & 5.2900 & 0.1340 & 0.1350 & 0.1335 & 3.24\err{ 4}{ 6} & 0.674\err{ 7}{ 7} & \\
  & 5.2900 & 0.1340 & 0.1350 & 0.1340 & 3.16\err{ 4}{ 6} & 0.656\err{ 7}{ 7} & \\
  & 5.2900 & 0.1340 & 0.1350 & 0.1345 & 3.08\err{ 4}{ 6} & 0.639\err{ 8}{ 8} & \\
  & 5.2900 & 0.1340 & 0.1350 & 0.1350 & 3.00\err{ 4}{ 6} & 0.623\err{ 8}{ 8} & \\
 &&&&&&&\\
\hline
 &&&&&&&\\
  & 5.9300 & 0.0000 & 0.1327 & 0.1327 & 3.05\err{1}{2} & 0.646\err{ 2}{ 3} & \\
  & 5.9300 & 0.0000 & 0.1332 & 0.1327 & 2.97\err{1}{2} & 0.629\err{ 3}{ 3} & \\
  & 5.9300 & 0.0000 & 0.1332 & 0.1332 & 2.88\err{ 2}{ 2} & 0.612\err{ 3}{ 3} & \\
  & 5.9300 & 0.0000 & 0.1334 & 0.1327 & 2.93\err{ 2}{ 2} & 0.622\err{ 3}{ 3} & \\
  & 5.9300 & 0.0000 & 0.1334 & 0.1332 & 2.85\err{ 2}{ 2} & 0.605\err{ 3}{ 3} & \\
  & 5.9300 & 0.0000 & 0.1334 & 0.1334 & 2.82\err{ 2}{ 2} & 0.598\err{ 3}{ 4} & \\
  & 5.9300 & 0.0000 & 0.1337 & 0.1337 & 2.72\err{ 2}{ 2} & 0.577\err{ 4}{ 4} & \\
  & 5.9300 & 0.0000 & 0.1339 & 0.1337 & 2.69\err{ 2}{ 2} & 0.570\err{ 5}{ 4} & \\
  & 5.9300 & 0.0000 & 0.1339 & 0.1339 & 2.66\err{ 3}{ 3} & 0.563\err{ 5}{ 5} & \\
\end{tabular}
\end{center}
\caption{Vector meson masses for all data sets.}
\protect\label{tb:vector}
\end{table}

%}}}
%{{{    table {tb:proton}
%% /research/python/staff/pyca/ukqcd/csw202/tex/M_proton.px
%% Tue May 22 16:23:29 BST 2001
\begin{table}[*htbp]
\begin{center}
 \begin{tabular}{ccccccc}
  & $\beta$ & $\ksea$ & $\kval$ & $r_0 M_N$ & $a M_N$ & \\
 &&&&&&\\
\hline
 &&&&&&\\
  & 5.2000 & 0.1355 & 0.1340 & 4.75\err{ 9}{ 6} & 0.942\err{12}{13} &\\
  & 5.2000 & 0.1355 & 0.1345 & 4.42\err{ 9}{ 6} & 0.876\err{15}{15} &\\
  & 5.2000 & 0.1355 & 0.1350 & 4.09\err{10}{ 7} & 0.81\err{ 2}{ 2} &\\
  & 5.2000 & 0.1355 & 0.1355 & 3.86\err{ 7}{ 5} & 0.766\err{10}{11} &\\
 &&&&&&\\
\hline
 &&&&&&\\
  & 5.2000 & 0.1350 & 0.1335 & 5.16\err{ 5}{ 6} & 1.086\err{ 8}{ 8} &\\
  & 5.2000 & 0.1350 & 0.1340 & 4.87\err{ 5}{ 6} & 1.024\err{ 8}{ 9} &\\
  & 5.2000 & 0.1350 & 0.1345 & 4.54\err{ 5}{ 7} & 0.954\err{ 8}{11} &\\
  & 5.2000 & 0.1350 & 0.1350 & 4.20\err{ 5}{ 7} & 0.883\err{10}{12} &\\
 &&&&&&\\
\hline
 &&&&&&\\
  & 5.2600 & 0.1345 & 0.1335 & 5.32\err{ 5}{ 9} & 1.125\err{ 8}{ 8} &\\
  & 5.2600 & 0.1345 & 0.1340 & 5.05\err{ 5}{ 9} & 1.068\err{ 9}{ 8} &\\
  & 5.2600 & 0.1345 & 0.1345 & 4.78\err{ 5}{ 9} & 1.011\err{10}{ 9} &\\
  & 5.2600 & 0.1345 & 0.1350 & 4.50\err{ 6}{ 9} & 0.951\err{10}{10} &\\
 &&&&&&\\
\hline
 &&&&&&\\
  & 5.2900 & 0.1340 & 0.1335 & 5.50\err{ 5}{ 9} & 1.143\err{ 8}{ 8} &\\
  & 5.2900 & 0.1340 & 0.1340 & 5.23\err{ 5}{ 8} & 1.086\err{ 9}{ 9} &\\
  & 5.2900 & 0.1340 & 0.1345 & 4.94\err{ 6}{ 9} & 1.027\err{10}{10} &\\
  & 5.2900 & 0.1340 & 0.1350 & 4.66\err{ 7}{ 9} & 0.968\err{13}{12} &\\
 &&&&&&\\
\hline
 &&&&&&\\
  & 5.9300 & 0.0000 & 0.1327 & 4.56\err{ 2}{ 3} & 0.968\err{ 5}{ 6} &\\
  & 5.9300 & 0.0000 & 0.1332 & 4.25\err{ 3}{ 4} & 0.902\err{ 5}{ 8} &\\
  & 5.9300 & 0.0000 & 0.1334 & 4.13\err{ 3}{ 4} & 0.876\err{ 6}{ 8} &\\
  & 5.9300 & 0.0000 & 0.1337 & 3.94\err{ 3}{ 5} & 0.836\err{ 7}{ 9} &\\
  & 5.9300 & 0.0000 & 0.1339 & 3.86\err{ 4}{ 4} & 0.818\err{ 7}{ 8} &\\
\end{tabular}
\end{center}
\caption{Nucleon masses for all data sets.}
\protect\label{tb:proton}
\end{table}

%}}}
%{{{    table {tb:delta}
%% /research/python/staff/pyca/ukqcd/csw202/tex/M_delta.px
%% Tue May 22 16:22:51 BST 2001
\begin{table}[*htbp]
\begin{center}
 \begin{tabular}{ccccccc}
  & $\beta$ & $\ksea$ & $\kval$ & $r_0 M_\Delta$ & $a M_\Delta$ & \\
 &&&&&&\\
\hline
 &&&&&&\\
  & 5.2000 & 0.1355 & 0.1340 & 5.12\err{10}{ 5} & 1.015\err{15}{12} &\\
  & 5.2000 & 0.1355 & 0.1345 & 4.87\err{10}{ 6} & 0.967\err{17}{15} &\\
  & 5.2000 & 0.1355 & 0.1350 & 4.64\err{11}{ 7} & 0.92\err{ 2}{ 2} &\\
  & 5.2000 & 0.1355 & 0.1355 & 4.30\err{15}{11} & 0.85\err{ 3}{ 2} &\\
 &&&&&&\\
\hline
 &&&&&&\\
  & 5.2000 & 0.1350 & 0.1335 & 5.57\err{ 6}{ 8} & 1.172\err{11}{11} &\\
  & 5.2000 & 0.1350 & 0.1340 & 5.31\err{ 6}{ 8} & 1.116\err{11}{12} &\\
  & 5.2000 & 0.1350 & 0.1345 & 5.02\err{ 7}{ 8} & 1.055\err{13}{15} &\\
  & 5.2000 & 0.1350 & 0.1350 & 4.75\err{ 8}{10} & 1.00\err{ 2}{ 2} &\\
 &&&&&&\\
\hline
 &&&&&&\\
  & 5.2600 & 0.1345 & 0.1335 & 5.61\err{ 5}{ 9} & 1.186\err{11}{10} &\\
  & 5.2600 & 0.1345 & 0.1340 & 5.36\err{ 5}{ 9} & 1.134\err{11}{11} &\\
  & 5.2600 & 0.1345 & 0.1345 & 5.11\err{ 6}{ 9} & 1.080\err{12}{11} &\\
  & 5.2600 & 0.1345 & 0.1350 & 4.83\err{ 7}{10} & 1.022\err{14}{13} &\\
 &&&&&&\\
\hline
 &&&&&&\\
  & 5.2900 & 0.1340 & 0.1335 & 5.80\err{ 6}{10} & 1.205\err{11}{10} &\\
  & 5.2900 & 0.1340 & 0.1340 & 5.56\err{ 6}{10} & 1.155\err{12}{11} &\\
  & 5.2900 & 0.1340 & 0.1345 & 5.33\err{ 6}{10} & 1.107\err{11}{13} &\\
  & 5.2900 & 0.1340 & 0.1350 & 5.09\err{ 7}{ 9} & 1.057\err{13}{12} &\\
 &&&&&&\\
\hline
 &&&&&&\\
  & 5.9300 & 0.0000 & 0.1327 & 5.09\err{ 3}{ 4} & 1.079\err{ 7}{ 8} &\\
  & 5.9300 & 0.0000 & 0.1332 & 4.84\err{ 4}{ 5} & 1.026\err{ 8}{ 9} &\\
  & 5.9300 & 0.0000 & 0.1334 & 4.74\err{ 4}{ 5} & 1.005\err{ 9}{ 9} &\\
  & 5.9300 & 0.0000 & 0.1337 & 4.58\err{ 5}{ 5} & 0.972\err{11}{11} &\\
  & 5.9300 & 0.0000 & 0.1339 & 4.47\err{ 6}{ 6} & 0.949\err{12}{11} &\\
\end{tabular}
\end{center}
\caption{Delta masses for all data sets.}
\protect\label{tb:delta}
\end{table}
%}}}

%{{{    table {tb:mpcac}
\begin{table}[*htbp]
\begin{center}
\begin{tabular}{cccccc}
$\beta$ &$\ksea$ & $\kvala$ & $\kvalb$ & $r_0 m_{\rm PCAC}$ & $a m_{\rm PCAC}$ \\
&&&&&\\
\hline
&&&&&\\
5.20 & 0.1355 & 0.1340 & 0.1340 & 0.329\err{4}{2} & 0.0652\err{2}{3} \\
5.20 & 0.1355 & 0.1345 & 0.1340 & 0.292\err{3}{2} & 0.0580\err{2}{3} \\
5.20 & 0.1355 & 0.1345 & 0.1345 & 0.256\err{3}{2} & 0.0508\err{2}{2} \\
5.20 & 0.1355 & 0.1350 & 0.1340 & 0.256\err{3}{2} & 0.0508\err{2}{3} \\
5.20 & 0.1355 & 0.1350 & 0.1345 & 0.221\err{3}{2} & 0.0438\err{2}{3} \\
5.20 & 0.1355 & 0.1350 & 0.1350 & 0.185\err{3}{2} & 0.0368\err{2}{3} \\
5.20 & 0.1355 & 0.1355 & 0.1340 & 0.221\err{3}{2} & 0.0438\err{2}{3} \\
5.20 & 0.1355 & 0.1355 & 0.1345 & 0.186\err{3}{2} & 0.0368\err{2}{3} \\
5.20 & 0.1355 & 0.1355 & 0.1350 & 0.151\err{2}{2} & 0.0299\err{3}{3} \\
5.20 & 0.1355 & 0.1355 & 0.1355 & 0.116\err{2}{2} & 0.0231\err{3}{3} \\
&&&&&\\
\hline
&&&&&\\
5.20 & 0.1350 & 0.1335 & 0.1335 & 0.424\err{3}{4} & 0.0893\err{2}{2} \\
5.20 & 0.1350 & 0.1340 & 0.1335 & 0.389\err{3}{4} & 0.0819\err{2}{2} \\
5.20 & 0.1350 & 0.1340 & 0.1340 & 0.355\err{2}{4} & 0.0746\err{2}{2} \\
5.20 & 0.1350 & 0.1345 & 0.1335 & 0.355\err{3}{4} & 0.0746\err{2}{2} \\
5.20 & 0.1350 & 0.1345 & 0.1340 & 0.320\err{2}{3} & 0.0674\err{2}{2} \\
5.20 & 0.1350 & 0.1345 & 0.1345 & 0.287\err{2}{3} & 0.0602\err{2}{3} \\
5.20 & 0.1350 & 0.1350 & 0.1335 & 0.320\err{2}{3} & 0.0674\err{2}{2} \\
5.20 & 0.1350 & 0.1350 & 0.1340 & 0.286\err{2}{3} & 0.0602\err{2}{3} \\
5.20 & 0.1350 & 0.1350 & 0.1345 & 0.253\err{2}{3} & 0.0532\err{2}{3} \\
5.20 & 0.1350 & 0.1350 & 0.1350 & 0.220\err{2}{2} & 0.0462\err{2}{3} \\
&&&&&\\
\hline
&&&&&\\
5.26 & 0.1345 & 0.1335 & 0.1335 & 0.491\err{3}{8} & 0.1038\err{3}{3} \\
5.26 & 0.1345 & 0.1340 & 0.1335 & 0.455\err{3}{7} & 0.0963\err{3}{3} \\
5.26 & 0.1345 & 0.1340 & 0.1340 & 0.420\err{3}{6} & 0.0888\err{3}{3} \\
5.26 & 0.1345 & 0.1345 & 0.1335 & 0.420\err{3}{6} & 0.0888\err{3}{3} \\
5.26 & 0.1345 & 0.1345 & 0.1340 & 0.385\err{3}{6} & 0.0815\err{3}{3} \\
5.26 & 0.1345 & 0.1345 & 0.1345 & 0.351\err{3}{6} & 0.0742\err{3}{3} \\
5.26 & 0.1345 & 0.1350 & 0.1335 & 0.385\err{3}{6} & 0.0814\err{3}{3} \\
5.26 & 0.1345 & 0.1350 & 0.1340 & 0.351\err{3}{6} & 0.0742\err{3}{3} \\
5.26 & 0.1345 & 0.1350 & 0.1345 & 0.317\err{2}{5} & 0.0670\err{3}{3} \\
5.26 & 0.1345 & 0.1350 & 0.1350 & 0.283\err{2}{5} & 0.0599\err{3}{3} \\
&&&&&\\
\hline
&&&&&\\
5.29 & 0.1340 & 0.1335 & 0.1335 & 0.530\err{4}{7} & 0.1101\err{3}{3} \\
5.29 & 0.1340 & 0.1340 & 0.1335 & 0.494\err{4}{7} & 0.1026\err{3}{3} \\
5.29 & 0.1340 & 0.1340 & 0.1340 & 0.458\err{4}{6} & 0.0952\err{3}{3} \\
5.29 & 0.1340 & 0.1345 & 0.1335 & 0.458\err{4}{6} & 0.0951\err{3}{3} \\
5.29 & 0.1340 & 0.1345 & 0.1340 & 0.423\err{4}{6} & 0.0878\err{3}{3} \\
5.29 & 0.1340 & 0.1345 & 0.1345 & 0.387\err{3}{5} & 0.0805\err{3}{3} \\
5.29 & 0.1340 & 0.1350 & 0.1335 & 0.422\err{4}{6} & 0.0877\err{3}{3} \\
5.29 & 0.1340 & 0.1350 & 0.1340 & 0.387\err{3}{5} & 0.0805\err{3}{3} \\
5.29 & 0.1340 & 0.1350 & 0.1345 & 0.353\err{3}{5} & 0.0733\err{3}{3} \\
5.29 & 0.1340 & 0.1350 & 0.1350 & 0.318\err{3}{5} & 0.0661\err{3}{3} \\
&&&&&\\
\hline
&&&&&\\
5.93 & 0.0000 & 0.1327 & 0.1327 & 0.3530\err{10}{12} & 0.07488\err{11}{11} \\
5.93 & 0.0000 & 0.1332 & 0.1327 & 0.3162\err{9}{11} & 0.06709\err{11}{11} \\
5.93 & 0.0000 & 0.1332 & 0.1332 & 0.2799\err{8}{10} & 0.05938\err{11}{11} \\
5.93 & 0.0000 & 0.1334 & 0.1327 & 0.3016\err{9}{11} & 0.06398\err{11}{12} \\
5.93 & 0.0000 & 0.1334 & 0.1332 & 0.2653\err{8}{10} & 0.05629\err{11}{12} \\
5.93 & 0.0000 & 0.1334 & 0.1334 & 0.2508\err{8}{10} & 0.05322\err{12}{11} \\
5.93 & 0.0000 & 0.1337 & 0.1337 & 0.2077\err{7}{8} & 0.04406\err{12}{11} \\
5.93 & 0.0000 & 0.1339 & 0.1337 & 0.1931\err{7}{8} & 0.04097\err{12}{11} \\
5.93 & 0.0000 & 0.1339 & 0.1339 & 0.1786\err{7}{8} & 0.03788\err{13}{12} \\
\end{tabular}
\end{center}
\caption{The quark mass $m_{\rm PCAC}$ as defined in eq.(\ref{eq:mpcac}) for
all the datasets.}
\protect\label{tb:mpcac}
\end{table}

%}}}
%{{{    table {tb:rs}
\begin{table}[*htbp]
\begin{center}
\begin{tabular}{cccccc}
$\beta$ & $\ksea$ & $\kvala$ & $\kvalb$ & $r(t)$ & $s(t)$ \\
&&&&&\\
\hline 
&&&&&\\
5.20 & 0.1355 & 0.1340 & 0.1340 & 0.0662\err{2}{3} & 0.1179\err{9}{8} \\
5.20 & 0.1355 & 0.1345 & 0.1340 & 0.0589\err{2}{3} & 0.1050\err{9}{8} \\
5.20 & 0.1355 & 0.1345 & 0.1345 & 0.0516\err{2}{3} & 0.0923\err{9}{8} \\
5.20 & 0.1355 & 0.1350 & 0.1340 & 0.0516\err{2}{2} & 0.0924\err{9}{8} \\
5.20 & 0.1355 & 0.1350 & 0.1345 & 0.0445\err{2}{3} & 0.0799\err{10}{8} \\
5.20 & 0.1355 & 0.1350 & 0.1350 & 0.0374\err{2}{3} & 0.0678\err{9}{8} \\
5.20 & 0.1355 & 0.1355 & 0.1340 & 0.0445\err{2}{3} & 0.0801\err{9}{8} \\
5.20 & 0.1355 & 0.1355 & 0.1345 & 0.0374\err{2}{3} & 0.0678\err{9}{8} \\
5.20 & 0.1355 & 0.1355 & 0.1350 & 0.0304\err{3}{3} & 0.0558\err{9}{8} \\
5.20 & 0.1355 & 0.1355 & 0.1355 & 0.0235\err{3}{3} & 0.0441\err{9}{7} \\
&&&&&\\
\hline
&&&&&\\
5.20 & 0.1350 & 0.1335 & 0.1335 & 0.0907\err{2}{2} & 0.1682\err{11}{9} \\
5.20 & 0.1350 & 0.1340 & 0.1335 & 0.0832\err{2}{2} & 0.1538\err{10}{9} \\
5.20 & 0.1350 & 0.1340 & 0.1340 & 0.0758\err{2}{2} & 0.1397\err{10}{10} \\
5.20 & 0.1350 & 0.1345 & 0.1335 & 0.0758\err{3}{2} & 0.1397\err{10}{10} \\
5.20 & 0.1350 & 0.1345 & 0.1340 & 0.0685\err{2}{3} & 0.1258\err{10}{9} \\
5.20 & 0.1350 & 0.1345 & 0.1345 & 0.0612\err{2}{3} & 0.1123\err{10}{9} \\
5.20 & 0.1350 & 0.1350 & 0.1335 & 0.0685\err{2}{3} & 0.1260\err{10}{10} \\
5.20 & 0.1350 & 0.1350 & 0.1340 & 0.0612\err{2}{3} & 0.1123\err{10}{10} \\
5.20 & 0.1350 & 0.1350 & 0.1345 & 0.0541\err{2}{3} & 0.0990\err{10}{10} \\
5.20 & 0.1350 & 0.1350 & 0.1350 & 0.0469\err{2}{3} & 0.0859\err{10}{9} \\
&&&&&\\
\hline
&&&&&\\
5.26 & 0.1345 & 0.1335 & 0.1335 & 0.1055\err{3}{3} & 0.1924\err{11}{9} \\
5.26 & 0.1345 & 0.1340 & 0.1335 & 0.0978\err{3}{3} & 0.1779\err{10}{9} \\
5.26 & 0.1345 & 0.1340 & 0.1340 & 0.0903\err{3}{3} & 0.1636\err{10}{8} \\
5.26 & 0.1345 & 0.1345 & 0.1335 & 0.0902\err{3}{3} & 0.1637\err{10}{8} \\
5.26 & 0.1345 & 0.1345 & 0.1340 & 0.0828\err{3}{3} & 0.1496\err{9}{8} \\
5.26 & 0.1345 & 0.1345 & 0.1345 & 0.0754\err{3}{3} & 0.1359\err{9}{8} \\
5.26 & 0.1345 & 0.1350 & 0.1335 & 0.0827\err{3}{3} & 0.1498\err{10}{8} \\
5.26 & 0.1345 & 0.1350 & 0.1340 & 0.0753\err{3}{3} & 0.1360\err{9}{8} \\
5.26 & 0.1345 & 0.1350 & 0.1345 & 0.0680\err{3}{3} & 0.1225\err{9}{8} \\
5.26 & 0.1345 & 0.1350 & 0.1350 & 0.0608\err{3}{3} & 0.1093\err{9}{8} \\
&&&&&\\
\hline
&&&&&\\
5.29 & 0.1340 & 0.1335 & 0.1335 & 0.1119\err{3}{3} & 0.2005\err{12}{11} \\
5.29 & 0.1340 & 0.1340 & 0.1335 & 0.1042\err{3}{3} & 0.1862\err{12}{12} \\
5.29 & 0.1340 & 0.1340 & 0.1340 & 0.0967\err{3}{3} & 0.1721\err{12}{11} \\
5.29 & 0.1340 & 0.1345 & 0.1335 & 0.0966\err{3}{3} & 0.1721\err{12}{12} \\
5.29 & 0.1340 & 0.1345 & 0.1340 & 0.0892\err{3}{3} & 0.1583\err{12}{12} \\
5.29 & 0.1340 & 0.1345 & 0.1345 & 0.0818\err{3}{3} & 0.1447\err{12}{11} \\
5.29 & 0.1340 & 0.1350 & 0.1335 & 0.0891\err{3}{3} & 0.1583\err{12}{12} \\
5.29 & 0.1340 & 0.1350 & 0.1340 & 0.0817\err{3}{3} & 0.1447\err{12}{12} \\
5.29 & 0.1340 & 0.1350 & 0.1345 & 0.0744\err{3}{3} & 0.1314\err{12}{11} \\
5.29 & 0.1340 & 0.1350 & 0.1350 & 0.0671\err{3}{3} & 0.1182\err{12}{11} \\
&&&&&\\
\hline
&&&&&\\
5.93 & 0.0000 & 0.1327 & 0.1327 & 0.07584\err{11}{12} & 0.1260\err{4}{5} \\
5.93 & 0.0000 & 0.1332 & 0.1327 & 0.06795\err{12}{12} & 0.1127\err{4}{5} \\
5.93 & 0.0000 & 0.1332 & 0.1332 & 0.06014\err{11}{12} & 0.0997\err{4}{5} \\
5.93 & 0.0000 & 0.1334 & 0.1327 & 0.06480\err{11}{12} & 0.1075\err{4}{5} \\
5.93 & 0.0000 & 0.1334 & 0.1332 & 0.05702\err{11}{12} & 0.0945\err{4}{5} \\
5.93 & 0.0000 & 0.1334 & 0.1334 & 0.05390\err{12}{12} & 0.0894\err{4}{5} \\
5.93 & 0.0000 & 0.1337 & 0.1337 & 0.04464\err{12}{11} & 0.0754\err{4}{4} \\
5.93 & 0.0000 & 0.1339 & 0.1337 & 0.04151\err{12}{11} & 0.0703\err{4}{4} \\
5.93 & 0.0000 & 0.1339 & 0.1339 & 0.03838\err{13}{12} & 0.0653\err{4}{4} \\
\end{tabular}
\end{center}
\caption{The values of $<r(t)>$ and $<s(t)>$ used to define $m_{\rm PCAC}$,
see eq.(\ref{eq:mpcac}).}
\protect\label{tb:rs}
\end{table}
%}}}

%{{{    table J {tb:J}
%% /research/python/staff/pyca/ukqcd/csw202/tex/J_approach1.px
%% Tue May 22 16:22:41 BST 2001
\begin{table}[*htbp]
\begin{center}
 \begin{tabular}{ccccc}
  & $\beta$ & $\ksea$ & $J$ & \\
 &&&&\\
\hline
 &&&&\\
\multicolumn{5}{c}{\bf First Approach} \\
 &&&&\\
  & 5.2000 & 0.1355 & 0.32\err{ 2}{ 4} & \\
 &&&&\\
\hline
 &&&&\\
  & 5.2000 & 0.1350 & 0.393\err{10}{ 9} & \\
  & 5.2600 & 0.1345 & 0.365\err{ 6}{ 6} & \\
  & 5.2900 & 0.1340 & 0.349\err{ 7}{ 8} & \\
 &&&&\\
\hline
 &&&&\\
  & 5.9300 & 0.0000 & 0.376\err{ 9}{12} & \\
%
%% /research/python/staff/pyca/ukqcd/csw202/tex/../unitary/J_approach2.px
%% Tue May 22 16:22:44 BST 2001
 &&&&\\
\hline
 &&&&\\
\multicolumn{5}{c}{\bf Second Approach} \\
 &&&&\\
  & - & - & 0.35\err{ 2}{ 2} & \\
%
%% /research/python/staff/pyca/ukqcd/csw202/tex/../matched_dyn/J_approach4.px_simp
%% Tue May 22 16:22:49 BST 2001
 &&&&\\
\hline
 &&&&\\
\multicolumn{5}{c}{\bf Third Approach} \\
 &&&&\\
  & - & - & 0.43\err{ 2}{ 2} & \\
\end{tabular}
\end{center}
\caption{$J$ values from the various approaches as described in the text.}
\protect\label{tb:J}
\end{table}
%}}}
%{{{    table aJ {tb:aJ}
%% /research/python/staff/pyca/ukqcd/csw202/tex/table_inva_J.px
%% Tue May 22 16:24:41 BST 2001
\begin{table}[*htbp]
\begin{center}
 \begin{tabular}{ccccc}
  & $\beta$ & $\ksea$ & $a$  [Fermi] & \\
 &&&&\\
\hline
 &&&&\\
  & 5.2000 & 0.1355 & 0.110\err{ 4}{ 3} & \\
 &&&&\\
\hline
 &&&&\\
  & 5.2000 & 0.1350 & 0.115\err{ 3}{ 3} & \\
  & 5.2600 & 0.1345 & 0.118\err{ 2}{ 2} & \\
  & 5.2900 & 0.1340 & 0.116\err{ 3}{ 4} & \\
 &&&&\\
\hline
 &&&&\\
  & 5.9300 & Quenched & 0.1186\err{17}{15} & \\
\end{tabular}
\end{center}
\caption{
Lattice spacing determined from the mesonic sector using the method of
\protect\cite{Allton:1997yv}.}
\protect\label{tb:aJ}
\end{table}
%}}}

%{{{    table M_PS / M_V {tb:rat_mass}

%% /research/python/staff/pyca/ukqcd/csw202/tex/../unitary2/mp:mv_unitary.px
%% Tue May 22 16:24:28 BST 2001
\begin{table}[*htbp]
\begin{center}
 \begin{tabular}{ccc}
  $\beta$ & $\kappa$ & $M_{PS}/M_V$ \\
 &&\\
\hline
 &&\\
  5.2000 & 0.1355 & 0.578\err{13}{19} \\
 &&\\
\hline
 &&\\
  5.2000 & 0.1350 & 0.700\err{12}{10} \\
  5.2600 & 0.1345 & 0.783\err{ 5}{ 5} \\
  5.2900 & 0.1340 & 0.835\err{ 7}{ 7} \\
\end{tabular}
\end{center}
\caption{The ratio $M_{PS}^{\rm unitary} / M_{V}^{\rm unitary}$ for the
dynamical data sets (i.e. with $\kappa \equiv \ksea \equiv \kval$).}
\protect\label{tb:rat_mass}
\end{table}

%}}}

%{{{    table pseudoq
%% /research/python/staff/pyca/ukqcd/csw202/tex/../all/table_mv_vs_mp^2.px
%% Tue May 29 14:59:23 BST 2001
 \begin{table}[*htbp]
 \begin{center}
 \begin{tabular}{lccccc}
hadron  & $\beta$ & $\ksea$ & $A$ & $B$ & \\
 &&&&&\\
\hline
 &&&&&\\
Vector Meson &&&&&\\
 &&&&&\\
  & 5.2000 & 0.1355 & 0.449\err{21}{15} & 0.65\err{ 6}{ 8} & \\
\hline
  & 5.2000 & 0.1350 & 0.457\err{11}{13} & 0.76\err{ 3}{ 3} & \\
  & 5.2600 & 0.1345 & 0.472\err{ 7}{ 8} & 0.69\err{ 2}{ 2} & \\
  & 5.2900 & 0.1340 & 0.470\err{15}{15} & 0.66\err{ 3}{ 3} & \\
\hline
  & 5.9300 & 0.0000 & 0.475\err{ 9}{ 7} & 0.70\err{ 2}{ 3} & \\
 \hline
%% /research/python/staff/pyca/ukqcd/csw202/tex/../all/table_mn_vs_mp^2.px
%% Tue May 29 14:59:32 BST 2001
 &&&&&\\
Nucleon &&&&&\\
 &&&&&\\
  & 5.2000 & 0.1355 & 0.653\err{15}{17} & 1.28\err{ 9}{10} & \\
\hline
  & 5.2000 & 0.1350 & 0.67\err{ 2}{ 2} & 1.32\err{ 6}{ 5} & \\
  & 5.2600 & 0.1345 & 0.72\err{ 2}{ 2} & 1.12\err{ 4}{ 4} & \\
  & 5.2900 & 0.1340 & 0.71\err{ 2}{ 2} & 1.13\err{ 4}{ 4} & \\
\hline
  & 5.9300 & 0.0000 & 0.653\err{12}{12} & 1.28\err{ 4}{ 4} & \\
 \hline
%% /research/python/staff/pyca/ukqcd/csw202/tex/../all/table_mdelta_vs_mp^2.px
%% Tue May 29 15:00:06 BST 2001
 &&&&&\\
Delta &&&&&\\
 &&&&&\\
  & 5.2000 & 0.1355 & 0.77\err{ 4}{ 3} & 1.12\err{14}{14} & \\
 \hline
  & 5.2000 & 0.1350 & 0.81\err{ 3}{ 3} & 1.13\err{ 7}{ 7} & \\
  & 5.2600 & 0.1345 & 0.80\err{ 2}{ 2} & 1.06\err{ 5}{ 5} & \\
  & 5.2900 & 0.1340 & 0.84\err{ 2}{ 2} & 0.95\err{ 4}{ 3} & \\
 \hline
  & 5.9300 & 0.0000 & 0.81\err{ 2}{ 2} & 1.08\err{ 5}{ 6} & \\
 \end{tabular}
 \end{center}
\caption{The fitting parameters for the partially quenched fit of eq.(\ref{eq:pseudoq})
for the hadronic masses.}
\protect\label{tb:pseudoq}
\end{table}
%}}}
%{{{    table kcrit
%% /research/python/staff/pyca/ukqcd/csw202/tex/kc.px_simp
%% Tue May 22 16:24:09 BST 2001
\begin{table}[*htbp]
\begin{center}
 \begin{tabular}{ccc}
$\beta$  & $\ksea$ & $\kcrit$ \\
 &&\\
\hline
 &&\\
5.20 & 0.1355 & 0.13645\err{ 3}{ 3} \\
\hline
5.20 & 0.1350 & 0.13663\err{ 5}{ 6} \\
5.26 & 0.1345 & 0.13709\err{ 3}{ 2} \\
5.29 & 0.1340 & 0.13730\err{ 3}{ 3} \\
\hline
5.93 & quenched & 0.135202\err{11}{11} \\
\end{tabular}
\end{center}
\caption{Values of $\kcrit$ obtained for all the datasets.}
\protect\label{tb:kcrit}
\end{table}
%}}}
%{{{    table unitary

%% /research/python/staff/pyca/ukqcd/csw202/tex/../unitary/mv_vs_mp^2_unitary.px_simp
%% Tue May 22 16:24:39 BST 2001
\begin{table}[*htbp]
\begin{center}
 \begin{tabular}{lccc}
hadron  & $A^{\rm unitary}$ & $B^{\rm unitary}$ & \\
 &&&\\
\hline
 &&&\\
Vector Meson & 0.476\err{14}{18} & 0.66\err{ 6}{ 5} & \\
 &&&\\
%% /research/python/staff/pyca/ukqcd/csw202/tex/../unitary/mn_vs_mp^2_unitary.px_simp
%% Tue May 22 16:24:25 BST 2001
Nucleon & 0.69\err{ 2}{ 3} & 1.20\err{ 9}{ 8} & \\
 &&&\\
%% /research/python/staff/pyca/ukqcd/csw202/tex//research/python/staff/pyca/ukqcd/csw202/tex/../unitary/mdelta_vs_mp^2_unitary.px_simp
%% Tue May 22 16:24:20 BST 2001
Delta  & 0.84\err{ 3}{ 3} & 0.94\err{12}{11} & \\
\end{tabular}
\end{center}
\caption{The fitting parameters for the ``unitary'' dataset fit of eq.(\ref{eq:unitary})
for the hadronic masses.}
\protect\label{tb:unitary}
\end{table}

%}}}
%{{{    table combined
%% /research/python/staff/pyca/ukqcd/csw202/tex/../matched/mv_planar.px_simp
%% Tue May 22 16:24:34 BST 2001
\begin{table}[*htbp]
\begin{center}
 \begin{tabular}{lccccc}
hadron  & $A_0$ & $A_1$ & $B_0$ & $B_1$ & \\
 &&&&&\\
\hline
 &&&&&\\
Vector Meson & 0.492\err{10}{ 9} & -0.004\err{ 2}{ 3} & 0.61\err{ 4}{ 4} &  0.015\err{ 9}{ 7} & \\
 &&&&&\\
%% /research/python/staff/pyca/ukqcd/csw202/tex/../matched/mproton_planar.px_simp
%% Tue May 22 16:24:31 BST 2001
%
Nucleon      & 0.663\err{13}{15} &  0.006\err{ 3}{ 4} & 1.23\err{ 6}{ 6} & -0.001\err{1}{1} & \\
 &&&&&\\
%% /research/python/staff/pyca/ukqcd/csw202/tex/../matched/mdelta_planar.px_simp
%% Tue May 22 16:24:14 BST 2001
%
Delta        & 0.84\err{ 2}{ 2}  & -0.002\err{ 5}{ 5} & 0.91\err{ 8}{ 9} &  0.02\err{ 2}{ 2} & \\
\end{tabular}
\end{center}
\caption{The fitting parameters for the combined fit of eq.(\ref{eq:combined})
for the hadronic masses.}
\protect\label{tb:combined}
\end{table}
%------------------------------------------------------------------------

%--------------------------------------------------------------------------
%}}}

%{{{    figure: various potential figures
%
%
% FIGURES
%
%
%
\begin{figure}
\centerline{\includegraphics[width=\columnwidth]{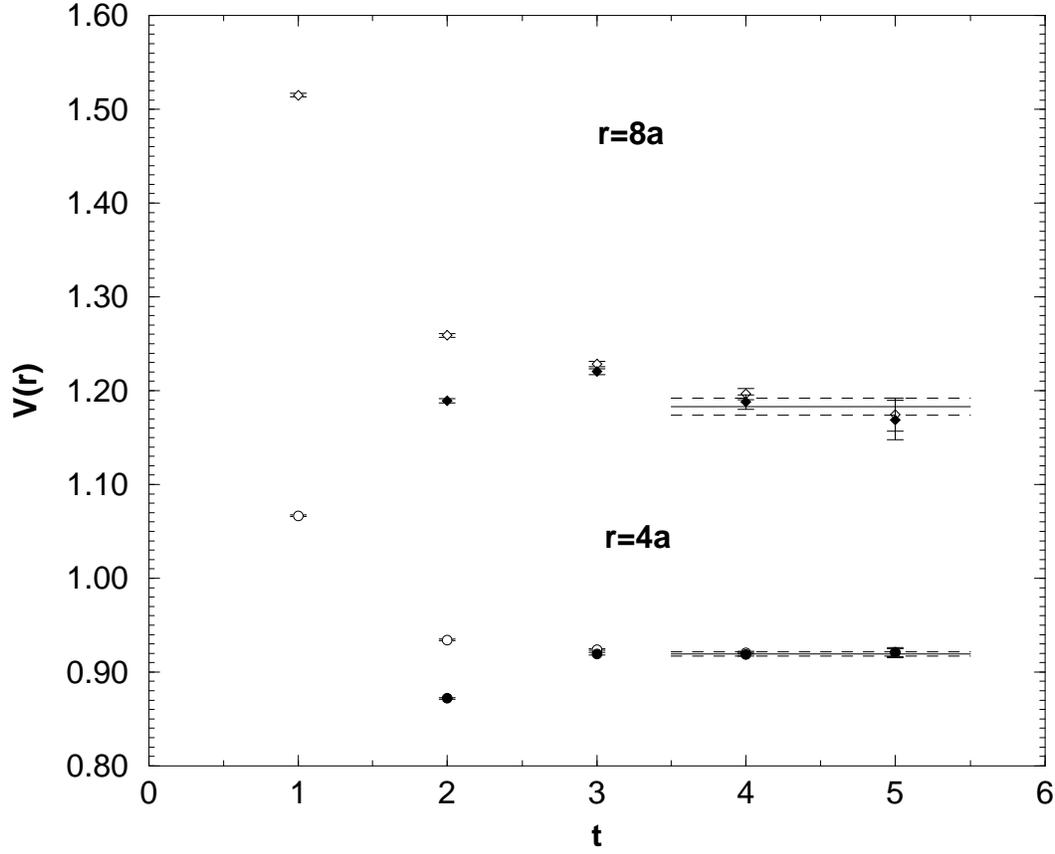}} 
\vspace*{8pt}
\caption{Effective potential energies
as a function of Euclidean time $t$ (open symbols). The
asymptotic estimates described in the text are shown as full symbols.
The final estimated potential $V(r$) is indicated by the lines
with error bands.
The data correspond to $(\beta,\ksea)=(5.20,0.1350)$ and
$\rvec=(4a,0,0)$ (circles) and 
$\rvec=(8a,0,0)$ (diamonds). 
}
\label{fg:Veff}
\end{figure}

%--------------------------------------------------------------------------
% r0V
\begin{figure}
\centerline{\includegraphics[width=\columnwidth]
{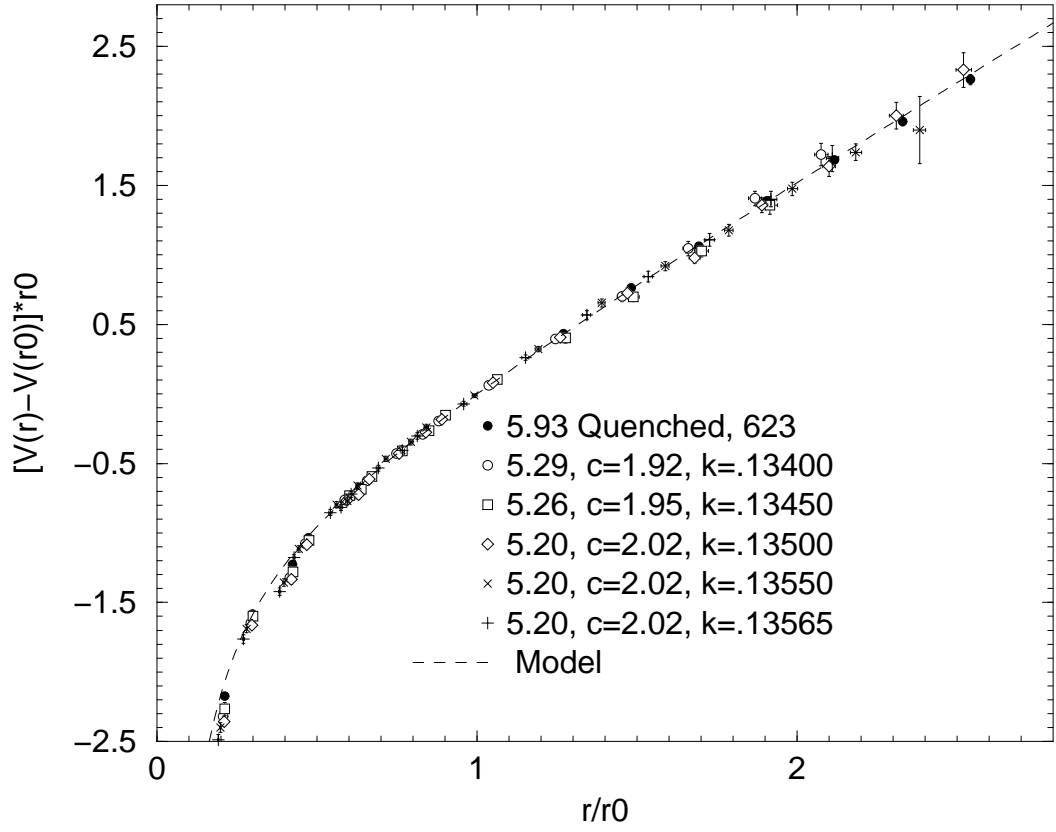}} 
\vspace*{8pt}
\caption{The static QCD potential expressed in units of $r_0$.
The dashed curve is a string model described in the text.
}
\label{fg:r0V}
\end{figure}
%
%------------------------------------------------------------------------
% r0V - string
\begin{figure}
\centerline{\includegraphics[width=\columnwidth]{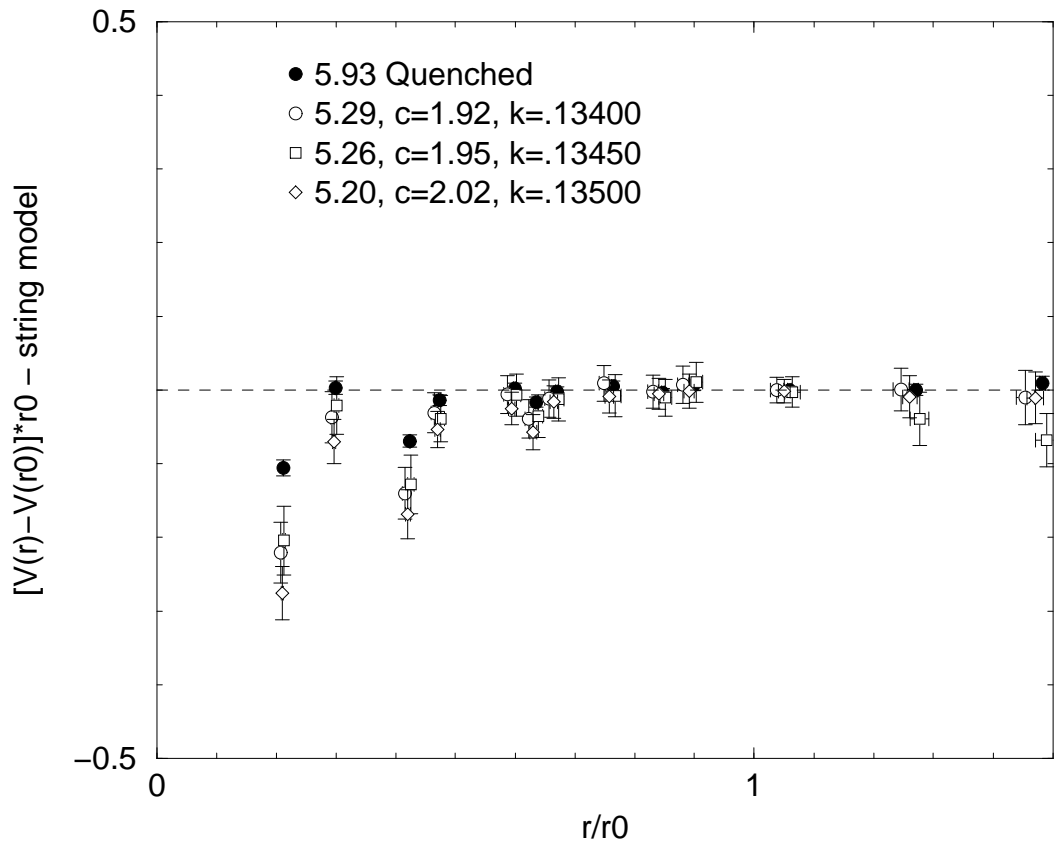}} 
\vspace*{8pt}
\caption{The difference between the static QCD potential expressed in
  physical units and the prediction of the string model described in
  the text. For clarity, only data from the matched ensembles are shown.}
\label{fg:PotDiff}
\end{figure}
%
%------------------------------------------------------------------------
%}}}

%{{{    figure {fig:effm}

\begin{figure}[htbp]
\begin{center}
\setlength{\unitlength}{1cm}
\setlength{\fboxsep}{0cm}
\begin{picture}(14.5,14.5)
\put(0,7.5){\begin{picture}(7,7)\put(-0.9,-0.4){\includegraphics[angle=0,width=3in]
{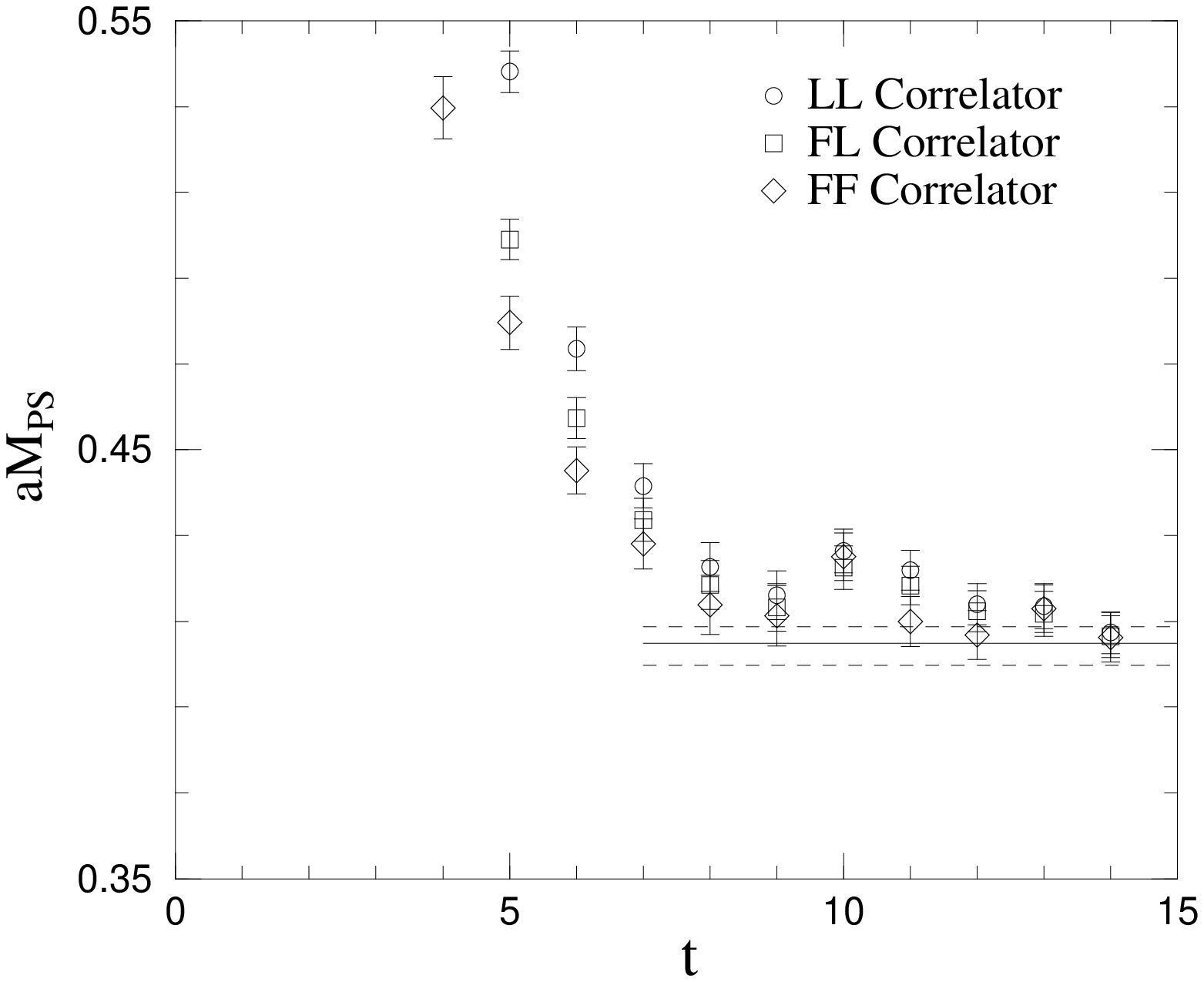}}\end{picture}}
\put(8.0,7.5){\begin{picture}(7,7)\put(-0.9,-0.4){\includegraphics[angle=0,width=3in]
{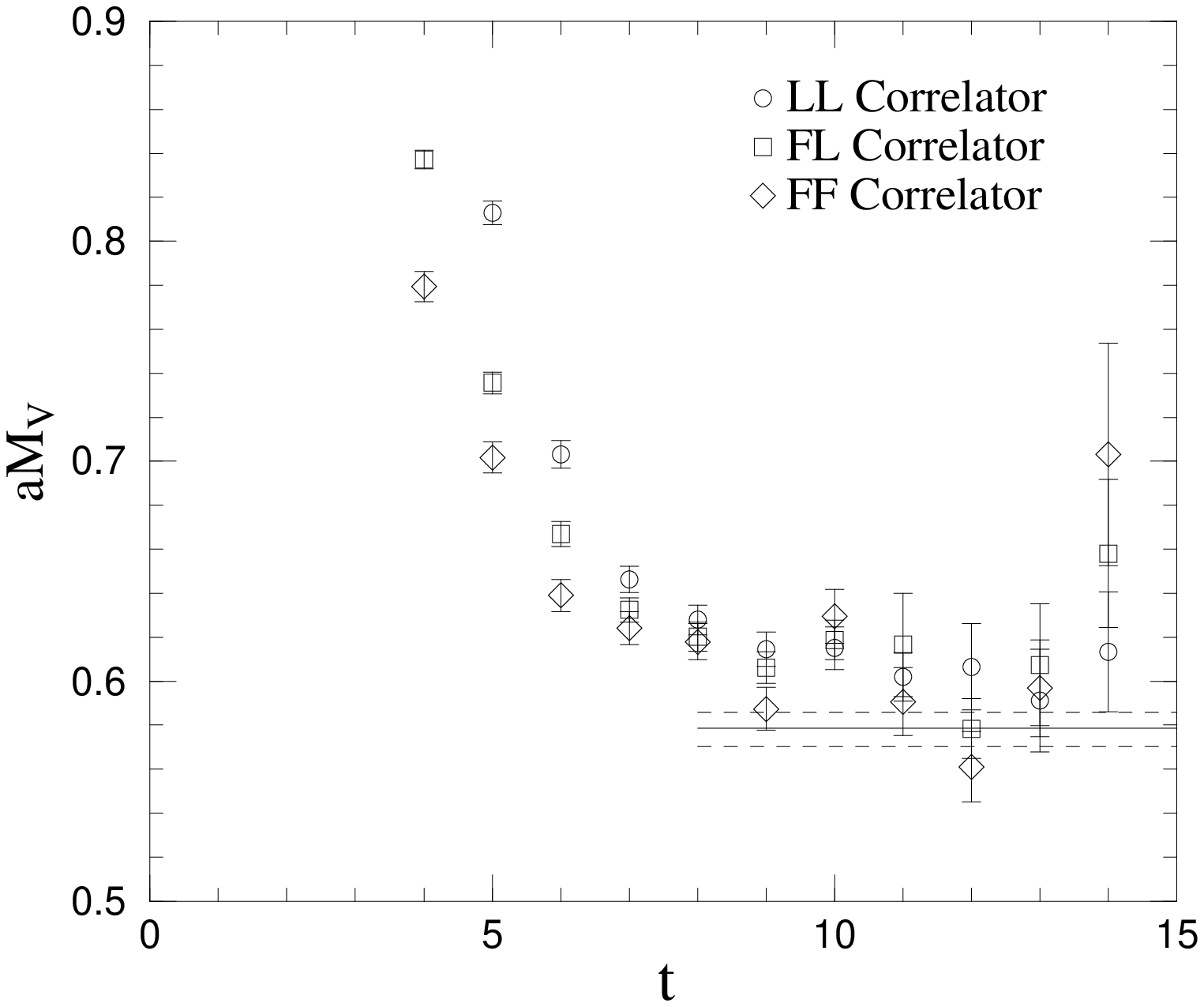}}\end{picture}}
\put(0,1){\begin{picture}(7,7)\put(-0.9,-0.4){\includegraphics[angle=0,width=3in]
{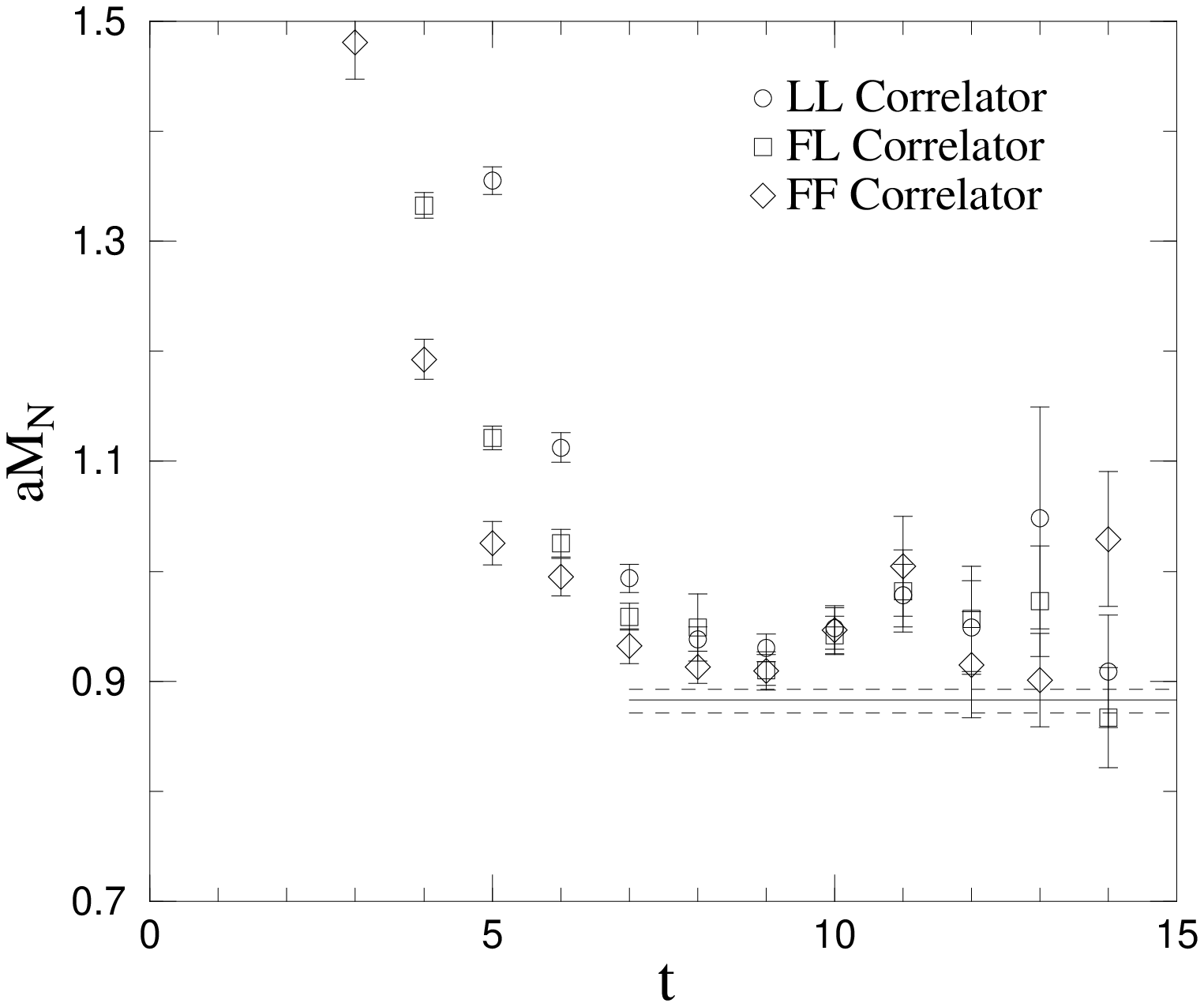}}\end{picture}}
\put(8.0,1){\begin{picture}(7,7)\put(-0.9,-0.4){\includegraphics[angle=0,width=3in]
{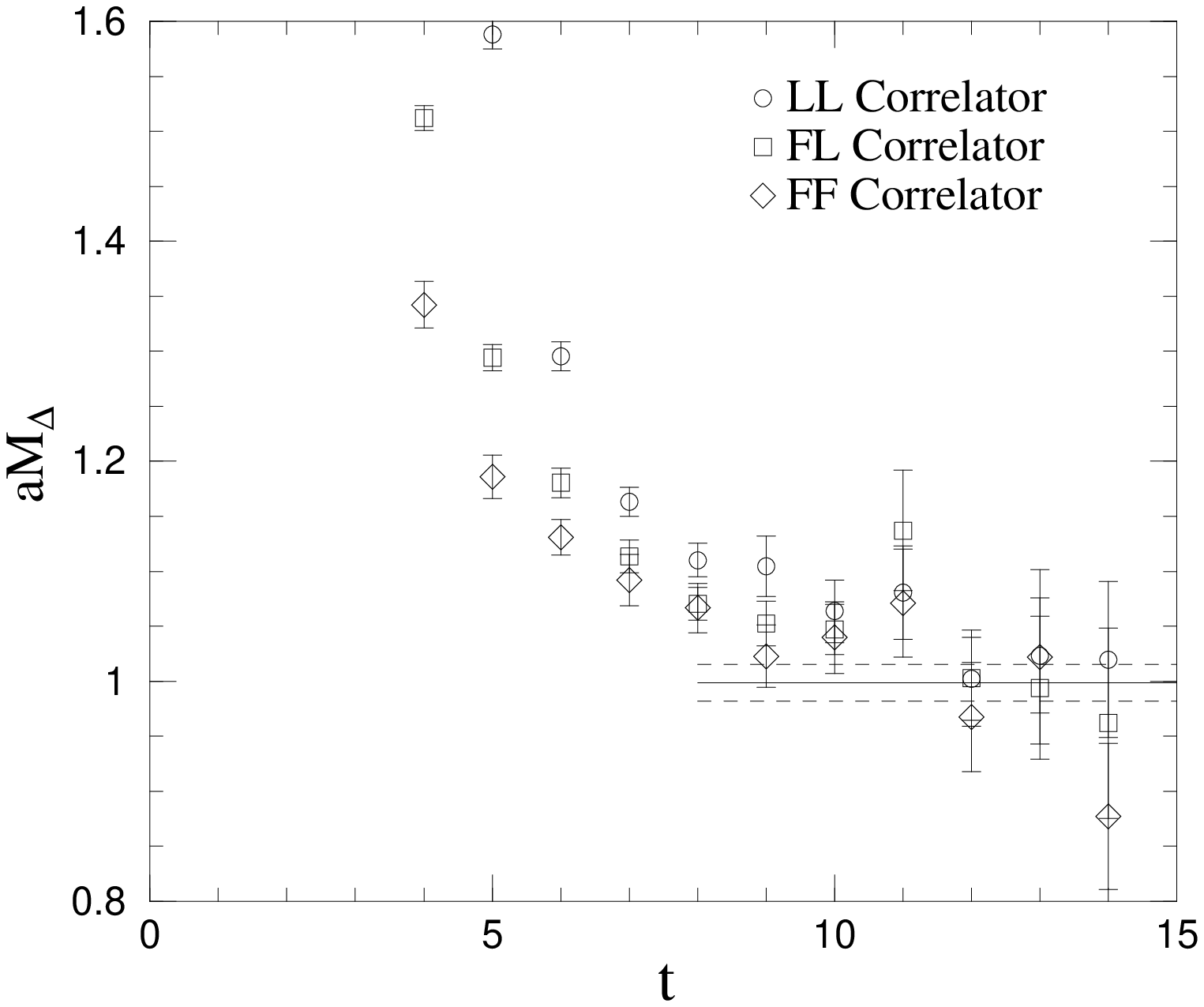}}\end{picture}}
\end{picture}
\end{center}
\caption{Effective mass plots for the pseudoscalar, vector, nucleon and delta
for the $\beta = 5.2$, $\ksea = 0.13500$ data set at $\kval = 0.13500$.
The horizontal lines show the fitted value for the mass (with error
bars) obtained by the fitting approach described in the text.
}
\label{fig:effm}
\end{figure}

%}}}

%{{{    figure {fig:mesons}

\begin{figure}[htp]
\centerline{\includegraphics[angle=0,width=\columnwidth]{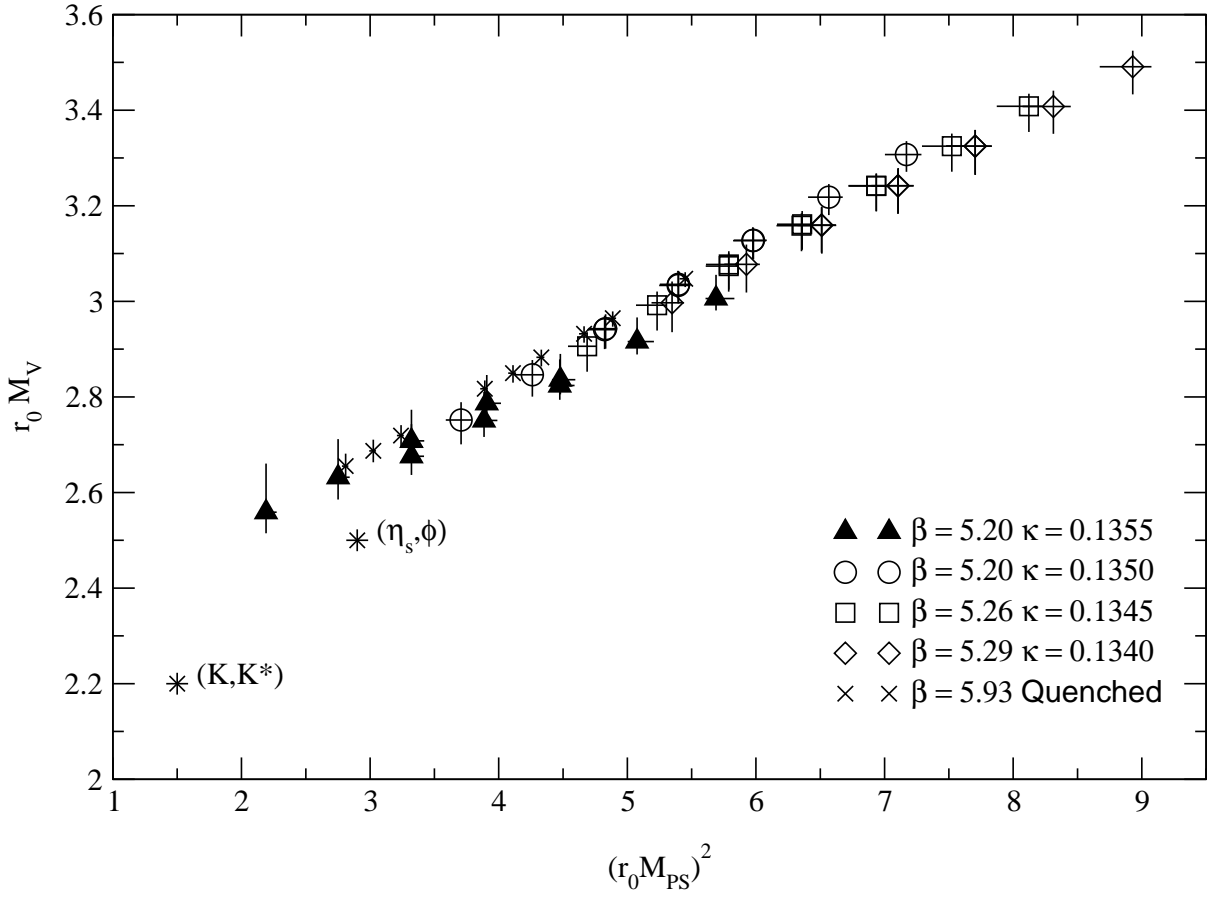}}
\vspace*{5mm}
\caption{
Vector mass plotted against pseudoscalar mass squared in units of $r_0$,
together with the experimental data points.
}
\label{fig:mesons}
\end{figure}

%}}}
%{{{    figure {fig:hyperfine}
\begin{figure}[htbp]
\centerline{\includegraphics[angle=0,width=\columnwidth]{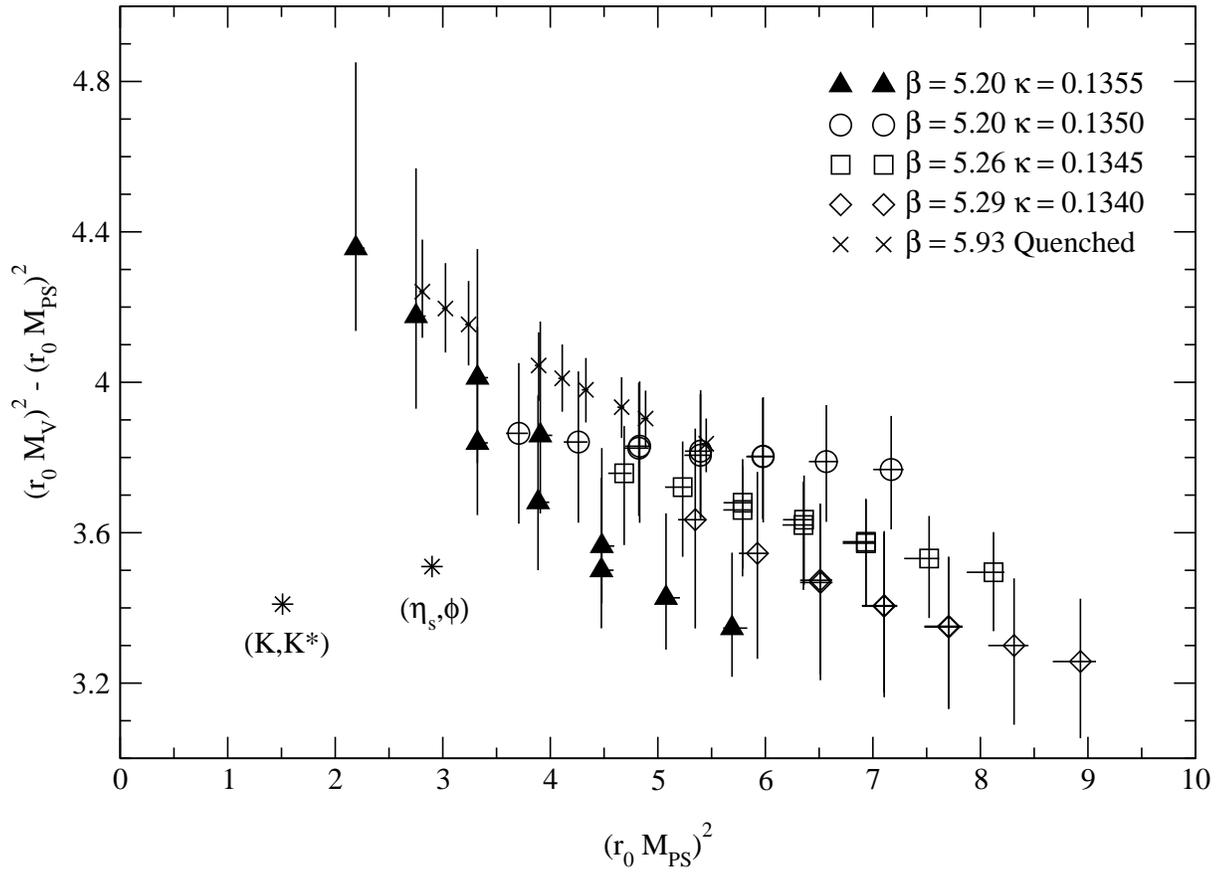}}
\vspace*{5mm}
\caption{
Vector-pseudoscalar hyperfine splitting in units of $r_0$.
}
\label{fig:hyperfine}
\end{figure}
%}}}
%{{{    figure {fig:J}

\begin{figure}
\centerline{\includegraphics[angle=0,width=\columnwidth]{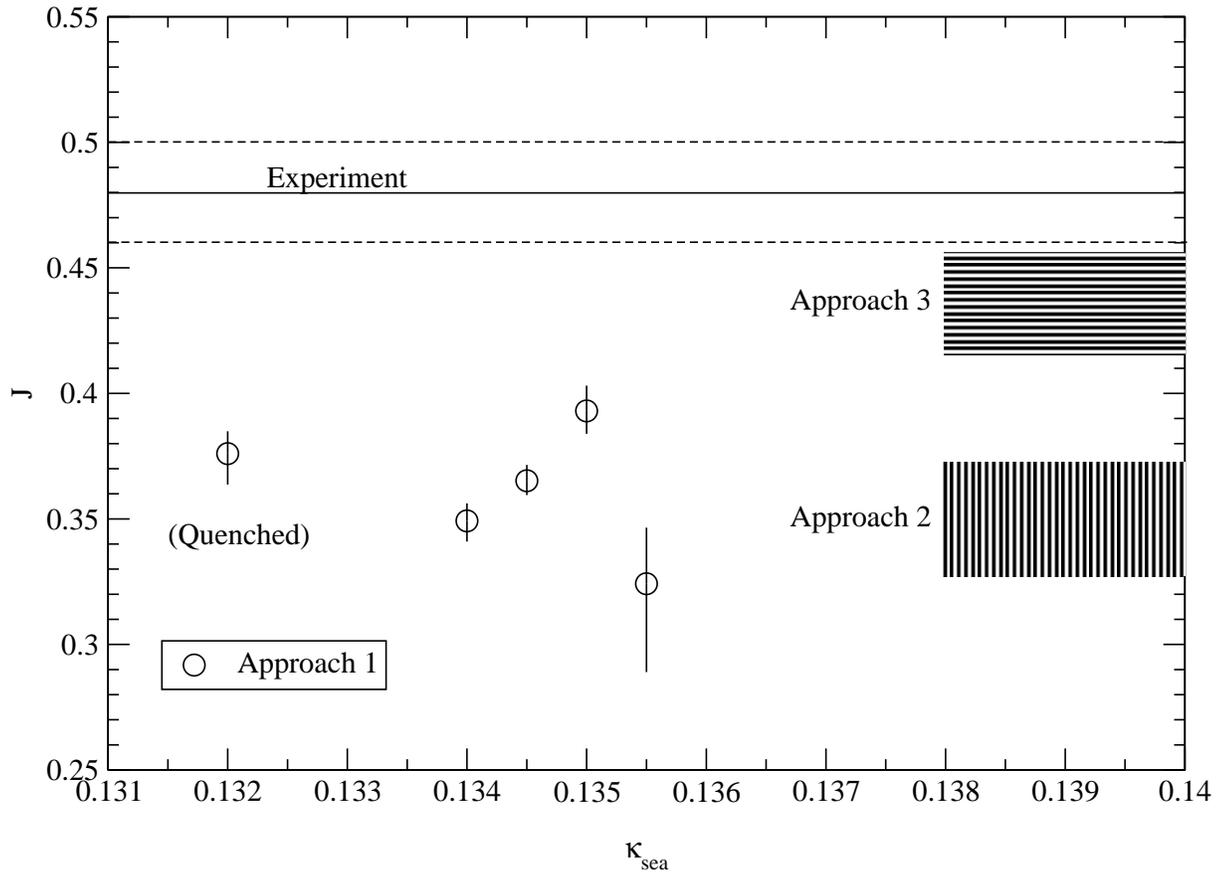}}
\vspace*{5mm}
\caption{
$J$ versus $\kappa^{sea}$ using the approaches as described in the text.
Note that the quenched data points have been plotted at $\ksea=0.132$
for convenience.
Approaches 2 \& 3 are obtained after a chiral extrapolation and are
shown as banded regions.
The experimental value $J = 0.48(2)$ is also shown.
}
\label{fig:J}
\end{figure}

%}}}
%{{{    figure {fig:delta}

\begin{figure}
\centerline{\includegraphics[angle=0,width=\columnwidth]{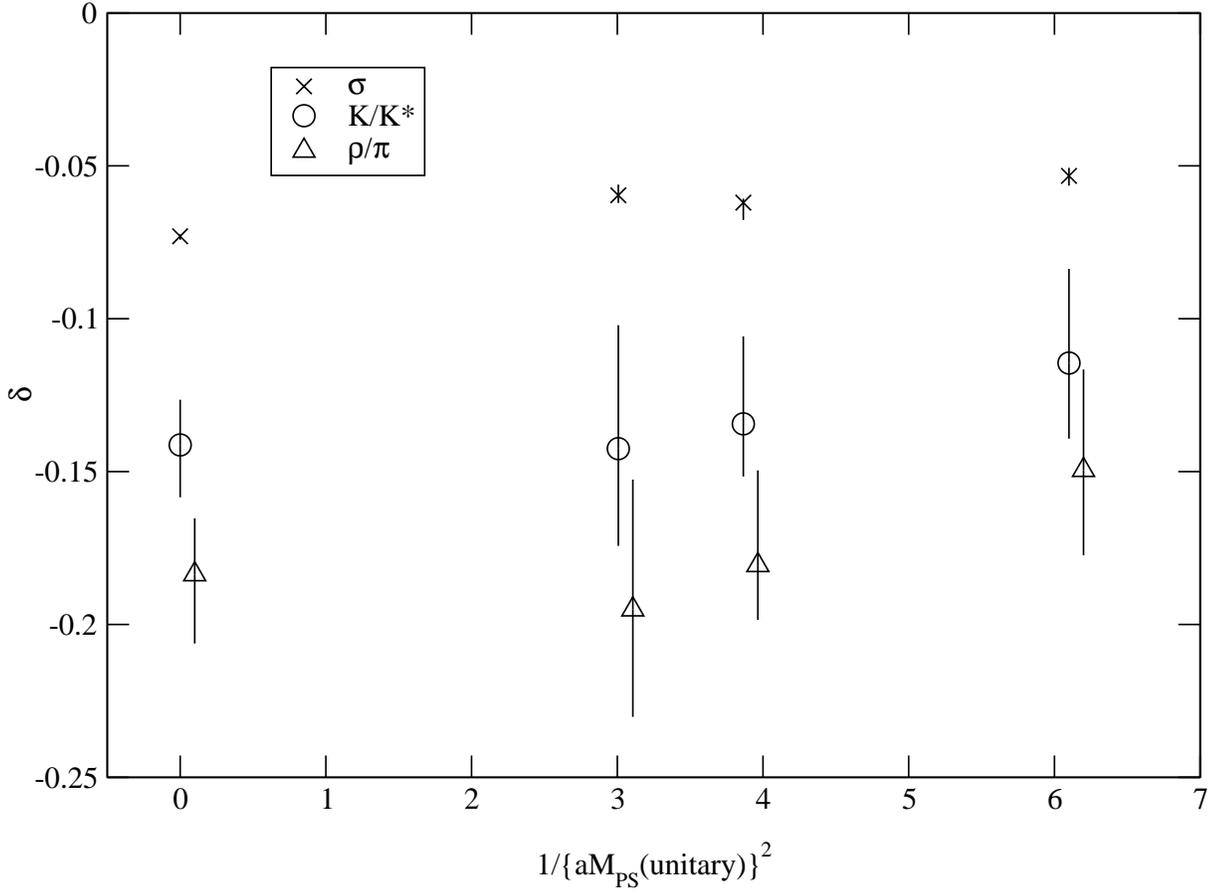}}
\vspace*{5mm}
\caption{
$\delta_i$ as a function of $(1/\hat{M}^{deg}_{PS})^2$ for $i =
\sqrt{\sigma}$ and the mass pairs $(M_K*,M_K)$ \& $(M_\rho,M_\pi)$.
$\delta_i$ is defined in eq.(\ref{eq:delta_ij}) with $j = r_0$.
}
\label{fig:delta}
\end{figure}

%}}}
%{{{    figure {fig:edin}
\begin{figure}
\centerline{\includegraphics[angle=0,width=\columnwidth]{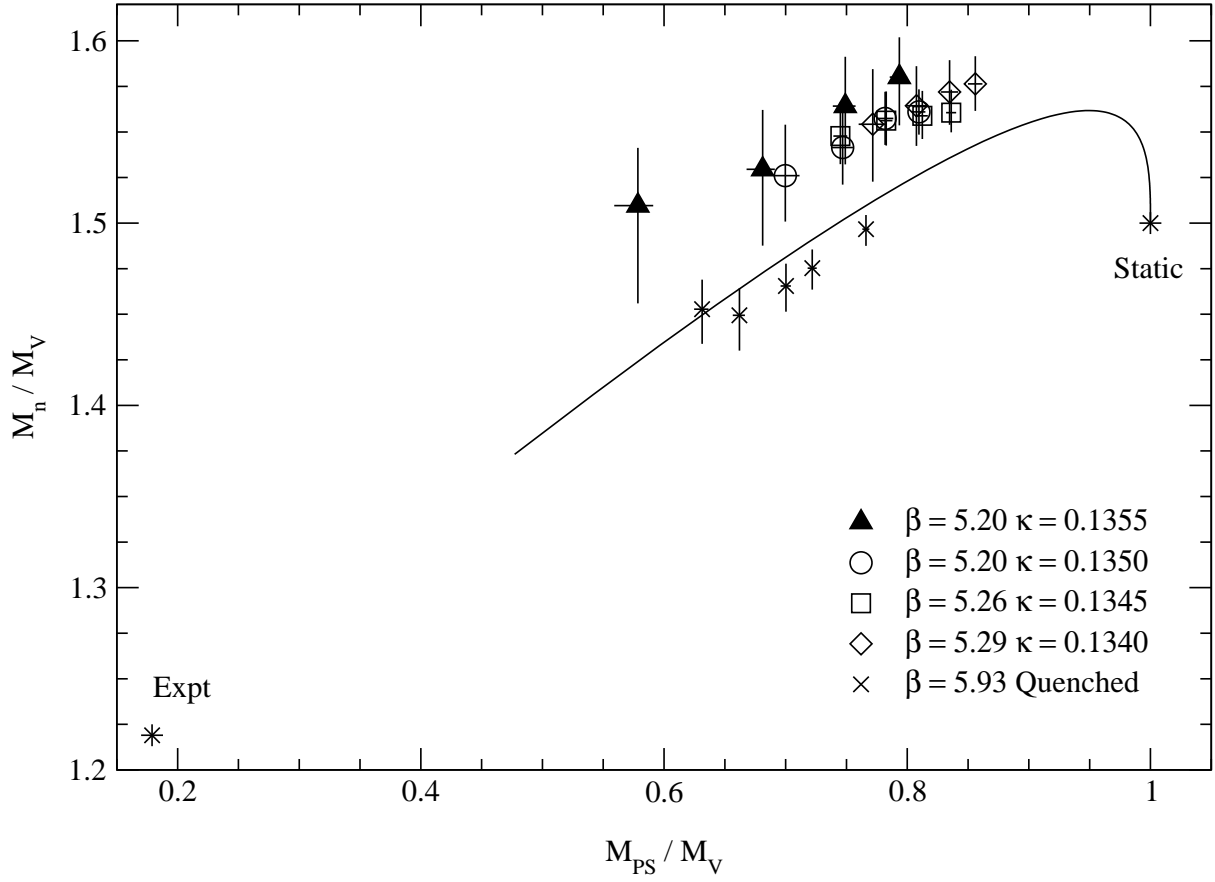}}
\vspace*{5mm}
\caption{
The Edinburgh plot for all the data sets. All degenerate $\kval$
correlators have been included. The phenomenological curve (from
\protect\cite{Ono:1978}) 
has been included as a guide to the eye.
}
\label{fig:edin}
\end{figure}
%}}}

%{{{    figure: various glueball, topo suscep and polyakov line etc
%-------------------------------------------------------------------

%
%
%
%
%
\begin{figure}[tb]
\begin{center}
\leavevmode
\epsfysize=300pt
\epsffile{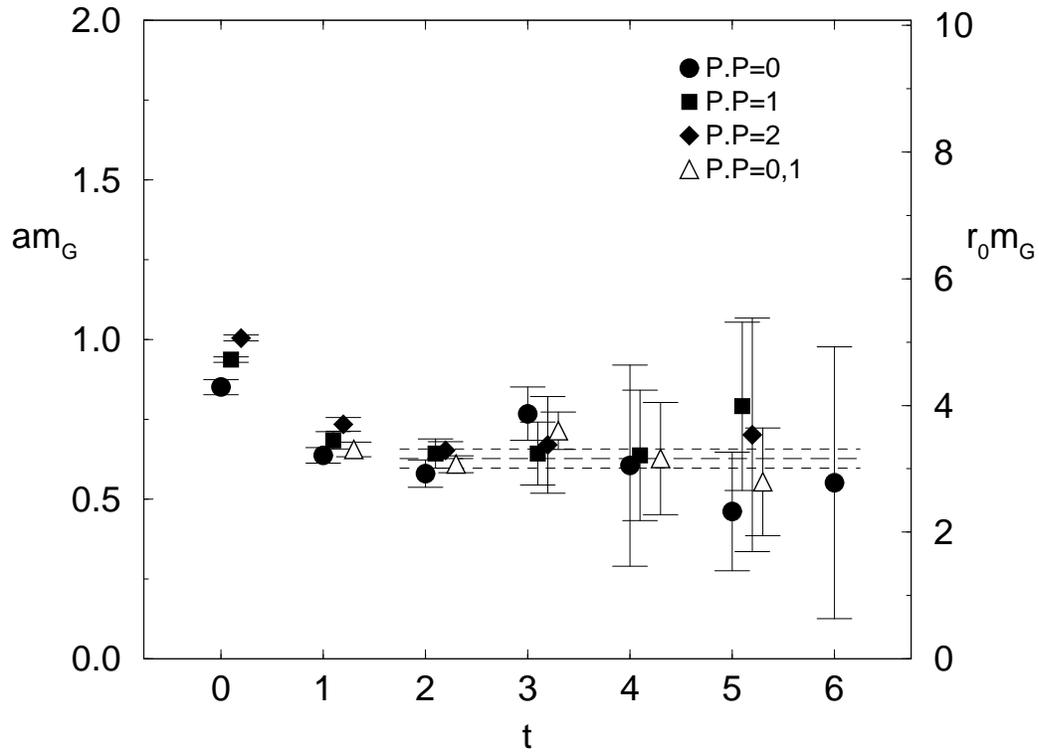}
\end{center}
%\vspace{-1.0cm}
\caption[]{\label{fig_glue_scal}
  Effective masses for the $A_1^{++}$ ground state on the
    $(\beta,\kappa) = (5.20,0.13550)$ ensemble.}
\end{figure}
\begin{figure}[tb]
\begin{center}
\leavevmode
\epsfysize=300pt
\epsffile{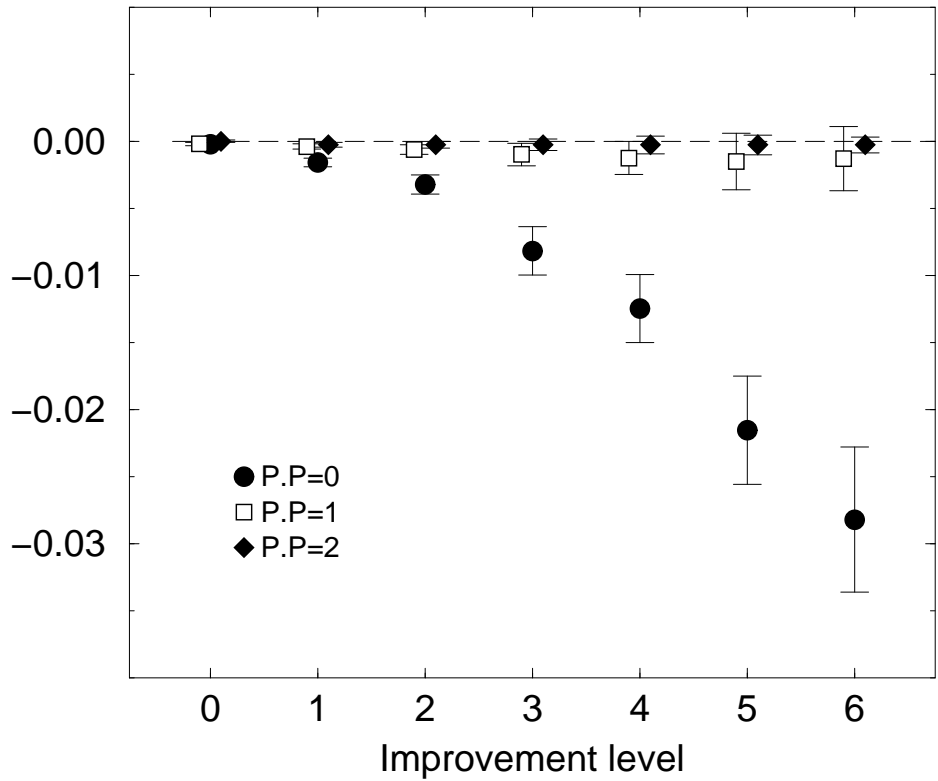}
\end{center}
%\vspace{-1.0cm}
\caption[]{\label{fig_poly_vevs}
  Vacuum expectation values for Polyakov loops at various
    blocking levels on the
    $(\beta,\kappa) = (5.20,0.13550)$ ensemble.}
\end{figure}
%-------------------------------------------------------------------

%
%
%
%
%
%
\begin{figure}[tb]
\begin{center}
\leavevmode
\epsfysize=510pt
\epsfbox[20 30 620 730]{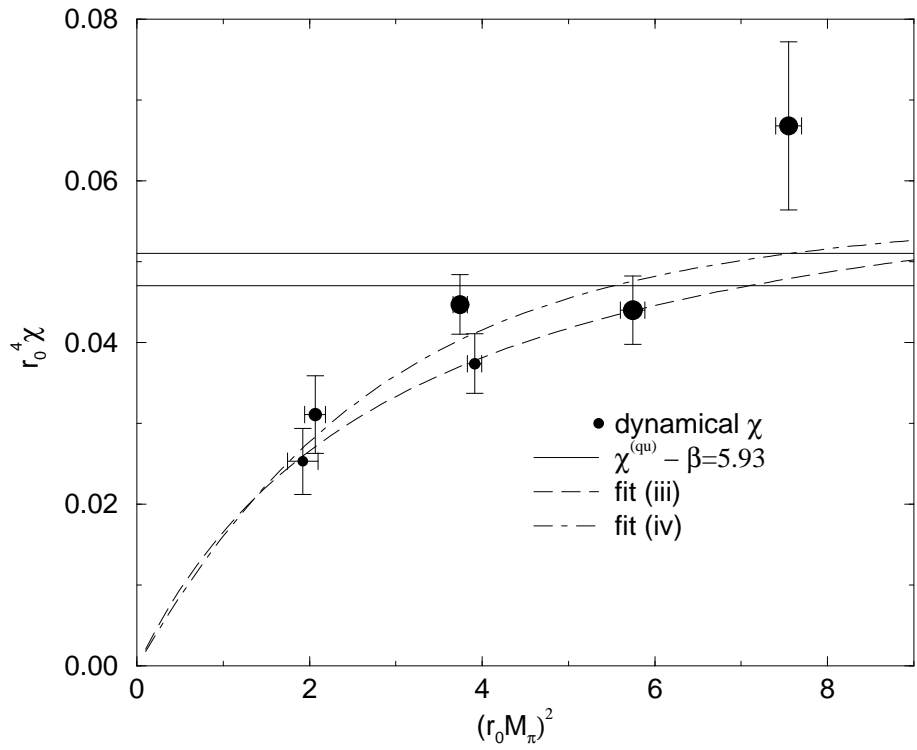}
\end{center}
\caption[]{\label{fig_r04} { The measured topological susceptibility,
    with interpolated quenched points at the same $\hat{r}_0$. The
    radius of the dynamical plotting points is proportional to
    $\hat{r}_0^{-1}$. The fits, independent of the quenched points,
    are: (iii)~Eqn.~(\ref{eqn_fit_nlge}) and
    (iv)~Eqn.~(\ref{eqn_fit_atan}).}}
\end{figure}
\begin{figure}[tb]
\begin{center}
\leavevmode
\epsfysize=510pt
\epsfbox[20 30 620 730]{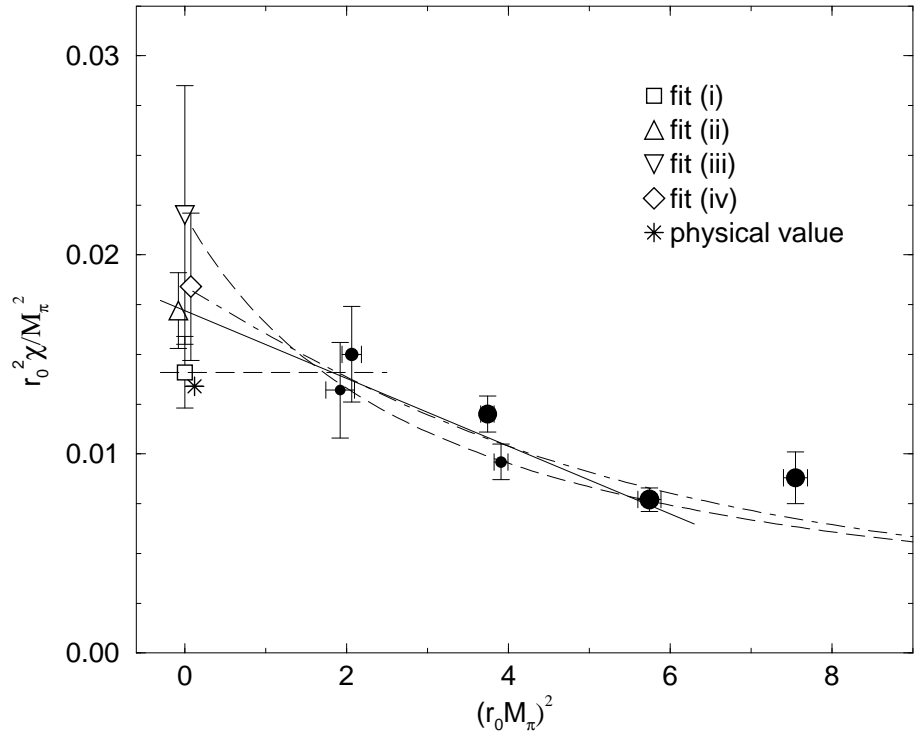}
\end{center}
\caption[]{\label{fig_r02} { The measured topological
    susceptibility. The radius of the dynamical plotting points is
    proportional to $\hat{r}_0^{-1}$. The fits, independent of the
    quenched points, are: (i)~Eqn.~(\ref{eqn_fit_fl}),
    (ii)~Eqn.~(\ref{eqn_fit_fl_lin}), (iii)~Eqn.~(\ref{eqn_fit_nlge})
    and (iv)~Eqn.~(\ref{eqn_fit_atan}).}}
\end{figure}
%----------------------------------------------------------------------

%}}}

\end{document}